\documentclass[useAMS,usenatbib]{mn2e}
\usepackage{longtable}
\usepackage{graphicx}
\usepackage{lscape}
\usepackage{rotating}
\newcommand{\mnras}{MNRAS}
\newcommand{\aj}{ApJ}
\newcommand{\apj}{ApJ}
\newcommand{\apjs}{ApJ}
\newcommand{\apjl}{ApJ}
\newcommand{\aap}{A\&A}

\newcommand{\gppr}{\stackrel{>}{\scriptstyle \sim}}
\newcommand{\gappr}{\raisebox{-0.4ex}{$\gppr$}}
\newcommand{\lppr}{\stackrel{<}{\scriptstyle \sim}}
\newcommand{\lappr}{\raisebox{-0.4ex}{$\lppr$}}

\newcommand{\Teff}{\mbox{$T_{\mathrm{eff}}$}}

\newcommand{\Porb}{\mbox{$P_{\mathrm{orb}}$}}

\newcommand{\Lines}[3]{\Ion{#1}{#2}\,$\lambda\lambda$\,#3}
\newcommand{\Ion}[2]{#1{\,\scriptsize #2}}

\newcommand{\kms}{\mbox{$\mathrm{km\,s^{-1}}$}}

\title[A catalogue of  white dwarf-main sequence binaries]
{Post-common envelope binaries from SDSS - VII: A catalogue of
  white dwarf-main sequence binaries}

\author[A. Rebassa-Mansergas et al.]{A. Rebassa-Mansergas$^{1,2}$,
B. T. G\"ansicke$^2$, M.R.Schreiber$^1$, D.Koester$^3$, 
\newauthor P. Rodr\'{\i}guez-Gil$^{4,5,6}$\\
$^{1}$ Departamento de F\'\i sica y Astronom\'\i a, Universidad de Valpara\'\i so, 
Avenida Gran Bretana 1111, Valpara\'\i so, Chile \\
$^{2}$ Department of Physics, University of Warwick, Coventry CV4 7AL, UK \\
$^{3}$ Institut f\"ur Theoretische Physik und Astrophysik, University of Kiel,
24098 Kiel, Germany\\
$^{4}$ Isaac Newton Group of Telescopes, Apdo. de Correos 321, E-38700, Santa Cruz
de La Palma, Spain\\
$^{5}$ Instituto de Astrof\'\i sica de Canarias, V\'\i a L\'actea, s/n,
La Laguna, E-38205, Tenerife, Spain\\
$^{6}$  Departamento  de  Astrof\'isica,  Universidad  de  La  Laguna,
E-38206, La Laguna, Tenerife, Spain
}

\begin{document}
\date{Accepted 2009. Received 2009; in original form 2009}
\pagerange{\pageref{firstpage}--\pageref{lastpage}} \pubyear{2009}
\maketitle

\begin{abstract}
We  present a  catalogue of  1602 white  dwarf-main  sequence binaries
(WDMS) from the spectroscopic Sloan  Digital Sky Survey Data Release 6
(SDSS DR6).   Among these  we identify 440  as new WDMS  binaries.  We
select WDMS  binary candidates from template fitting  all 1.27 million
DR6  spectra,  using  combined  constraints  in  both  $\chi^{2}$  and
signal-to-noise ratio.  In addition,  we use Galaxy Evolution Explorer
(GALEX) and  UKIRT Infrared Sky  Survey (UKIDSS) magnitudes  to search
for  objects in which  one of  the two  components dominates  the SDSS
spectrum.   We use  a decomposition/fitting  technique to  measure the
effective temperatures, surface gravities, masses and distances to the
white  dwarfs, as  well as  the spectral  types and  distances  to the
companions in our catalogue.   Distributions and density maps obtained
from these stellar parameters are  then used to study both the general
properties  and the  selection effects  of WDMS  binaries in  SDSS.  A
comparison between the distances measured  to the white dwarfs and the
main  sequence  companions  shows  $d_\mathrm{sec}>d_\mathrm{wd}$  for
$\sim$1/5 of  the systems,  a tendency found  already in  our previous
work. The hypothesis that  magnetic activity raises the temperature of
the inter-spot  regions in  active stars that  are heavily  covered by
cool spots,  leading to  a bluer optical  colour compared  to inactive
stars, remains the best explanation  for this behaviour.  We also make
use of SDSS-GALEX-UKIDSS magnitudes to investigate the distribution of
WDMS binaries, as well as their white dwarf effective temperatures and
companion  star  spectral types,  in  ultraviolet  to infrared  colour
space.  We show  that WDMS binaries can be  very efficiently separated
from single main sequence stars and white dwarfs when using a combined
ultraviolet, optical, and infrared colour selection.  Finally, we also
provide  radial   velocities  for  1068  systems   measured  from  the
\Lines{Na}{I}{8183.27,8194.81} absorption doublet and/or the H$\alpha$
emission line.  Among the  systems with multiple SDSS spectroscopy, we
find   five  new  systems   exhibiting  significant   radial  velocity
variations,   identifying    them   as   post-common-envelope   binary
candidates.
\end{abstract}


\label{firstpage}

\section{Introduction}
\label{s-intro}

Binaries  containing  a  white  dwarf  primary plus  a  main  sequence
companion  were initially  main sequence  binaries in  which  the more
massive  star evolved  through  the  giant phase  and  became a  white
dwarf. In  the majority  of cases the  initial separation of  the main
sequence binary is wide enough to allow the evolution of both stars as
if they were single.  A small  fraction is believed to undergo a phase
of dynamically unstable mass transfer once the more massive star is on
the giant  branch or  the asymptotic giant  branch \citep{webbink84-1,
dekool92-1, willems+kolb04-1}.  As a consequence of this mass-transfer
the envelope of the giant will engulf its core and the companion star,
i.e.   the  system is  entering  a  common  envelope phase  \citep[CE,
e.g.][]{livio+soker88-1,      iben+livio93-1,      taam+sandquist00-1,
webbink07-1}.  Friction  inside this envelope causes  a rapid decrease
of  the  binary separation.   Henceforth  orbital  energy and  angular
momentum are extracted from the  binary orbit and lead to the ejection
of the  envelope, exposing a post-common-envelope binary  (PCEB). As a
consequence  the  orbital  period  distribution of  WDMS  binaries  is
clearly bi-modal,  with PCEBs  concentrated at short  orbital periods,
and   wide  WDMS   binaries  (non-PCEBs)   at  long   orbital  periods
\citep{willems+kolb04-1}.  After  the ejection of  the envelope, close
WDMS  binaries  evolve  to  shorter orbital  periods  through  angular
momentum  loss via  magnetic braking  and/or through  the  emission of
gravitational  waves.  WDMS  binaries  include progenitors  of a  wide
range  of  astronomical  objects  such as  cataclysmic  variables  and
super-soft X-ray  sources, with some of those  objects likely evolving
at later stages into type Ia supernova.

Population  synthesis models  have  been developed  for  a variety  of
binary   stars   undergoing   CE   evolution   \citep{dewi+tauris00-1,
nelemans+tout05-1,      politano+weiler06-1,      politano+weiler07-1,
davisetal08-1, davisetal09-1}.  However, the theoretical understanding
of  both  CE  evolution  and  magnetic  braking  is  currently  poorly
constrained   by  observations   \citep{schreiber+gaensicke03-1},  and
progress on this front is most  likely to arise from the analysis of a
large  sample of  PCEBs that  are  well-understood in  terms of  their
stellar  components.   WDMS binaries  appear  most  promising in  that
respect, as  their stellar components  are relatively simple,  and the
SDSS \citep{yorketal00-1, stoughtonetal02-1, abazajianetal09-1} offers
the possibility  to dramatically increase the number  of WDMS binaries
available for detailed follow-up studies.  Up to date there exist four
compilations of SDSS  WDMS binaries, namely \citet{eisensteinetal06-1}
(obtained   as    part   of    their   search   of    white   dwarfs),
\cite{silvestrietal07-1}          (which         contains         also
\citet{raymondetal03-1,silvestrietal06-1} as a subset, and was claimed
to be complete for the SDSS DR5), \citet{augusteijnetal08-1} (obtained
from   SDSS  DR5  using   colour  cuts   plus  proper   motions),  and
\citet{helleretal09-1} (obtained  from a search  for sdB stars  in the
SDSS DR6).

Here we initiate  a comprehensive study to compile  a master sample of
spectroscopic WDMS  binaries from SDSS.  In this  first publication we
make      use       of      the      SDSS       spectroscopic      DR6
\citep{adelman-mccarthyetal08-1}  to create a  catalogue of  1602 WDMS
binaries and candidates that were serendipitously observed.  This list
is  not  only  a superset  of  all  those  previous SDSS  WDMS  binary
compilations, but also includes 440 new WDMS binaries.  In addition we
provide a coherent  analysis of the system parameters  of both stellar
components, as well  as an extension to GALEX/UKIDSS  colours to study
the properties  of SDSS WDMS binaries.   In a parallel  effort, and as
part  of SEGUE  (the  SDSS Extension  for  Galactic Understanding  and
Exploration), some of  us carried out a dedicated  program to identify
$\sim300$    WDMS    binaries    containing    cold    white    dwarfs
\citep{schreiberetal07-1},  a population  clearly  underrepresented in
previous samples of  WDMS binaries.  A detailed analysis  of the SEGUE
population  of  WDMS  is  in  preparation (Schreiber  et  al.   2009).
Finally,  we plan  to summarise  our  efforts by  presenting the  WDMS
binary     content    of     the    final     SDSS     data    release
\citep[DR7,][]{abazajianetal09-1}. This last paper will be accompanied
by a public online data base of WDMS binaries.

The final  master sample of  ``all'' spectroscopic SDSS  WDMS binaries
will  form a  superb database  for  future follow-up  studies of  WDMS
binaries.    In   particular,   analysing   the   fraction   and   the
characteristics  of the  PCEBs  among the  WDMS  binaries may  provide
strong  constraints on  current theories  of compact  binary evolution
\citep{rebassa-mansergasetal07-1,                    schreiberetal08-1,
rebassa-mansergasetal08-1,        pyrzasetal09-1,       nebotetal09-1,
schwopeetal09-1}.

The structure of the paper  is as follows.  In Sect.\,\ref{s-ident} we
present  our method  of identifying  WDMS binaries,  and  estimate the
completeness  of   the  sample  in   Sect.\,\ref{s-completeness}.   In
Sect.\,\ref{s-final} we provide our final catalogue.  Using a spectral
decomposition/model  atmosphere   analysis,  we  derive   white  dwarf
effective  temperatures,  surface  gravities, masses,  companion  star
spectral  types,  and distances  in  Sect.\,\ref{s-param}. We  compare
these  stellar  parameters  to  those  obtained in  other  studies  in
Sect.\,\ref{s-comparison},  and study  the selection  effects  of WDMS
binaries  within  SDSS  in  Sect.\,\ref{s-effects}.  We  provide  then
colour-colour  diagrams and  colour-colour cuts  for WDMS  binaries in
Sect.\,\ref{s-colors}   and  Sect.\,\ref{s-cuts},   respectively.   We
finally measure  the \Lines{Na}{I}{8183.27,8194.81} absorption doublet
and/or the H$\alpha$  emission radial velocities in Sec.\,\ref{s-rvs},
and summarise our results in Sect.\,\ref{s-conc}.

\begin{figure*}
\begin{center}
\includegraphics[angle=-90,width=\textwidth]{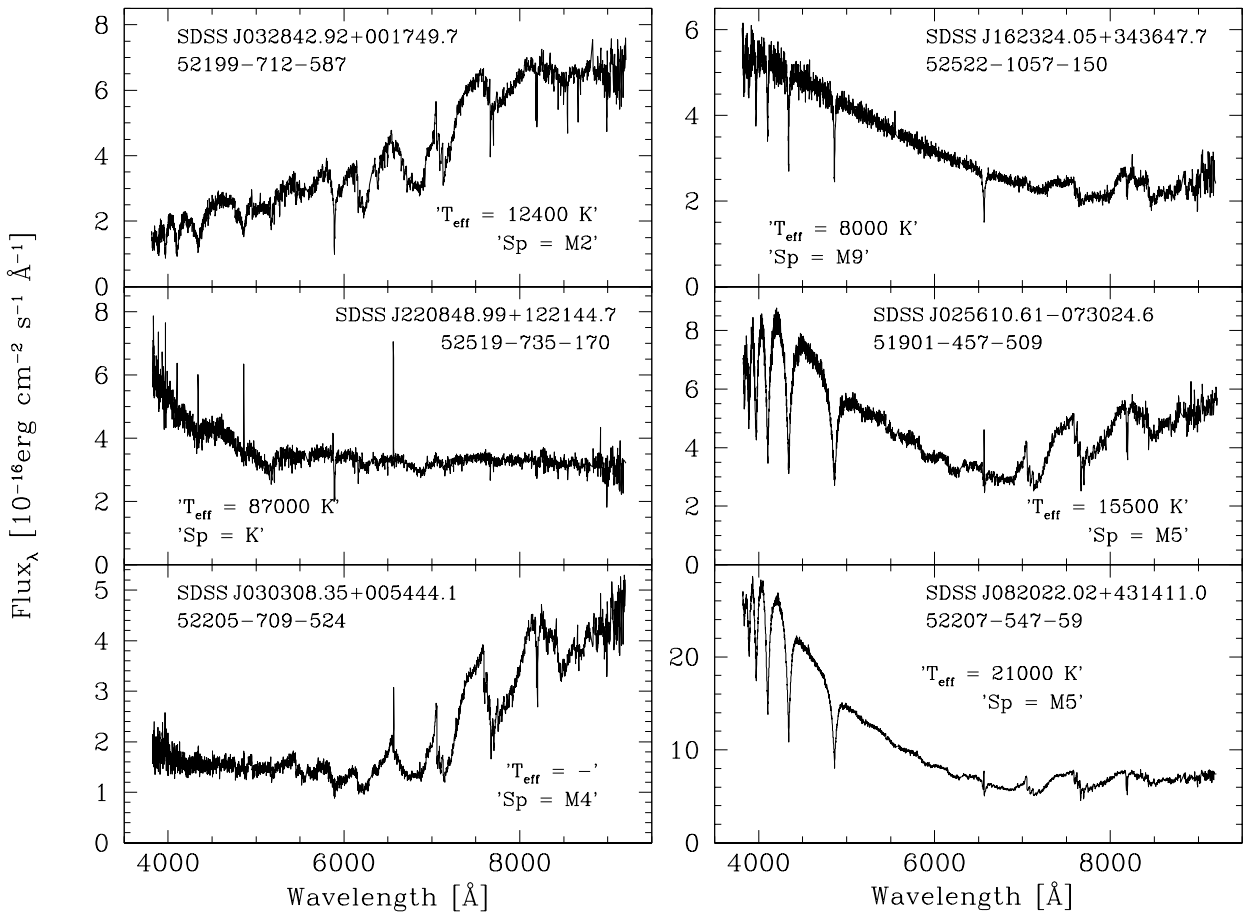}
\caption{Six examples  of previously known WDMS binaries  used in this
work as WDMS binary  templates. SDSS names and MJD-PLT-FIB identifiers
are indicated for each of them. White dwarf effective temperatures and
spectral  type of  the companions  (see Sec.\,\ref{s-param})  are also
indicated in each panel.}
\label{f-templ}
\end{center}
\end{figure*}

\section{Identification of WDMS binaries in SDSS}
\label{s-ident}
\subsection{Computational method}
\label{s-comp}

We have developed a procedure  based on $\chi^{2}$ template fitting in
order to  automatically identify WDMS binary candidates  from the SDSS
DR6  spectroscopic  data  base \citep{adelman-mccarthyetal08-1}.   Our
initial  template  set consisted  of  several  dozen  SDSS spectra  of
confirmed   WDMS    binaries   from   \citet{eisensteinetal06-1}   and
\citet{silvestrietal07-1}. These spectra were chosen to sample a broad
range in  white dwarf  temperatures and subtypes  (DA, DB, and  DC) as
well  as   companion  star   spectral  types,  and   to  be   of  high
signal-to-noise  ratio (S/N).   A set  of representative  templates is
shown in  Fig.\,\ref{f-templ}. In  addition, we compiled  a set  of 17
single      DA     white      dwarf     template      spectra     from
\citeauthor{eisensteinetal06-1}'s       (\citeyear{eisensteinetal06-1})
list, covering  the entire observed range  of \Teff\ and  $\log g$, as
well as the M0-M9 \citet{bochanskietal07-1} M-dwarf templates.

\begin{figure*}
\includegraphics[angle=-90,width=\columnwidth]{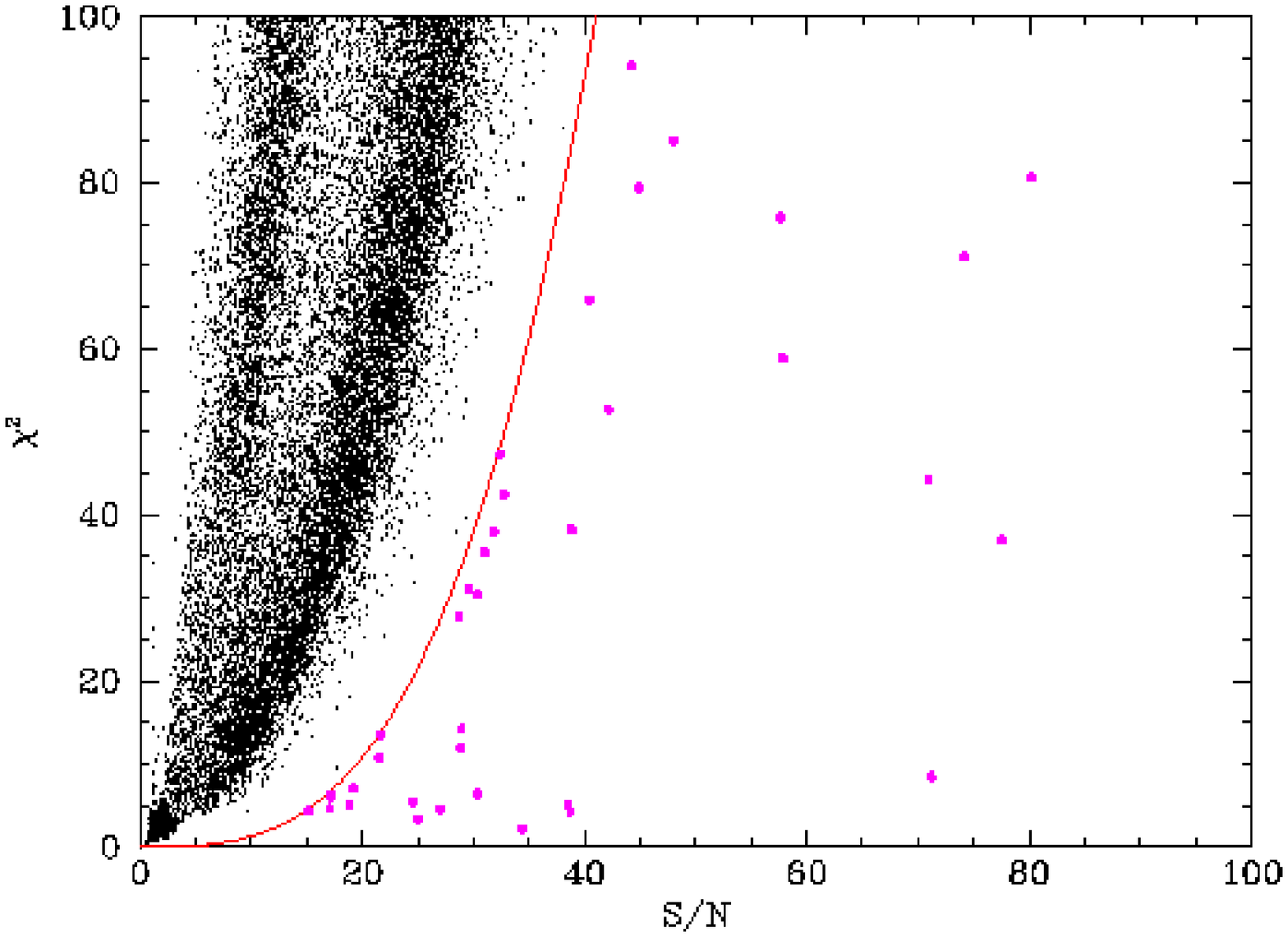}
\includegraphics[angle=-90,width=\columnwidth]{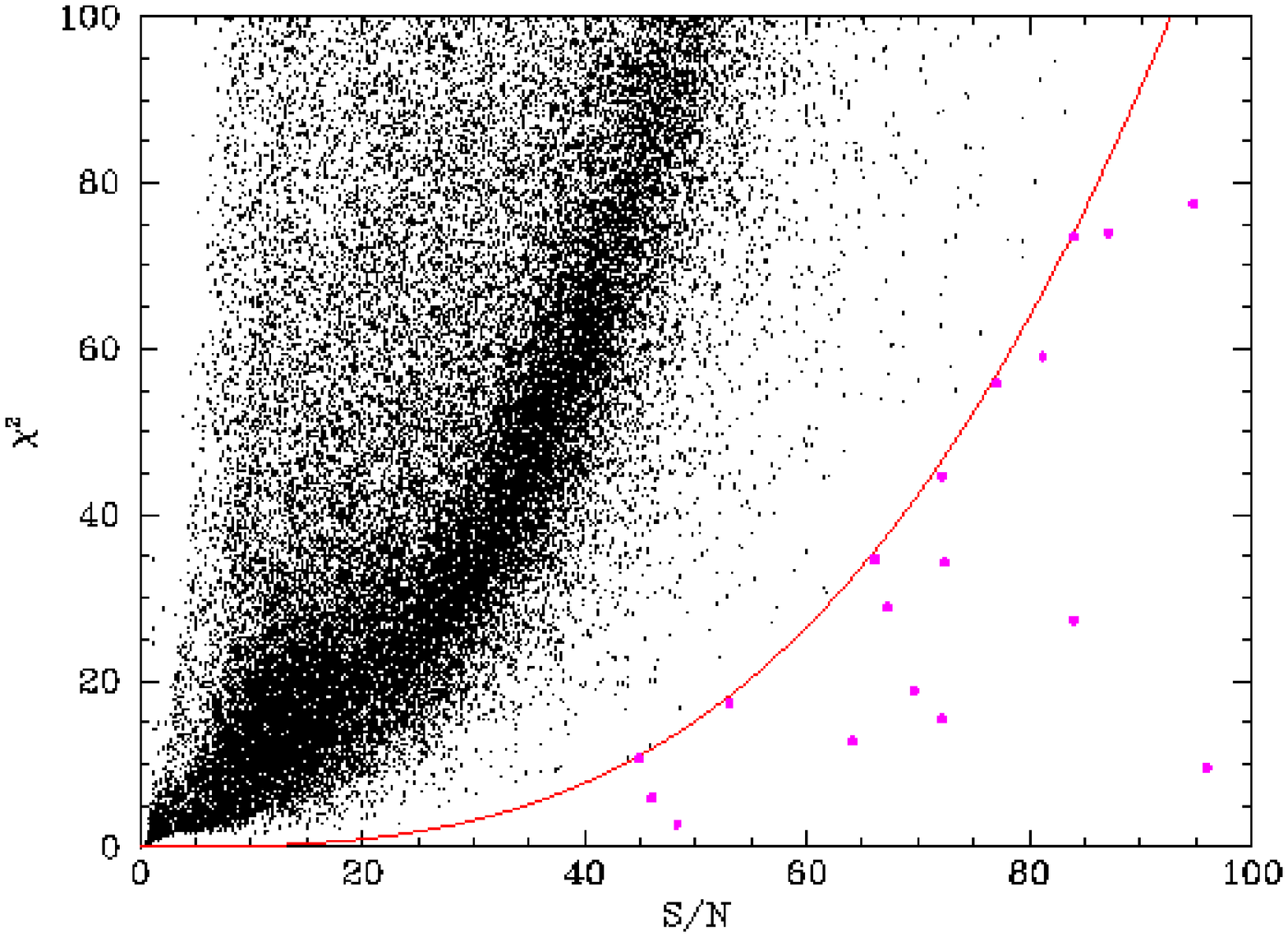}
\caption{$\chi^{2}-\mathrm{S/N}$  plane obtained  fitting  two of  our
WDMS    binary   templates   (SDSS\,J103121.97+202315.1,    left   and
SDSS\,J204431.44-061440.2, right) to the entire SDSS spectra database.
Objects      falling       in      the      area       defined      by
$\chi^2_\mathrm{spec}<a\times\mathrm{S/N}^{b}$  were  considered  WDMS
binary              candidates.               Left              panel:
$\chi^2_\mathrm{spec}<0.001\times\mathrm{S/N}^{3.1}$  .   Right panel:
$\chi^2_\mathrm{spec}<0.0001\times\mathrm{S/N}^{3.05}$.   WDMS  binary
candidates   are    shown   in   magenta,   in    red   the   equation
$\chi^2_\mathrm{spec}<a\times\mathrm{S/N}^{b}$    defined   for   each
template.
\label{f-const}}
\end{figure*}

Each of  these WDMS  binary, white dwarf,  and M-dwarf  templates were
then fitted to the full 1.27  million spectra in DR6. In this process,
the  template  spectrum was  normalised  to  the  SDSS spectrum  under
scrutiny, and a reduced $\chi^2$  was calculated using the flux errors
of  the two  spectra added  in quadrature.   In practice,  our fitting
procedure  produced for  each of  the  WDMS binary,  white dwarf,  and
M-dwarf templates a list  of spectrum identifier (MJD-PLT-FIB), S/N of
the spectrum, and $\chi^2$ for all  SDSS DR6 spectra.  For each of the
templates,  we plotted $\chi^2$  as a  function of  S/N of  the target
spectrum (see  Fig\,\ref{f-const}), and  defined a power  law relation
$\chi^2_\mathrm{max}=a\times\mathrm{S/N}^{b}$.    We   considered  any
spectrum with
\begin{equation}
\chi^2_\mathrm{spec}<\chi^2_\mathrm{max}
\end{equation}
as a candidate WDMS binary,  white dwarf, or M-dwarf (depending on the
current  template).  The  S/N-dependent form  of $\chi^2_\mathrm{max}$
accounts for the increase of  $\chi^2$ for higher values of S/N.  This
constraint had to  be defined individually for each  of the templates,
as  the  different spectral  shapes  resulted  in  a large  spread  of
$\chi^2$  distributions.   The ($\chi^2$,  S/N)  planes obtained  from
fitting  the SDSS  spectra are  shown  for two  different WDMS  binary
templates in Fig.\,\ref{f-const}.

After  a  first run  through  the  DR6  spectra, we  complemented  the
template set  with the  spectra of a  number of newly  identified WDMS
binaries, and re-run  the fitting for those new  templates again. That
process  was  repeated  until  no  new  WDMS  binary  candidates  were
found~--~at  which point we  had used  a total  of 163  different WDMS
binary template spectra.

Even  though  the  above  method efficiently  identifies  WDMS  binary
candidates    among   the    spectra   in    DR6,   the    choice   of
$\chi_\mathrm{max}^{2}$ alone does  not avoid completely the filtering
of other  astronomical objects, such as quasars,  main sequence stars,
and galaxies.   In addition, for  templates that are dominated  by the
white dwarf (M-dwarf), the list of candidates will unavoidably contain
a  substantial number  of single  white dwarfs  (M-dwarfs).   Hence we
first visually  inspected all WDMS  binary candidates, as well  as the
white  dwarf and  M-type  star  subsamples (a  total  of $\sim$  70000
spectra), and  removed those  objects that were  not of  our interest,
i.e.  neither  WDMS binary, white  dwarf, or M-dwarf  candidates.  The
final result of  the template fitting were a list  of 1491 WDMS binary
candidates,  8368  single white  dwarf  candidates,  and 15379  single
M-dwarf candidates.  It is worth mentioning that we found here 36 WDMS
binaries  that were  observed  only/first by  SEGUE, and  consequently
decided to include them in  the SEGUE list of WDMS binaries (Schreiber
et al. 2009, in preparation).

\subsection{Red and blue excess in SDSS spectra: help from GALEX and UKIDSS}
\label{s-excess}

While the template  fitting proved to be a robust  method to find WDMS
binaries in  which both stellar components  contribute clearly visible
amounts of  flux, the procedure  is prone to  mis-classify white dwarf
dominated WDMS binaries as  single white dwarfs, and M-dwarf dominated
WDMS binaries as  single M-dwarfs. We therefore decided  to probe more
specifically for the presence of excess  flux at the red (blue) end of
the  SDSS spectra  in  objects classified  initially  as single  white
dwarfs (M-dwarfs).

For the search of red flux excess in single white dwarf candidates, we
fitted synthetic white dwarf  spectra computed with the code described
by \citet{koesteretal05-1}  to the  SDSS spectra, and  then calculated
the  reduced $\chi^2$  over the  wavelength  ranges $4000-7000$\,\AA\,
($\chi_\mathrm{b}^2$)  and  $7000-9000$\,\AA\,  ($\chi_\mathrm{r}^2$).
Objects   with   $\chi_\mathrm{r}^2/\chi_\mathrm{b}^2   >$  1.5   were
``promoted''  from  single  white  dwarf  candidates  to  WDMS  binary
candidates.

The search for blue flux  excess proceeded in an analogous fashion for
the single M-dwarf  candidates, only that we used the  set of high S/N
M-dwarf  templates from  \citet{rebassa-mansergasetal07-1}  instead of
model spectra, and calculated the reduced $\chi^2$ over the wavelength
ranges   $4000-5000$\,\AA\,   and   $7000-9000$\,\AA.   Objects   with
$\chi_\mathrm{b}^2/\chi_\mathrm{r}^2  >  1.5$  were ``promoted''  from
single M-dwarf candidates to WDMS binary candidates.

\begin{figure*}
\begin{center}
\includegraphics[width=\columnwidth]{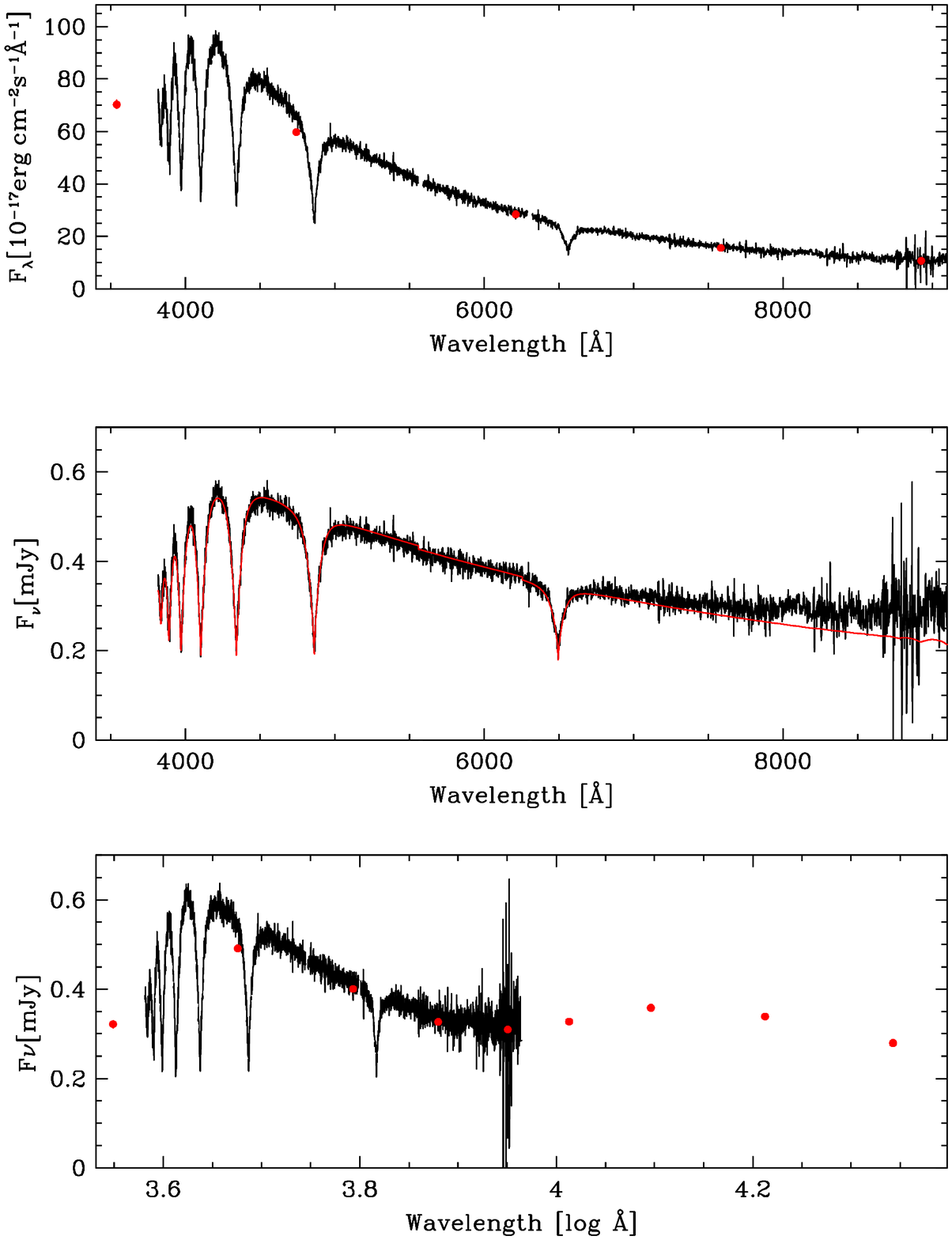}
\includegraphics[width=\columnwidth]{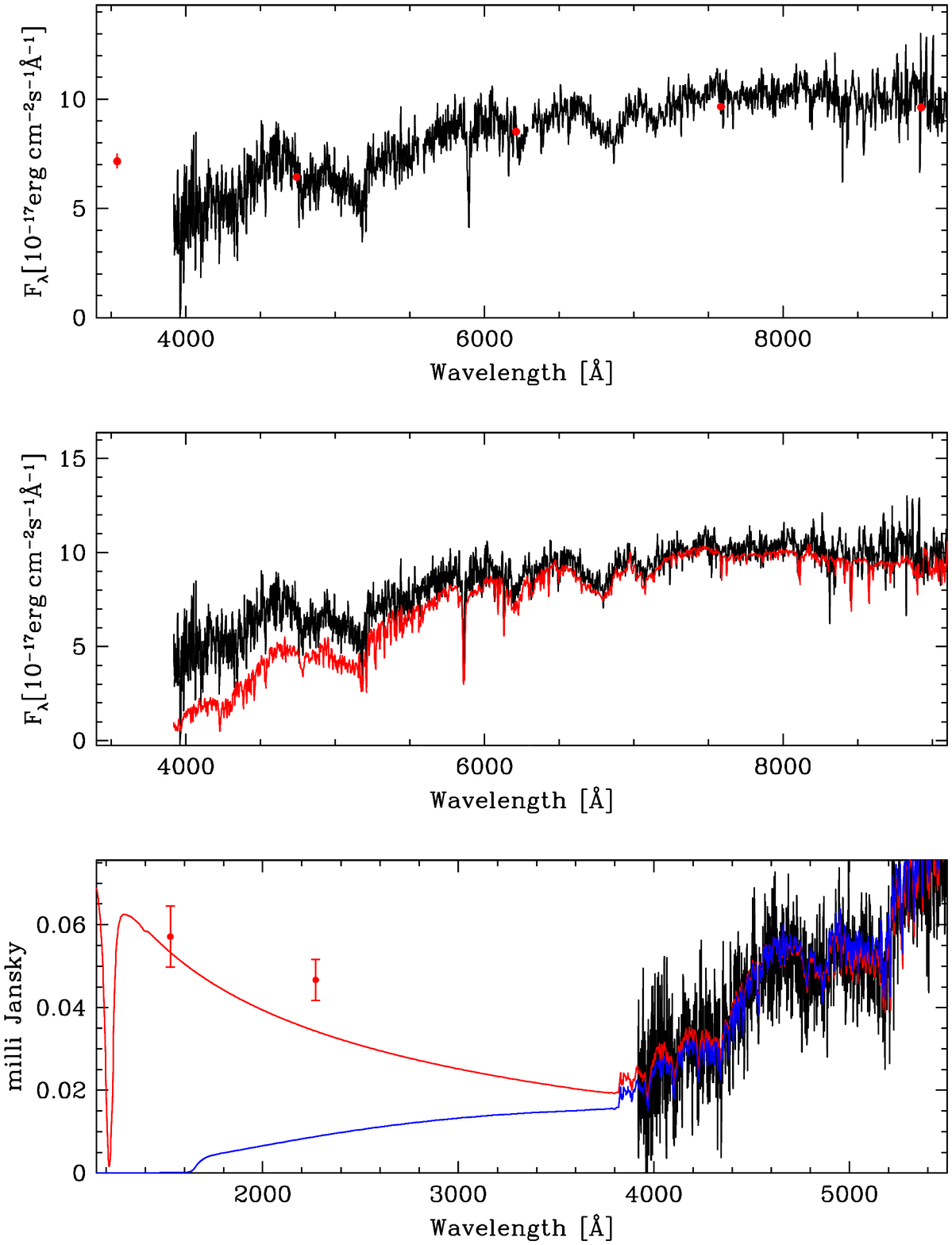}
\caption{\label{f-excessIR}    Top     left:    SDSS    spectrum    of
SDSSJ\,132925.21+123025.5,   a   WDMS   binary   candidate   initially
catalogued as single DA white  dwarf.  The red dots represent the SDSS
magnitudes.   Middle left  panel: the  best white  dwarf model  fit is
superimposed in  red, unambiguously identifying the red  excess of the
binary. Bottom left panel:  SDSS and UKIDSS magnitudes superimposed to
the  SDSS spectrum.   Again, the  UKIDSS magnitudes  clearly  show the
presence  of a  low-mass companion.   Top  right panel:  the same  for
SDSSJ\,131928.80+580634.2, an initially  catalogued early M-type star.
Middle  and  bottom  right  panels:   the  best  M-type  fit  and  the
ultraviolet GALEX  magnitudes clearly confirm the presence  of a white
dwarf  primary,  respectively.   The   red  and  blue  straight  lines
represent the  white dwarf  solutions (red for  the hot, blue  for the
cold)   obtained   from    decomposing/fitting   the   spectrum   (see
Sect\,\ref{s-param}).}
\label{f-excess}
\end{center}
\end{figure*}

On the left  and right top panels of  Fig.\,\ref{f-excess} we show the
SDSS  spectra   (black  line)  and  SDSS  magnitudes   (red  dots)  of
SDSSJ\,132925.21+123025.5  and  SDSSJ\,131928.80+580634.2, along  with
the best-fit white dwarf model and M-dwarf template (red lines, middle
panels).  These two objects  were initially classified by our template
fitting procedure as single white  dwarf and single M star candidates,
respectively, but  ``promoted'' to WDMS binary candidates  by the flux
excess measurement as described above. The flux excess is more obvious
when plotting $F_\nu$ (middle  left panel) instead of $F_\lambda$ (top
left panel). However,  in several cases, the detection  of blue or red
flux excess is rather marginal.

As  a   final  step  in  our   search  for  WDMS   binaries,  we  have
cross-correlated our entire list  of WDMS binary candidates with GALEX
\citep{martinetal05-1,  morrisseyetal05-1} DR\,4, providing  near- and
far-ultraviolet  ($nuv$,  $fuv$) magnitudes,  and  with  the DR\,4  of
UKIDSS    \citep{dyeetal06-1,    hewettetal06-1,    lawrenceetal07-1},
providing infrared  $yJHK$ magnitudes. We then  inspected the observed
ultraviolet-optical spectral energy distribution of all secondary star
dominated  WDMS binary candidates,  and the  optical-infrared spectral
energy   distribution  of  all   white-dwarf  dominated   WDMS  binary
candidates.  Objects where a  clear ultraviolet or infrared excess was
detected were then included in our WDMS binary sample.

For  SDSSJ\,132925.21+123025.5,  the  UKIDSS magnitudes  unambiguously
confirm the  existence of a  low-mass companion (bottom left  panel of
Fig.\,\ref{f-excess}).  Similarly,  the  ultraviolet GALEX  magnitudes
clearly   confirm  the   presence  of   a  white   dwarf   primary  in
SDSSJ\,131928.80+580634.2       (bottom      right       panel      of
Fig.\,\ref{f-excess}).

\subsection{SDSS images}
\label{s-images}

As  a final  check on  the nature  of the  WDMS binary  candidates, we
inspected their SDSS DR6  images for morphological problems, and found
primarily two types of issues.

\begin{figure*}
\begin{center}
\hspace*{5ex}
\includegraphics[width=0.62\columnwidth]{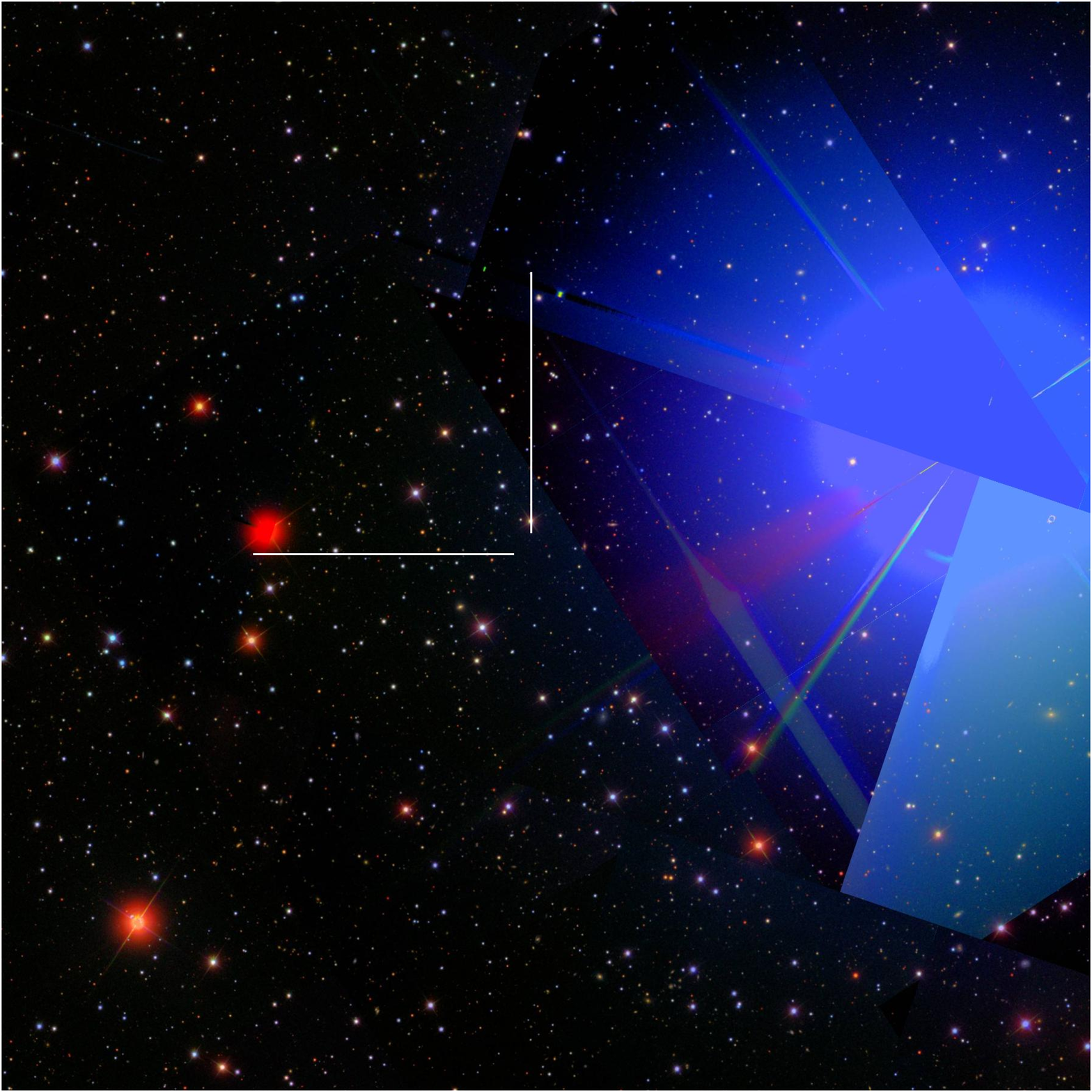}
\hspace*{0.5ex}
\includegraphics[width=0.62\columnwidth]{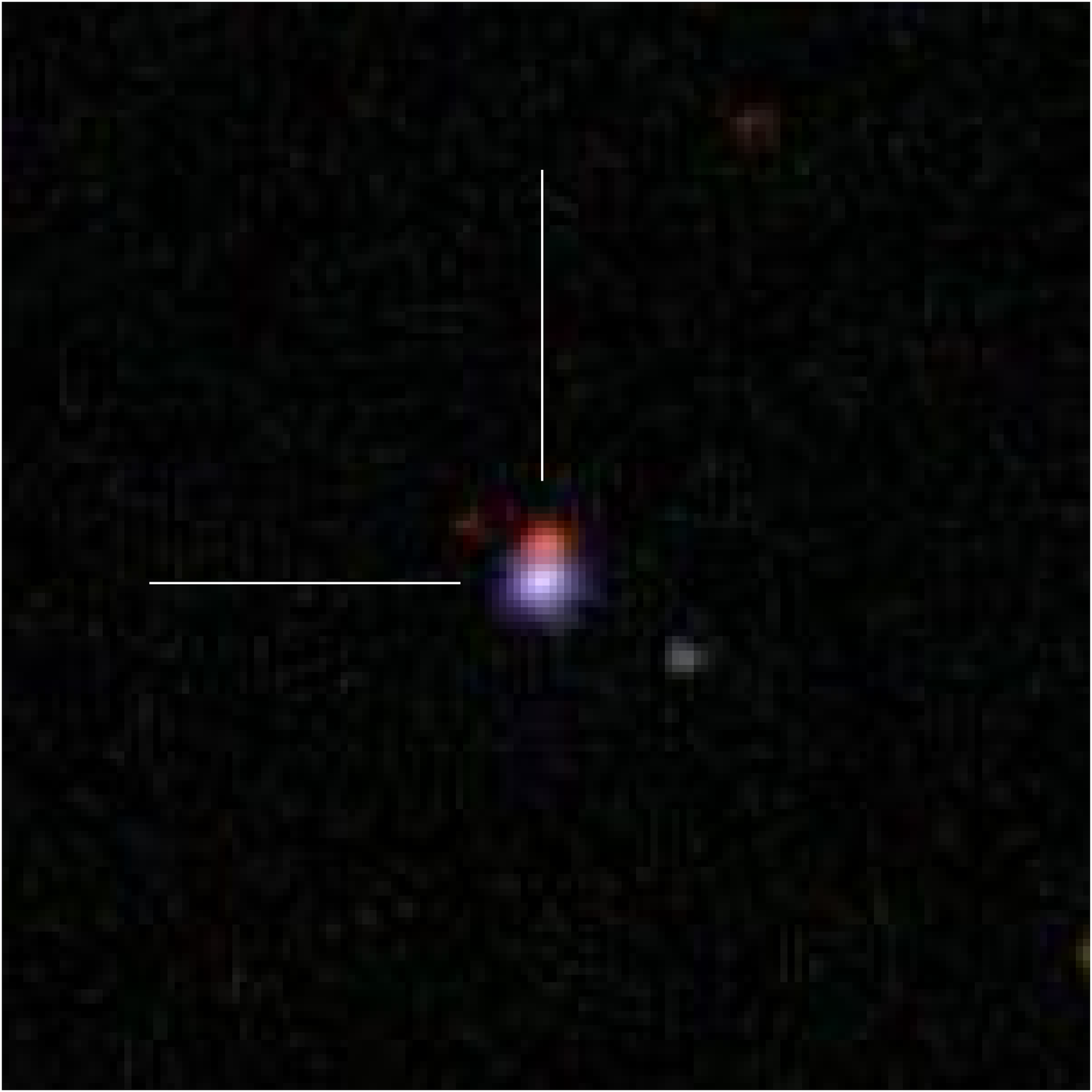}
\hspace*{0.5ex}
\includegraphics[width=0.622\columnwidth]{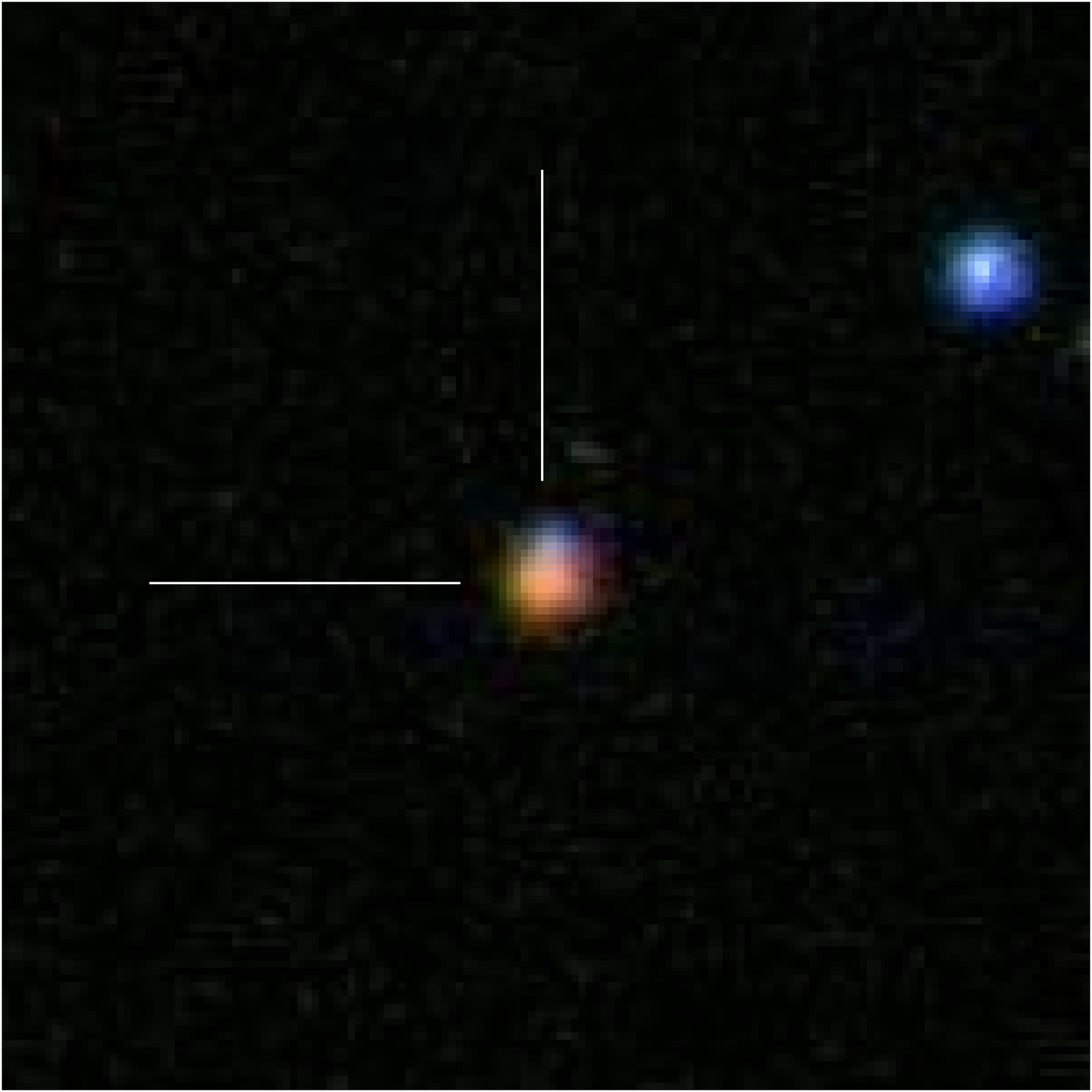}
\includegraphics[angle=-90,width=\textwidth]{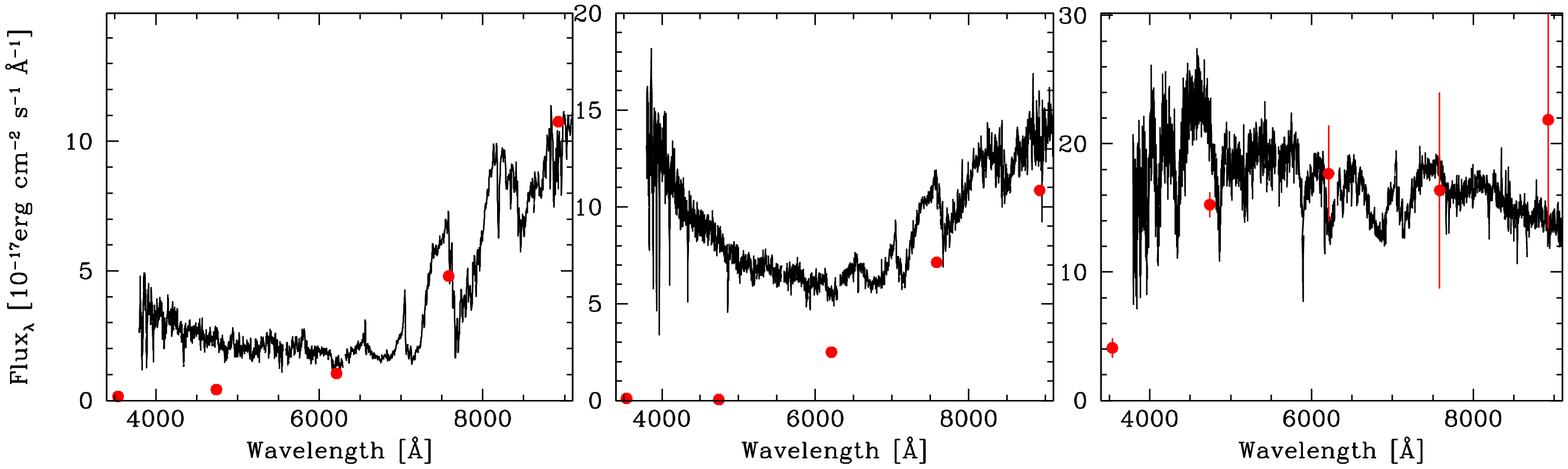}
\caption{\label{f-exclude}   SDSS   images   of  three   WDMS   binary
candidates.  Top  left panel: SDSS  image of SDSSJ\,073531.86+315015.2
($30  \arcmin  \times  30  \arcmin$),  a  single  red  star  initially
considered  as  a WDMS  binary  candidate.   Bottom  left panel:  SDSS
magnitudes (red  dots) and spectrum  (black line) of the  same system.
The  light  from  the  saturated  bright star  (Castor  A/B)  is  also
dispersed  in the  spectrum.   The magnitudes  are  consistent with  a
single    red   star.     Top    middle   panel:    SDSS   image    of
SDSSJ\,005827.24+005642.6 ($1  \arcmin \times 1  \arcmin$).  The image
suggests  a  resolved WDMS  binary  pair.   Middle  bottom panel:  the
detection of the Balmer lines typical  of an early F star in the blue,
together  with  the  typical  spectral  features of  a  low-mass  main
sequence star in  the red (black solid line),  indicate that these are
two single stars superimposed in the same image rather than a resolved
WDMS binary  pair.  SDSS magnitudes  are indicated with red  dots, and
are consistent with  those of a low-mass star.   Top right panel: SDSS
image of  SDSSJ\,025306.37+001329.2 ($1 \arcmin \times  1 \arcmin$), a
resolved  WDMS  binary  in  our  sample.   Bottom  right  panel:  SDSS
magnitudes (red  dots) and spectrum  (black line) of the  same system.
Whilst  the SDSS  spectrum  clearly shows  both  components, the  SDSS
magnitude errors are untypically  large due to unsuccessful deblending
of the close pair.}
\end{center}
\end{figure*}

Firstly, single  white dwarfs  (M-dwarfs) may occasionally  be located
close  to  very bright  M-dwarfs  (white  dwarfs  or A-stars)  causing
scattered  light to  enter  the spectroscopic  fibre  resulting in  an
(apparent)   two-component  spectrum.    A   spectacular  example   is
SDSSJ\,073531.86+315015.2 (left  panels of Fig.\,\ref{f-exclude}): the
SDSS spectrum  clearly exhibits  an M-dwarf at  red wavelengths  and a
blue component  with strong Balmer  lines in the  blue~--~however, the
SDSS image reveals  that this is a single M-dwarf at  a distance of 12
arc-min of Castor A/B~--~two  $V=1.88$ and $V=2.96$ A-stars.  The SDSS
magnitudes (red  dots) are superimposed  to the SDSS  spectrum (black)
and are consistent  with a single red star.   Single M-dwarfs are also
likely to  be found superimposed  with single early-type stars  in the
same figure.  An example  is SDSSJ\,005827.24+005642.6 (see middle top
panel of Fig.\,\ref{f-exclude}).  At first glance one could be tempted
to consider  it a  spatially resolved WDMS  binary pair.  However, the
SDSS  spectrum (middle  bottom panel  of  the same  figure) shows  the
typical Balmer lines  of an early F star in the  blue, while at redder
wavelengths the  typical spectral features  of a low-mass star  can be
seen.  The large  difference in  absolute magnitudes  between  the two
spectral types implies that these two stars are a chance superposition
along the line of sight, and not a physical binary.

Secondly,  SDSS images can  help identifying  WDMS binaries  among our
sample that  are spatially resolved,  but close enough that  flux from
both stars will enter into the spectroscopic fibre. In such cases, the
SDSS  magnitudes are  often discrepant  with the  flux-calibrated SDSS
spectrum, and/or have large  errors as consequence from the deblending
applied  by the  photometric pipeline.   Figure.\,\ref{f-exclude} (top
right panel) shows the  SDSS image of SDSSJ\,025306.37+001329.2, which
reveals a  spatially resolved  pair of red  and blue stars.   The SDSS
spectrum of SDSSJ\,025306.37+001329.2  contains the typical signatures
of a  WDMS binary, i.e.  broad  Balmer lines from the  white dwarf and
TiO absorption bands from the M-dwarf, however, the errors on the SDSS
magnitudes  are untypically  large, and  do  not match  well the  flux
calibrated SDSS spectrum.

\subsection{Cross-checks with previous WDMS binary catalogues}
\label{s-finalcat}

A  total number  of 1491  WDMS  binary candidates  were identified  in
Sect.\,\ref{s-comp}.     From   the    analysis    carried   out    in
Sec.\,\ref{s-excess}  and Sec.\,\ref{s-images}  we have  identified 94
and 89 WDMS binaries by  their blue and red excess respectively, while
115 WDMS  binary candidates were  removed after inspecting  their SDSS
images. This increased  the number of systems to  1559 WDMS candidates
in the spectroscopic SDSS DR6 data base.
 
In order to  evaluate the efficiency of our  procedure we compared our
results to five previously published lists of WDMS binaries from SDSS,
namely   \citet{vandenbesselaaretal05-1},  \citet{eisensteinetal06-1},
\citet{silvestrietal07-1}    (which   includes   the    systems   from
\citet{raymondetal03-1}  and  \cite{silvestrietal06-1}  as a  subset),
\citet{augusteijnetal08-1},    and    \citet{helleretal09-1}.     This
comparison is summarised in Table\,\ref{t-crosscheck}.

\begin{table}
\centering
\caption{\label{t-crosscheck}   Comparison   of   the  WDMS   binaries
identified  here   with  those  provided   in  previous  publications.
N$_\mathrm{WDMS}$    gives    the    number   of    objects    listed,
N$_\mathrm{ident}$  the number  of  WDMS binaries  we have  identified
within the  list, N$_\mathrm{remov}$ the  number of objects we  do not
consider  being WDMS  binaries,  and N$_\mathrm{lost}$  the number  of
systems that  we have missed  (the percentage is given  in brackets).}
\setlength{\tabcolsep}{1.1ex}
\begin{small}
\begin{tabular}{lcccr}
\hline
\hline
    Publication  & N$_\mathrm{WDMS}$ & N$_\mathrm{ident}$ &  N$_\mathrm{remov}$ & N$_\mathrm{lost}$ \\
\hline
\citet{vandenbesselaaretal05-1} &    15  &  15     &  -      &   - (0\%)  \\
\citet{silvestrietal07-1}       &  1225  & 996     & 204     &  25 (2\%)  \\
\citet{augusteijnetal08-1}      &   130  & 110     &  16     &   4 (3\%)  \\
\citet{helleretal09-1}          &   636  & 554     &  81     &   1 (0.1\%)   \\
\hline
\end{tabular}
\end{small}
\end{table}

The  15 WDMS  binaries  containing DB  (13)  and DC  (2) white  dwarfs
presented  by \citet{vandenbesselaaretal05-1}  have  been successfully
identified as WDMS binaries with our template fitting algorithm.

The  WDMS  binary sample  presented  by \citet{eisensteinetal06-1}  is
almost  entirely contained in  the WDMS  binary catalogue  provided by
\citet{silvestrietal07-1} (see below for a detailed comparison).  Only
ten objects of  \citet{eisensteinetal06-1} with available SDSS spectra
are  not listed  in \citet{silvestrietal07-1},  and our  algorithm has
successfully identified these ten systems as WDMS binaries.

\citet{silvestrietal07-1}  claim that  their  catalogue contains  1253
objects but in fact only  1228 spectra (corresponding to 1225 objects)
are listed in  the electronic edition of their  paper, and 996 systems
of those  appear in our catalogue. One  is a SEGUE object  and will be
included  in  Schreiber et  al.   (2009,  in  preparation).  The  vast
majority (i.e.  204)  of the remaining 229 objects  are not classified
as  WDMS  binaries  by  us,  either  because  of  their  spectroscopic
appearance, or  because of morphological problems in  the SDSS images.
Two examples of mis-classified WDMS binaries are the $z = 0.21$ quasar
SDSS\,J032428.78-004613.8    and    the     $z    =    0.11$    galaxy
SDSS\,J114334.70+455134.2.   An  updated  classification  of  the  204
objects  is given  in  Table\,\ref{t-sil}, whilst  the  full table  is
available in the electronic edition of this paper.  The spectra of the
25 genuine  WDMS binaries identified  by \citet{silvestrietal07-1} but
overlooked  by our template  fitting algorithm,  are dominated  by the
emission of  one of  the stellar components.   We added these  25 WDMS
binaries to our sample.

\begin{table}
\centering
\caption{\label{t-sil} Updated classification  of the 204 objects from
\citet{silvestrietal07-1} which are not considered as WDMS binaries by
us. The complete  table can be found in the  electronic edition of the
paper.} \setlength{\tabcolsep}{1.1ex}
\begin{small}
\begin{tabular}{lr}
\hline
\hline
   Object  & Classification \\
\hline
SDSSJ001324.33-085021.4   &   M star + ? \\
SDSSJ003839.36+260258.5   &   G star \\
SDSSJ005714.52-000755.8   &   M star \\
SDSSJ005827.24+005642.6	  &   MS+MS superposition \\
SDSSJ012516.98-010944.2   &   M star \\
... & ... \\
\hline
\end{tabular}
\end{small}
\end{table}

\citet{augusteijnetal08-1}  developed  a  WDMS  binary  identification
algorithm  based  on SDSS  imaging  and  the  proper motion  catalogue
provided by \citet{gould+kollmeier04-1}.  This way they identified 651
WDMS binary candidates  with SDSS $ugriz$ photometry, of  which 95 are
contained in the SDSS  DR5 spectroscopic data base.  Cross-correlating
their full list of 651 objects against the DR6 spectroscopic data base
we find 176 spectra for 130 objects.  All but 20 of these objects were
in our list.  Again, the majority  of those systems (16) that have not
been  identified  by  our  template  fitting procedure  are  not  WDMS
binaries.  We identify nine  quasars, three cataclysmic variables, one
F star, one M-dwarf, one DA  white dwarf, and one DC white dwarf.  The
four  remaining  WDMS binaries  have  been  overlooked  by our  method
because  they contain  rather cold  white dwarfs  and the  spectra are
dominated by the companions.  We added the four systems to our sample.

Recently \citet{helleretal09-1} discussed the properties of 636 colour
selected WDMS binaries  from SDSS DR6. Most of  their systems had been
identified previously.   Comparing their sample with  our catalogue we
find 82 objects  missing in our list.  Inspecting  the SDSS images and
spectra (if  available) of  these 82  systems we find  81 of  them not
being  WDMS  binaries.   Two  examples  are  the  low-redshift  galaxy
SDSSJ\,122953.46+473150.3    \citep[U7639,][]{makarovaetal98-1},   and
again the $z=0.21$  quasar SDSSJ\,032428.78-004613.8, also included in
\citet{silvestrietal07-1}.   An updated  classification  for these  81
objects  is  provided in  Table\,\ref{t-heller},  whilst the  complete
table is provided  in the electronic edition of  this paper.  Only one
system analysed by \citet{helleretal09-1} but not in our sample turned
out to be a WDMS binary, and we added it to our catalogue.

In summary,  comparing the sample  of WDMS binaries obtained  with our
template  fitting procedure  with previously  published  catalogues of
WDMS binaries shows that our  method represents a robust and efficient
tool to identify WDMS binaries (Table\,\ref{t-crosscheck}).

\section{Completeness of the sample}
\label{s-completeness}

\begin{figure*}
\begin{center}
\includegraphics[angle=-90,width=\textwidth]{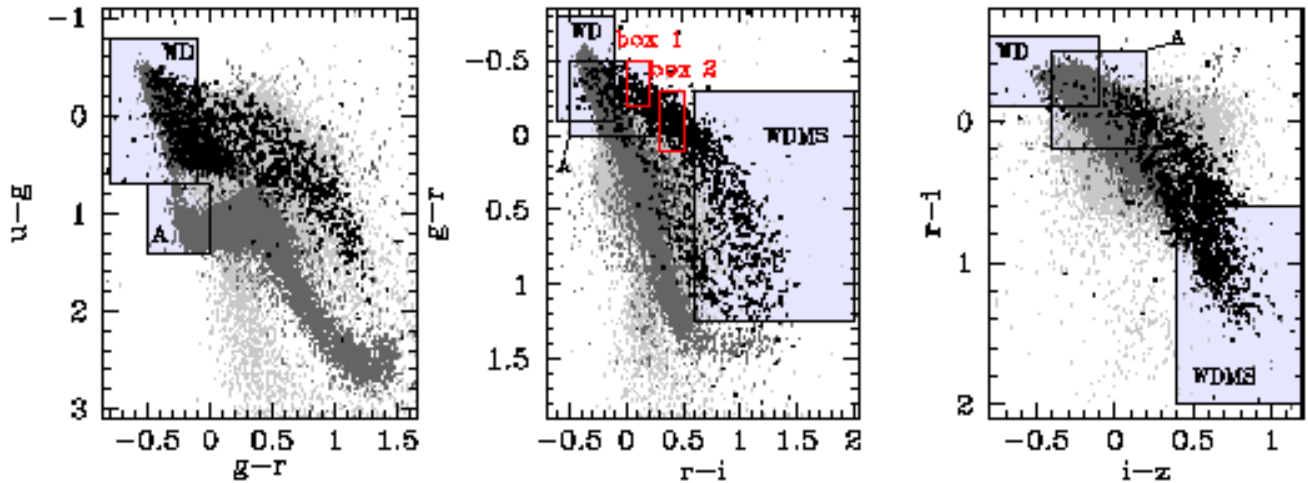}
\caption{From  left  to  right  the ($u-g$,$g-r$),  ($g-r$,$r-i$)  and
($r-i$,$i-z$)  colour diagrams  for the  WDMS binaries  in  our sample
(black; excluding WDMS binary candidates), stellar sources (dark ray),
and  quasars (light  gray). The  white dwarf,  A-star and  WDMS binary
exclusion  boxes  defined by  \citet{richardsetal02-1}  are shaded  in
light blue.  The regions within the two red boxes in the ($g-r$,$r-i$)
plane,     and    the    WDMS     binary    exclusion     region    of
\citet{richardsetal02-1} in the ($g-r$,$r-i$) and ($r-i$,$i-z$) planes
are used to  estimate the completeness of our  WDMS binary sample (see
Sect.\,\ref{s-completeness}).}
\label{f-completeness}
\end{center}
\end{figure*}

In  the   previous  section  we  showed  that   our  selection  method
successfully recovered the vast majority of previously identified SDSS
WDMS  binaries. In  this  section, we  investigate  both the  internal
completeness  of our catalogue,  i.e.  the  fraction of  WDMS binaries
contained in the DR6 spectroscopic  data base that has been identified
by our algorithm, and the external completeness, i.e.  the fraction of
point sources for which SDSS  $ugriz$ photometry is available that has
been  spectroscopically observed  by SDSS.   A detailed  study  of the
external (spectroscopic) completeness of  WDMS binaries within SDSS is
currently  underway, and  we  use this  paper  to provide  preliminary
results on this issue.  To that  end, we first define three regions in
SDSS colour-colour  planes: the WDMS binary exclusion  region that has
been defined  by \citet{richardsetal02-1} as part of  the quasar fibre
allocation  algorithm,   and  two  small  rectangular   boxes  in  the
$(g-r,r-i)$  plane  that  sample  a  significant  population  of  WDMS
binaries in  colour space. Particularly, these two  boxes were defined
to  test  how the  external  completeness  varies  between inside  and
outside the WDMS  exclusion box.  The $(u-g, g-r)$,  $(g-r, r-i)$, and
$(r-i,      i-z)$       colour-colour      diagrams      shown      in
Fig.\,\ref{f-completeness} illustrate  the locations of  these regions
(WDMS, box 1, box 2).  The WDMS binaries in our sample excluding those
systems    that    are   classified    as    candidates   only    (see
Sect.\,\ref{s-final}) are shown in black, stellar sources are in gray,
and quasars  in light gray.  Highlighted  in light blue  are the white
dwarf (WD),  A-star (A) and  WDMS binary exclusion regions  defined by
\citet{richardsetal02-1}.  Both the A-star  and the WDMS binary region
in the colour  planes are combined with a logical  ``and''.  Box 1 and
box   2   are    defined   by   ($-0.5<g-r<-0.2$,   $0<r-i<0.2$)   and
($-0.3<g-r<0.1$, $0.3<r-i<0.5$) respectively.

To determine  the completenesses we  visually classified all  SDSS DR6
spectra of point-sources with $g\le20$, and used the casjobs interface
\citep{li+thakar08-1}
\footnote{http://casjobs.sdss.org/CasJobs/} to  the SDSS data  base to
determine  the   number  of  photometric  point   sources  with  clean
photometry in each of the regions defined above.  The fraction of WDMS
binaries identified  by the template matching routine  among the total
number  of  spectroscopically  identified  WDMS  binaries  inside  the
selected region  gives an estimate of the  internal completeness.  The
external   completeness   is  simply   given   by   the  fraction   of
spectroscopically observed point  sources inside a given colour-colour
region.   In order  to evaluate  the  impact of  the brightness  limit
applied  in the  quasar selection  algorithm \citep{richardsetal02-1},
i.e. a  de-reddened $i<19.1$,  we additionally performed  the external
completeness  analysis for the  ranges $i\le19.1$  and $19.1<i\le20.1$
outside  the WDMS  exclusion box,  i.e.   for box  1 and  box 2.   The
results are  given in Table\,\ref{t-complet} and can  be summarised as
follows.

\begin{table}
\centering
\caption{\label{t-heller}  Updated classification  of  the 82  objects
from \citet{helleretal09-1} which are  not considered as WDMS binaries
by us.  The complete table can  be found in the  electronic edition of
the paper.} \setlength{\tabcolsep}{1.1ex}
\begin{small}
\begin{tabular}{lr}
\hline
\hline
   Object  & Classification \\
\hline
SDSSJ013007.13+002635.3 &  carbon star \\
SDSSJ020538.10+005835.3	&  no available spectra, WDMS? \\
SDSSJ023906.69+002916.6	&  DA \\
SDSSJ025622.18+330944.8	&  O star? \\
SDSSJ032428.78-004613.8	&  quasar \\
... & ... \\
\hline
\end{tabular}
\end{small}
\end{table}

\begin{itemize}

\item SDSS  DR6 contains 8002 photometric point  sources with $g\le20$
inside the  \citet{richardsetal02-1} WDMS binary  exclusion box.  SDSS
spectra are available for 575 (or $7\%$ of the point sources) systems,
and we visually classify 389 of these as WDMS binaries.  The remaining
186  objects are  mostly  single M  stars  but we  also identify  four
cataclysmic   variables  and  eight   quasars.   The   small  external
spectroscopic completeness of SDSS DR6  ($7\%$) in the WDMS binary box
as  well as  the  large number  of  WDMS binaries  ($67\%$) among  the
spectroscopically observed  objects is not surprising,  as this region
has  been  explicitly  excluded  in  the  quasar  selection  algorithm
\citep{richardsetal02-1}.  All  but 12 of  the 389 WDMS  binaries have
been  successfully  identified  by  our  automated  template  matching
algorithm, which gives an  internal completeness of $97\%$.  Among the
12 missing objects  nine contain cool (probably DC)  white dwarfs, one
contains a clear  DA white dwarf, one a clear  DC white dwarf primary,
and   one   we   identify   as  a   low-accretion-rate   polar,   LARP
\citep[SDSS\,J204827.90+005008.9,][]{schmidtetal05-1}.  In  all the 12
cases  the spectra  are dominated  by  the emission  of the  companion
stars.

\item Box 1 contains 708 point  sources with $g<20$ in DR6 and for 247
of these at least one  SDSS spectrum is available, corresponding to an
external completeness  of $35\%$.  As  expected, we find  the external
completeness  to  significantly  change  at  the  $i$-magnitude  limit
incorporated  in the  quasar search  algorithm. While  it  increases to
$\sim60\%$ for $i\le19.1$ it drops to $\sim15\%$ for $19.1<i\le20.1$.

Among the  247 objects  are 67 ($27\%$)  WDMS binaries.  The remaining
systems are mainly quasars, a few single white dwarfs, early-type main
sequence  stars,  and  two  cataclysmic  variables. All  the  67  WDMS
binaries  in this  box have  been identified  by our  search algorithm
(Sect.\,\ref{s-ident})  equivalent  to  an  internal  completeness  of
$100\%$.

\item SDSS DR 6 contains 6689 SDSS point sources in box 2 and for 2280
of  them  SDSS spectroscopy  is  available,  which  gives an  external
completeness  of  $34\%$ using $g<20$.  

\begin{table}
\centering
\caption{\label{t-complet}  We  give here  the  number  of SDSS  point
sources   N$_\mathrm{p}$,  the  number   of  available   SDSS  spectra
N$_\mathrm{spec}$,   the  number   of   spectroscopic  WDMS   binaries
N$_\mathrm{WDMS}$, and the number  of WDMS binaries we have identified
N$_\mathrm{ident}$  for each  of the  WDMS, box  1 and  box  2 regions
defined in Sect.\,\ref{s-completeness}.   The external completeness is
estimated    as    N$_\mathrm{spec}$/N$_\mathrm{p}$,   the    internal
completeness         as         N$_\mathrm{ident}$/N$_\mathrm{WDMS}$.}
\setlength{\tabcolsep}{1.1ex}
\begin{small}
\begin{tabular}{llcccccc}
\hline
\hline
& Reg. &  N$_\mathrm{p}$  & N$_\mathrm{spec}$ & N$_\mathrm{WDMS}$ &
N$_\mathrm{ident}$ & Ext.C. & Int.C. \\
\hline
$g\le20$  & WDMS  & 8002 &  575 & 389 & 377 &  7\% & 97\%  \\
 & Box 1 &  708 &  247 &  67 &  67 & 35\% & 100\% \\
 & Box 2 & 6689 & 2280 & 135 & 131 & 34\% & 97\%  \\
\hline
$i\le19.1$  & Box 1 &  313 &  188 & &&60\% \\
 & Box 2 & 2822 & 1672 & &&60\% \\
\hline
$i\le20.1$ & Box 1 &  389   &  59 &&&15\% \\
$i > 19.1$ & Box 2 & 13110   & 1770 &&&14\% \\
\hline
\end{tabular}
\end{small}
\end{table}

As  for   box  1,  the   quasar  limit  on  the   reddening  corrected
$i$-magnitude   of  19.1   is   very  much   affecting  the   external
completeness.    It  increases   significantly  to   $\sim\,60\%$  for
$i\le19.1$ but decreases to  $\sim\,14\%$ for objects fainter than the
quasar limit,  i.e.  for $19.1<i\le20.1$.  At first  glance, one might
consider the  spectroscopic completeness of  $\sim$60\% for $i\le19.1$
to be  surprisingly low  compared to the  $\sim95$\% obtained  for the
quasar  selection algorithm \citep{vandenberketal05-1}.   However, one
should  keep in  mind  that  ``outside the  exclusion  boxes'' is  not
equivalent  to  inside  the  (rather complex)  quasar  selection.   In
addition,   in   DR6   the   area   of   the   footprint   followed-up
spectroscopically was only $\simeq82\%$  of the imaging area, implying
a  systematically lower spectroscopic  completeness than  expected for
DR7.   As   mentioned  earlier,  a  more  complete   analysis  of  the
completeness of SDSS  WDMS binaries is underway and  will be presented
in a subsequent publication.

From the 2280 spectra that are  available in box\,2 only 135 (6\%) are
WDMS binaries.  The vast majority of the remaining objects are quasars
but  we also  identify some  few single  main sequence  stars  and one
cataclysmic variable.  All but four of the 135 WDMS binaries have been
found  by our template  fitting method,  corresponding to  an internal
completeness  of  $97\%$.   The   four  systems  that  have  not  been
identified by our algorithm are SDSS\,J110539.77+250628.6, which is in
fact the semi-detached  magnetic cataclysmic variable ST\,LMi observed
during  a deep  low state;  SDSS\,J124959.76+035726.6, a  typical WDMS
binary  previously  listed  as  a cataclysmic  variable  candidate  by
\citet{szkodyetal04-1}; SDSS\,J150954.40+243449.3,  which has a broken
SDSS spectrum; and SDSS J204218.52-065638.4, a spatially resolved main
sequence dominated WDMS binary.
 
\end{itemize}

\begin{table*}
\label{t-magall}
\caption{\label{t-magall}  The  complete  catalogue.  Coordinates  and
GALEX-SDSS-UKIDSS magnitudes for the 1602 WDMS binaries and candidates
are also  included.  The entire table (including  also the photometric
errors) is  provided in the electronic  edition of the  paper.  We use
``-''    to    indicate    that    no   magnitude    is    available.}
\setlength{\tabcolsep}{1.5ex}
\begin{small}
\begin{tabular}{lrrccccccccccc}
\hline
\hline
     SDSS\,J & ra[$^{\circ}$] & dec[$^{\circ}$] & $nuv$ & $fuv$ &  $u$ &  $g$ &  $r$ &  $i$ &  $z$ &  $y$ &  $J$ &  $H$ &  $K$  \\      
\hline
000152.09+000644.7   & 0.46704 &  0.11242 &  18.45 &  17.90 & 19.03 &  18.61 &  17.94 &  17.50 &  17.25 &  16.51 &  16.05 &  15.40 &  15.28 \\
000442.00--002011.6  & 1.17500 & -0.33656 &      - &      - & 23.72 &  20.38 &  19.13 &  18.65 &  18.28 &      - &      - &      - &      - \\
000611.94+003446.5   & 1.54975 &  0.57958 &  21.78 &      - & 21.38 &  20.92 &  20.12 &  19.00 &  18.38 &  17.53 &  17.05 &  16.58 &  16.20 \\
001029.87+003126.2   & 2.62446 &  0.52394 &  20.17 &  19.96 & 21.92 &  20.85 &  19.97 &  19.00 &  18.42 &  17.65 &  17.14 &  16.52 &  16.36 \\
001247.18+001048.7   & 3.19658 &  0.18019 &  20.50 &  20.71 & 20.73 &  20.21 &  19.66 &  18.63 &  17.96 &  17.09 &  16.60 &  16.13 &      - \\
001339.20+001924.3   & 3.41333 &  0.32342 &  16.41 &  19.73 & 15.94 &  15.56 &  15.55 &  15.63 &  15.89 &      - &      - &      - &      - \\
001359.39--110838.6  & 3.49749 &-11.14405 &  17.77 &  17.42 & 18.30 &  18.43 &  18.31 &  20.75 &  22.82 &      - &      - &      - &      - \\
001549.02+010937.3   & 3.95425 &  1.16036 &  20.97 &  20.68 & 21.23 &  20.86 &  20.60 &  19.85 &  19.27 &  18.46 &  17.86 &  17.45 &  17.12 \\
001726.63--002451.1  & 4.36096 & -0.41419 &  19.71 &  20.30 & 19.67 &  19.28 &  19.02 &  18.18 &  17.54 &  16.60 &  16.07 &  15.56 &      - \\
001733.59+004030.4   & 4.38996 &  0.67511 &  20.83 &  22.43 & 22.09 &  20.79 &  19.58 &  18.17 &  17.38 &  16.37 &  15.84 &  15.27 &  14.97 \\
001749.24--000955.3  & 4.45517 & -0.16536 &  15.87 &  15.40 & 16.56 &  16.86 &  17.03 &  16.78 &  16.47 &  15.75 &  15.33 &  14.76 &  14.56 \\
001853.79+005021.5   & 4.72412 &  0.83931 &  20.46 &  20.27 & 21.00 &  20.38 &  19.64 &  18.80 &  18.35 &  17.52 &  17.09 &  16.51 &  16.27 \\
001855.19+002134.5   & 4.72996 &  0.35958 &  22.42 &  22.12 & 21.60 &  20.60 &  19.87 &  18.97 &  18.38 &  17.54 &  17.09 &      - &      - \\
002143.78--001507.9  & 5.43242 & -0.25219 &  22.20 &      - & 22.58 &  19.63 &  18.39 &  17.02 &  16.30 &  15.40 &  14.87 &  14.31 &  14.05 \\
            ...     &   ...     &     ...   &  ...   &  ...   &  ...  &  ...   &  ...   &  ...   &  ...   &  ...   &  ...   &  ...   & ...     \\
\hline
\end{tabular}
\end{small}
\end{table*}

Three main conclusions  can be drawn from the  analysis carried out in
Fig.\,\ref{f-completeness}.   Firstly, the  spectroscopic completeness
in SDSS is  much larger in quasar dominated  colour-colour space, i.e.
34\% in box\,2, compared to only  7\% in the WDMS binary exclusion box
of \citet{richardsetal02-1}.  The low  completeness in the WDMS binary
exclusion box, combined with the  high fraction of WDMS binaries among
all objects  in this region ($67\%$)  implies that the  number of SDSS
WDMS binaries could be  dramatically increased by additional follow-up
spectroscopy  of  point  sources  located  in  this  box.   Since  the
spectroscopic completeness  in the WDMS  binary exclusion box  is 7\%,
and only  389 WDMS  binaries benefit from  spectra, this  implies then
that  SDSS  did not  target  $\sim$5500  WDMS  binaries for  follow-up
spectroscopy  within this  region.  As  partners of  SEGUE some  of us
performed such  a program and  identified $\sim300$ new  WDMS binaries
\citep{schreiberetal07-1}.  Secondly, outside  the exclusion boxes the
external   completeness  drops   significantly   from  $\sim60\%$   to
$\sim15\%$  at the  $i$-magnitude  limit of  19.1  implemented in  the
quasar search  algorithm. Finally, and most importantly,  only 16 WDMS
binaries (four  in box\,2  and 12 in  the WDMS binary  exclusion box),
have not  been identified by  our template fitting  routine.  Assuming
that the three analysed colour boxes are representative for the entire
WDMS  binary  bridge,  we  derive  an  internal  completeness  of  our
catalogue  of  $\sim98\%$.   Virtually  all  the  systems  our  search
algorithm  failed to  identify are  dominated by  the emission  of the
secondary  star.  Such  systems  are therefore  expected  to form  the
remaining $\sim2\%$  of WDMS binaries  contained in the SDSS  DR6 data
base.

The 16 WDMS binaries that  we previously overlooked have been added to
our  sample that  now  forms our  final  SDSS DR6  WDMS catalogue,  as
described in the following section.

\section{The final catalogue}
\label{s-final}

From    the   analysis    described   in    Sect.\,\ref{s-ident}   and
Sect.\,\ref{s-completeness} a total number  of 1602 WDMS binaries have
been identified.  These  systems form our final catalogue  of SDSS DR6
WDMS binaries and WDMS binary candidates.  We describe in this section
the main tables  characterising our sample.  An excerpt  of each table
is  given  here,  while  the  complete  tables can  be  found  in  the
electronic edition of the paper.
  
In Table\,\ref{t-magall} we list  the coordinates, GALEX DR4, SDSS DR6
and  UKIDSS  DR4 magnitudes.   Occasionally  multiple  SDSS and  GALEX
magnitudes  are available  for  one  system. In  these  cases we  give
averaged  magnitudes.   We  used  PSF  (point  spread  function)  SDSS
magnitudes   when  available,   fibre  magnitudes   otherwise.   GALEX
magnitudea are available for 1327 WDMS binaries, UKIDSS magnitudes for
466.

\begin{figure}
\begin{center}
\includegraphics[angle=-90,width=\columnwidth] {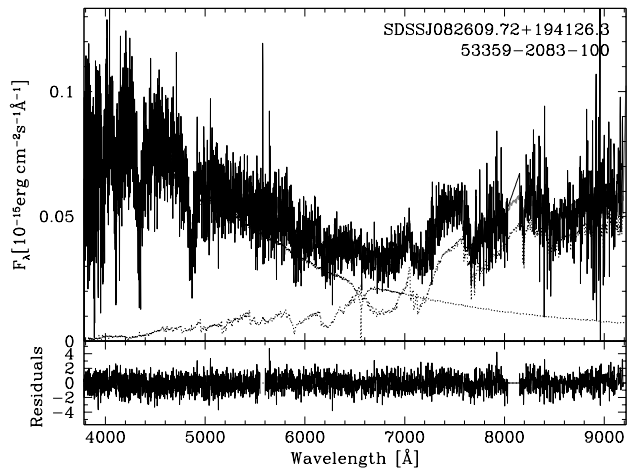}
\caption{\label{f-decomp}         Two-component         fits        to
SDSSJ\,082609.72+194126.3, a WDMS binary  in our sample. The top panel
shows the WDMS  binary spectrum as black line,  and the two templates,
white dwarf and M-dwarf, as  dotted lines.  The bottom panel shows the
residuals (in units of the standard deviation) from the fit.  The SDSS
spectrum  identifiers MJD,  PLT and  FIB  are given  below the  object
name. See also Figs.\,\ref{f-wdfit},\ref{f-galex}.}
\end{center}
\end{figure}

\begin{table}
\caption{\label{t-systems}  The catalogue  divided into  the different
WDMS binary subtype, as defined in Sec.\,\ref{s-final}.}
\setlength{\tabcolsep}{2ex}
\begin{center}
\begin{small}
\begin{tabular}{lrlr}
\hline
\hline
type  &  number & type & number \\
\hline
DA/M  & 1176   & DB/M:     &   1   \\                  
DB/M  &   45   & DA:/K     &   5   \\       	       
DC/M  &   34   & DA/K:     &  10   \\                   
DA/K  &   49   & DH/M      &   1(x)\\                  
DB/K  &   3    &  LARP     &   1(*)\\                  
WD/M  &   135  & PG1159/M  &   1(+)\\                  
WD/K  &   26   & (WD/M)    &  32   \\                  
WD/M: &    1   & (WD/K)    &   2   \\                  
DA:/M &    27  & (DA/M)    &   4   \\                  
DB:/M &    1   & (DC/M)    &   1   \\ 
DC:/M &    3   & (DA/K)    &   1   \\
DA/M: &   43   &           &       \\
\hline
\end{tabular}
\end{small}
\end{center}
\begin{minipage}{\columnwidth}
(x) \citet{szkodyetal09-1}; (*) \citet{schmidtetal05-1}; (+) \citet{nageletal06-1}
\end{minipage}
\end{table}

In Table\,\ref{t-systems}  we provide relative numbers  of the stellar
components in our WDMS binary  catalogue.  For the white dwarfs we use
the flags DA, DB, DC, WD (if  the white dwarf type is unknown), DH (if
the  white  dwarf is  magnetic),  and  PG1159 (hot  hydrogen-deficient
pre-white   dwarf).   We   use  the   acronym  LARP   to   indicate  a
low-accretion-rate polar.  The  secondary stars are flagged as  M or K
according to their spectral type.  If the flag is followed by a colon,
the classification of the stellar component is uncertain.  Finally, we
list in  brackets the WDMS  binary candidates.  We  consider basically
two  types   of  candidates:(1)  systems  with  very   low  S/N  ratio
spectroscopy that does not allow a clear classification of the stellar
components in their spectra, and  (2) systems with marginal blue (red)
excess in their SDSS  spectra, without morphological problems in their
SDSS  images   (Sec.\,\ref{s-images}),  and  with   $\chi^{2}$  values
(Sec.\,\ref{s-excess}) favouring a  binary classification but no GALEX
(UKIDSS) magnitudes  available to confirm the existence  of the second
component.

\section{Stellar parameters}
\label{s-param}

In         \citet{rebassa-mansergasetal07-1,        schreiberetal08-1,
rebassa-mansergasetal08-1}      we      presented      a      spectral
decomposition/fitting technique and a  M-dwarf spectral type -- radius
relation  to  determine  the   stellar  parameters  of  WDMS  binaries
spectroscopically  identified  by  the  SDSS.  Our  method  allows  to
estimate the  effective temperature, surface gravity,  mass and radius
of the  white dwarf, as  well as the  spectral type and radius  of the
main  sequence  companion.   In  addition,  two  independent  distance
estimates  can be  obtained by  estimating the  best-fit  flux scaling
factors of  the two components  (see \citet{rebassa-mansergasetal07-1}
for  details).    We  here  basically  use  the   same  procedure  but
incorporate two modifications.

Firstly, we compiled an additional  library of 222 high S/N spectra of
DB  white dwarfs from  SDSS DR\,4  \citep{eisensteinetal06-1} covering
the entire  range of observed \Teff.  Decomposing  the binary spectrum
and  fitting   the  stellar   components  in  the   same  way   as  in
\citet{rebassa-mansergasetal07-1}  but  using  the just  mentioned  DB
templates instead of DA templates  allows to estimate the DB effective
temperatures in our catalogue.

\begin{table*}
\label{t-param}
\caption{\label{t-paramcat}    White     dwarf    masses,    effective
temperatures, surface  gravities, spectral types and  distances of the
1602  WDMS binaries  in  our catalogue,  as  determined from  spectral
modelling. We list both the  hot and the cold solutions, with preferred
solution given in the first line for each spectrum. The other solution
is given for completeness. The complete table, including also notes for
individual  systems, can  be found  in the  electronic edition  of the
paper.  We use  the flags $e$, $s$, $a$, and $h$  and ``re'' for those
systems     which     have      been     studied     previously     by
\citet{eisensteinetal06-1},                  \citet{silvestrietal07-1},
\citet{augusteijnetal08-1}  and \citet{helleretal09-1}, and  which are
resolved WDMS binary pairs,  respectively.  Again, we indicate that no
stellar          parameters         are          measured         with
``-''.}. \setlength{\tabcolsep}{1.0ex}
\begin{small}
\begin{tabular}{lccccccccccccccccc}
\hline
\hline
                SDSS\,J   &    type  &      MJD  &      PLT   &     FIB   &    \Teff[k]   &   err  &     $\log$ g   &    err    &    M$_\mathrm{wd} [M_\mathrm{\bigodot}]$   &    err  &      d$_\mathrm{wd}$[pc]   &  err   &      Sp   &    d$_\mathrm{sec}$[pc] &  err  & flag &    \\
\hline
...                 &  ...    &  ...    &  ... & ...   &  ...    &   ...  &   ...    &  ...    &  ...   &  ...     &  ...     &   ...  &     ... &  ...     &  ...   &  ...   \\
013356.07-091535.1 &   DA/M &   53612 &  1915 &  431 &   12392 &    1324 &    7.47 &    0.36 &    0.36  &   0.16 &     531 &     112   &     8  &    247  &    100 &    &  \\
013356.07-091535.1 &   DA/M &   53612 &  1915 &  431 &   12250 &    1328 &    7.49 &    0.36 &    0.36  &   0.16 &     523 &     110   &     8  &    247  &    100 &    &  \\
013418.52+010100.0 &   DA/M &   53741 &  1502 &  517 &   22811 &    2601 &    7.75 &    0.41 &    0.50  &   0.22 &     912 &     224   &     1  &   1146  &    225 & re &  \\
013418.52+010100.0 &   DA/M &   53741 &  1502 &  517 &    9844 &     473 &    8.66 &    0.46 &    1.01  &   0.25 &     201 &      73   &     1  &   1146  &    225 &    &  \\
013441.30-092212.7 &   DA/M &   52147 &   662 &  477 &   12110 &    2657 &    7.06 &    1.13 &    0.24  &   0.48 &    2464 &    1314   &     2  &   1914  &    456 &    &  \\
013441.30-092212.7 &   DA/M &   52147 &   662 &  477 &   11699 &    2808 &    7.14 &    1.19 &    0.26  &   0.56 &    2315 &    1285   &     2  &   1914  &    456 &    &  \\
013441.30-092212.7 &   WD/M &   52178 &   662 &  468 &       - &       - &       - &       - &       -  &      - &       - &       -   &     2  &   1900  &    453 &    &  \\
013441.30-092212.7 &   WD/M &   52178 &   662 &  468 &       - &       - &       - &       - &       -  &      - &       - &       -   &     2  &   1900  &    453 &    &  \\
013504.31-085919.0 &   DA/M &   53612 &  1915 &  464 &    9187 &     480 &    8.82 &    0.69 &    1.10  &   0.36 &     218 &     134   &     4  &    471  &    139 &  a &  \\
013504.31-085919.0 &   DA/M &   53612 &  1915 &  464 &   25891 &    4599 &    7.40 &    0.77 &    0.38  &   0.36 &    2119 &     919   &     4  &    471  &    139 &  a &  \\
013716.08+000311.3 &   DA/M &   52203 &   698 &  465 &   18756 &    3667 &    8.52 &    0.70 &    0.94  &   0.37 &     586 &     295   &     2  &    856  &    204 &s/e &  \\
013716.08+000311.3 &   DA/M &   52203 &   698 &  465 &   15601 &    5326 &    8.88 &    0.95 &    1.14  &   0.49 &     364 &     326   &     2  &    856  &    204 &s/e &  \\
013750.03+003237.6 &   DA/M &   52203 &   698 &  470 &   22292 &     894 &    8.38 &    0.14 &    0.86  &   0.08 &     271 &      29   &     2  &    272  &     65 &  s &  \\
013750.03+003237.6 &   DA/M &   52203 &   698 &  470 &   16717 &    1890 &    9.24 &    1.16 &    1.32  &   0.60 &      91 &      31   &     2  &    272  &     65 &  s &  \\
013801.04+230329.3 &   DA/M &   53341 &  2064 &  216 &   11835 &     610 &    8.13 &    0.21 &    0.69  &   0.14 &     329 &      47   &     7  &    218  &    120 &s/h &  \\
013801.04+230329.3 &   DA/M &   53341 &  2064 &  216 &   15246 &     772 &    7.77 &    0.18 &    0.49  &   0.09 &     518 &      56   &     7  &    218  &    120 &s/h &  \\ 
013841.09+131257.6 &   WD/K &   51882 &   426 &  138 &       - &       - &       - &       - &       -  &      - &       - &       -   &     -  &      -  &      - &    &  \\
013841.09+131257.6 &   WD/K &   51882 &   426 &  138 &       - &       - &       - &       - &       -  &      - &       - &       -   &     -  &      -  &      - &    &  \\
...                 &  ...    &  ...    &  ... & ...   &  ...    &   ...  &   ...    &  ...    &  ...   &  ...     &  ...     &   ...  &     ... &  ...     &  ...   &  ...      \\
\hline
\hline
\end{tabular}
\end{small}
\end{table*}

\begin{figure}
\begin{center}
\includegraphics[width=\columnwidth] {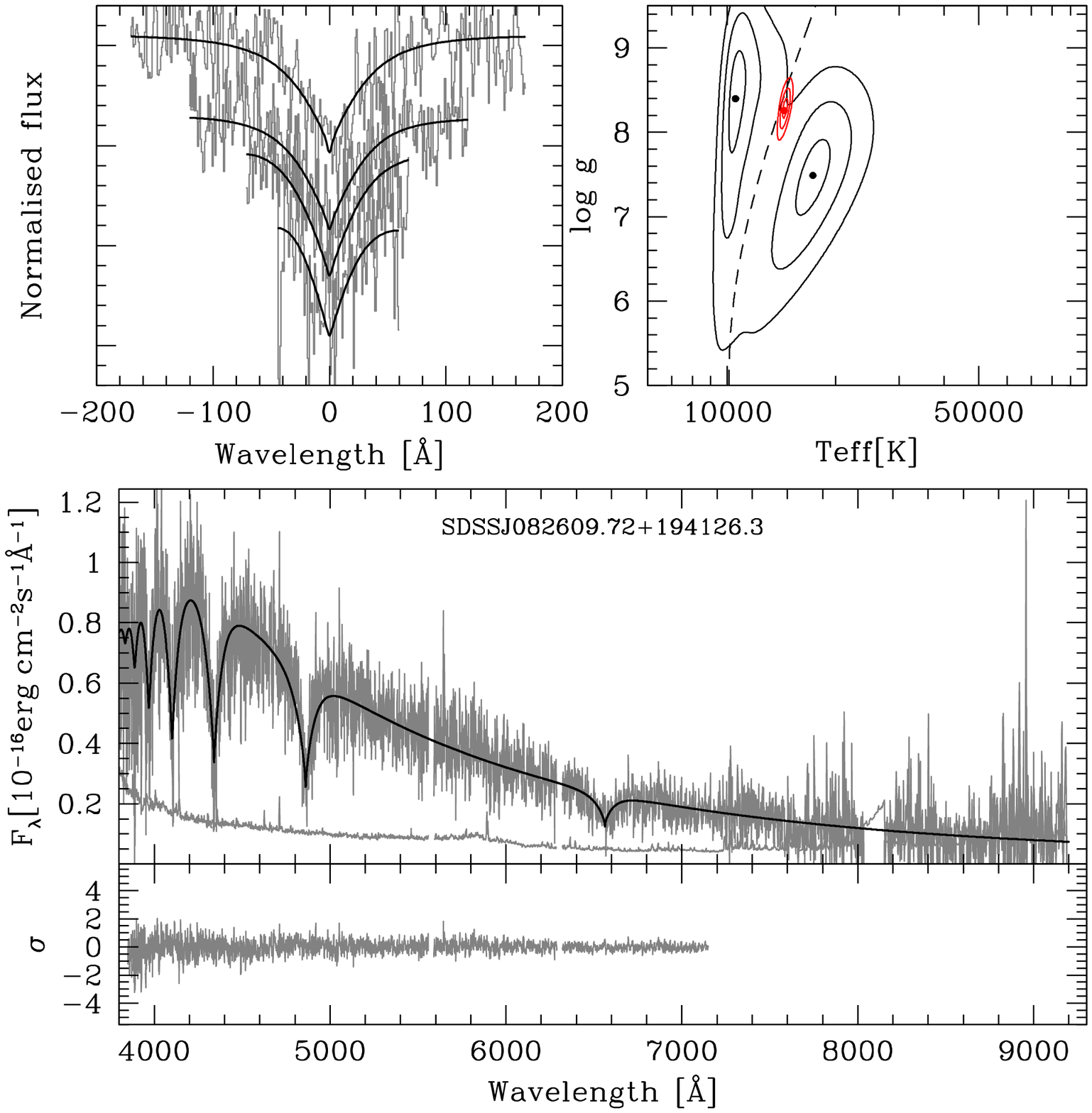}
\caption{\label{f-wdfit}  Spectral  model   fit  to  the  white  dwarf
component of SDSSJ\,082609.72+194126.3  obtained after subtracting the
best-fit M-dwarf template.  Top  left panel: best-fit (black lines) to
the  normalised H$\beta$ to  H$\epsilon$ (gray  lines, top  to bottom)
line  profiles.   Top right  panels:  3,  5,  and 10$\sigma$  $\chi^2$
contour plots in the $\Teff-\log g$ plane. The black contours refer to
the best  line profile fit, the red  contours to the fit  of the whole
spectrum.   The  dashed  line  indicates  the  occurrence  of  maximum
H$\beta$ equivalent width.  The best ``hot'' and ``cold'' line profile
solutions  are indicated  by black  dots, the  best fit  to  the whole
spectrum is indicated  by a red dot. Bottom  panel: the residual white
dwarf spectra resulting from the spectral decomposition and their flux
errors (gray lines)  along with the best-fit white  dwarf model (black
line)  and the  residuals of  the fit  (gray line,  bottom).  See also
Figs.\ref{f-decomp},\ref{f-galex}.}
\end{center}
\end{figure}

\begin{figure}
\begin{center}
\includegraphics[angle=-90,width=\columnwidth] {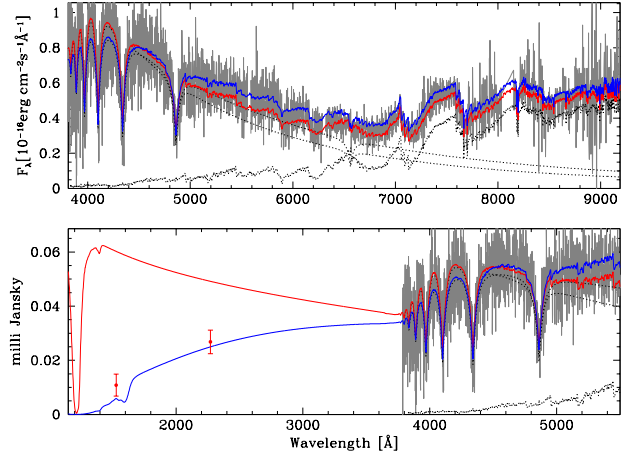}
\caption{\label{f-galex} Top:  in blue  the cold white  dwarf solution
obtained  in Fig.\,\ref{f-wdfit}  plus the  best-fit  M-dwarf template
obtained  in Fig.\,\ref{f-decomp};  in red  the same  but for  the hot
solution; in  gray the SDSS spectrum  of SDSSJ\,082609.72+194126.3; in
black dotted line  the best-fit M-dwarf template, and  the white dwarf
models that satisfy the cold and hot solutions in Fig.\,\ref{f-wdfit}.
Bottom:  the same but  including the  near- and  far-ultraviolet GALEX
magnitudes     of    this    object     (red    dots).     See    also
Figs.\ref{f-decomp},\ref{f-wdfit}.}
\end{center}
\end{figure}

As a  second modification we  take into account the  ultraviolet GALEX
magnitudes if the spectral fitting  does not provide a unique solution
for   the  white  dwarf   temperatures.   An   example  is   given  in
Figs.\,\ref{f-decomp}-\ref{f-galex}.     Performing    the    spectral
decomposition  of SDSSJ\,082609.72+194126.3 into  the white  dwarf and
main  sequence  components  (Fig.\,\ref{f-decomp}) and  modelling  the
Balmer lines of  the white dwarf model fit  provides two solutions for
the  white dwarf  temperature, the  so called  cold and  hot solutions
(black dots  in the top  right panel of Fig\,\ref{f-wdfit}).   In most
cases an additional  fit to the entire spectrum  breaks this ambiguity
and clearly indicates which of  the two solutions has to be preferred.
However,  in  the  case  of  SDSSJ\,082609.72+194126.3  the  ambiguity
remains as the fit to the  entire spectrum results in values that fall
exactly on the line of maximum H$\beta$ equivalent width.  As shown in
the  bottom panel of  Fig.\,\ref{f-galex}, the  ultraviolet magnitudes
measured by GALEX clearly exclude the hot solution.

The  stellar  parameters and  distances  obtained  for  the 1602  WDMS
binaries in  our catalogue are given  in Table\,\ref{t-paramcat}. Both
the cold and  the hot solutions are provided  for each spectrum, while
the solution preferred by us is given in the first line.  The complete
table is available in the electronic version of the paper.

\section{Distribution of stellar parameters}
\label{s-histo}

We  present  in this  section  distributions  of  surface gravity  and
effective temperature  of the  white dwarfs, as  well as  the spectral
type  distribution  of  the  companion  stars  in  our  catalogue.  To
facilitate  the comparison  with previous  works we  also  provide the
distribution of  white dwarf  masses.  Our template  fitting algorithm
covers  secondary star  spectral  types M0-M9  and  the spectral  type
distribution shown  here includes  only clear classifications  in this
range. For the distributions of  the white dwarf stellar parameters we
only considered  systems with relative errors smaller  than 25\%. This
resulted in 1433,  1198, 1127, and 558 WDMS  binaries for the spectral
type,  surface gravity,  effective temperature,  and white  dwarf mass
distributions, respectively (see Fig.\,\ref{f-histocat}).

In general terms  the distributions are similar to  those presented in
\citet{rebassa-mansergasetal07-1}:  the   most  frequent  white  dwarf
temperatures range between 10000--20000\,K, white dwarf masses cluster
around  $M_\mathrm{wd}\simeq0.5$\,M$_\odot$, $\log  g \simeq$  7.8 for
the vast  majority of the white  dwarfs, and the spectral  type of the
companion stars are typically  M3--4.  In Fig.\,\ref{f-kshist} we show
the  \Teff, $M_\mathrm{wd}$,  $\log g$,  and spectral  type cumulative
distributions  obtained from the  WDMS binaries  studied in  this work
(blue        lines),       the        systems        analysed       in
\citet{rebassa-mansergasetal07-1} (red lines)  and of a volume-limited
sample of single white dwarfs \citep[black lines,][]{holbergetal08-1}.
Kolmogorov-Smirnov  (KS) tests  were applied  to compare  the  sets of
effective  temperatures, white  dwarf masses,  and $\log  g$.   As the
secondary spectral  type distribution  consists of discrete  values we
performed  a $\chi^2$  test in  this case.   We briefly  describe each
parameter distribution in the following sub-sections.

\subsection{White dwarf temperature}

We obtain a $40\%$ probability (called KS probability in what follows)
for the maximum vertical  distance between the white dwarf temperature
distribution    obtained     here    and    the     distribution    in
\citet{rebassa-mansergasetal07-1}  being equal to  or larger  than the
maximum  vertical distance  between  the two  cumulative white  dwarf
temperature distributions.   Hence, there are no  indications that the
two distributions are  not drawn form the same  parent population.  In
contrast,   comparing  the   WDMS  binary   white   dwarf  temperature
distribution with  the temperature distribution  of the volume-limited
sample    of   single   white    dwarfs   \citep[][black    lines   in
Fig.\,\ref{f-kshist}]{holbergetal08-1} we  obtain a KS  probability of
10$^{-15}$. This is straight forward  to understand as the presence of
the  secondary  star makes  the  identification  of  cold white  dwarf
primaries rather difficult.

\subsection{White dwarf mass and surface gravity}

Comparing  the  white dwarf  mass  and  surface gravity  distributions
obtained      here     with      the      sample     described      in
\citet{rebassa-mansergasetal07-1} we obtain  KS probabilities of 50\%,
and  2\%  respectively.  The  low  KS  probability  for the  $\log  g$
distributions  is  caused  by  the significantly  higher  fraction  of
low-mass    white    dwarfs     in    the    present    sample    (see
Sect.\ref{s-compreb}). This is most likely an effect of the systematic
changes  in the  flux  calibration pipeline  of  SDSS spectra.   These
changes especially  affect the blue  part in the spectrum  and, hence,
the fit  to the continuum that  allows to distinguish  between the hot
and cold  solutions.  In addition,  the decision between hot  and cold
solution is affected by  taking into account GALEX ultraviolet fluxes,
one  of  the  modifications  incorporated  in  this  paper  (see  also
Sect.\,\ref{s-compreb}).

\begin{figure}
\begin{center}
\includegraphics[width=\columnwidth]{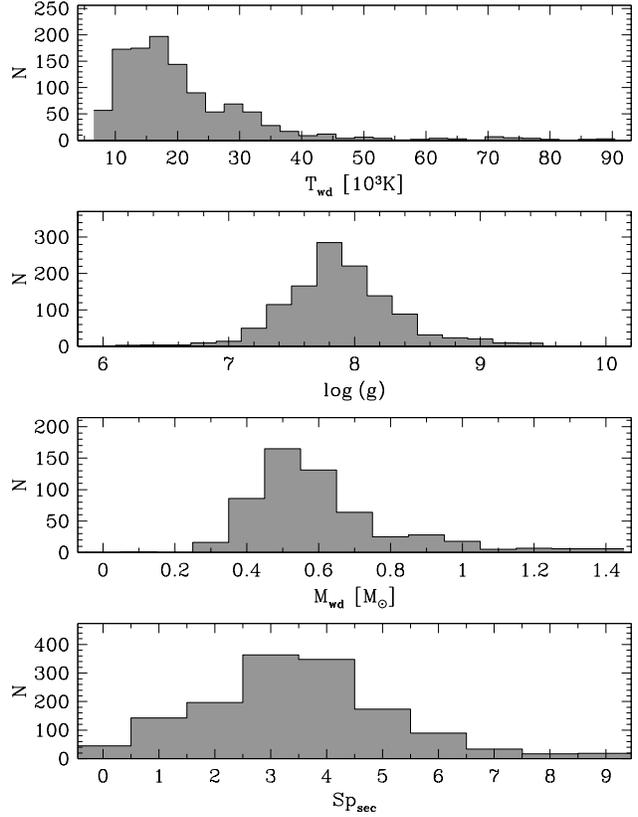}
\caption{\label{f-histocat} White dwarf effective temperature, surface
gravity  and mass, and  spectral type  of the  secondary distributions
obtained from the SDSS WDMS binaries in our catalogue.}
\end{center}
\end{figure}

Comparing  the  obtained  distributions  with  those  derived  from  a
volume-limited sample  of single white dwarfs  \citep[][black lines in
Fig.\,\ref{f-kshist}]{holbergetal08-1}  we obtain KS  probabilities of
10$^{-12}$ for the  white dwarf masses and 10$^{-15}$  for the surface
gravities.  The  white dwarf mass  distributions of WDMS  binaries and
single white  dwarfs are indeed  expected to be different.   The white
dwarf mass distribution of WDMS  binaries is composed of two different
types of  systems, i.e.  WDMS  binaries with an initial  main sequence
binary  separation large  enough  to avoid  mass  transfer during  the
evolution of  the more massive star,  and systems that  suffer from CE
evolution  (PCEBs).    The  white  dwarf  mass   distribution  of  the
non-interacting WDMS binary population  is expected to be identical to
the  mass distributions of  single white  dwarfs, i.e.   clustering at
$M_\mathrm{wd}$  $\simeq$0.6\,M$_\odot$, while  PCEBs are  expected to
contain  a large  number  of low-mass  He-core  white dwarfs  ($M_{\rm
wd}\leq\,0.47$M$_{\odot}$).   For  example,  \citet{dekool+ritter93-1}
estimate the fraction of such  systems among PCEBs to be about $50\%$.
As $\sim30\%$ of  the presented sample of WDMS  binaries from the SDSS
have evolved through a  CE phase \citep{schreiberetal08-1}, the rather
large   fraction   of  white   dwarfs   with  $M_\mathrm{wd}   \simeq$
0.4\,M$_\odot$  in   Fig.\,\ref{f-histocat}  is  not   surprising  but
expected.  We  are currently pursuing a  large-scale follow-up project
of WDMS  binaries to identify the  PCEBs in our sample  and to measure
their orbital parameters.  A detailed analysis of the white dwarf mass
distribution   of   PCEB   and   wide  WDMS   binaries   is   underway
(Rebassa-Mansergas et al.  2009a, in preparation).

\subsection{Secondary star spectral type}
 
\begin{figure}
\begin{center}
\includegraphics[angle=-90,width=\columnwidth]{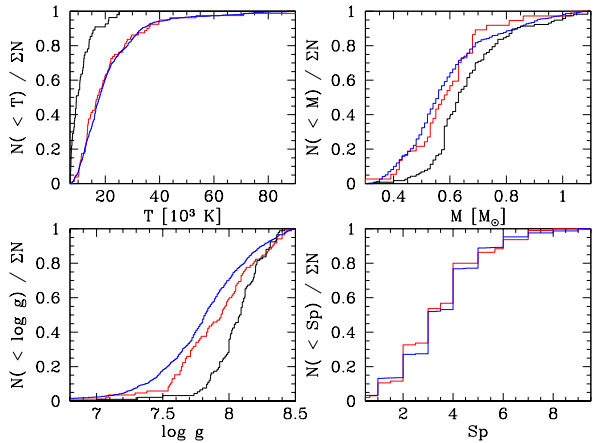}
\caption{\label{f-kshist}  Effective  temperature  (top  left),  white
dwarf  mass (top right),  surface gravity  (bottom left)  and spectral
type (bottom right) cumulative distributions obtained from the stellar
parameters of the WDMS  binaries presented here (blue lines), analysed
in   \citet{rebassa-mansergasetal07-1}  (red   lines),   and  from   a
volume-limited   sample   of    single   white   dwarfs   \citep[black
lines,][]{holbergetal08-1}.}
\end{center}
\end{figure}

The distribution  of spectral  types is shown  in the bottom  panel of
Fig.\,\ref{f-histocat}.  The  cut-off at early spectral  types is very
likely a consequence of selection  effects as discussed in more detail
in  Sect.\,\ref{s-effects}.  The  strongly decreasing  number  of WDMS
binaries with  late type (M7-M9)  companions might also be  related to
selection effects, as  late-type stars are faint and  harder to detect
against  a moderately hot  white dwarf.   Nevertheless, SDSS  covers a
much  broader colour space  than previous  surveys and,  in principle,
should be  able to identify  more WDMS binaries containing  cool white
dwarfs plus  very late-type companions.   It is worth  mentioning also
that \citet{farihietal05-1} have constructed the relative distribution
of spectral  types in  the local M/L  dwarf distribution,  which peaks
around  M3--4,  and steeply  declines  towards  later spectral  types,
suggesting that late-type companions to white dwarfs are intrinsically
rare  (see  Sect.\,\ref{s-effects} for  more  details).  The  $\chi^2$
comparison  of the  WDMS binary  spectral type  distribution presented
here with the one described in \citet{rebassa-mansergasetal07-1} gives
a probability  of $82\%$  that both distributions  are drawn  from the
same parent distribution (see also Fig.\,\ref{f-kshist}).

\subsection{Distances}
\label{s-dist}

\begin{figure*}
\begin{center}
\includegraphics[angle=-90,width=\columnwidth]{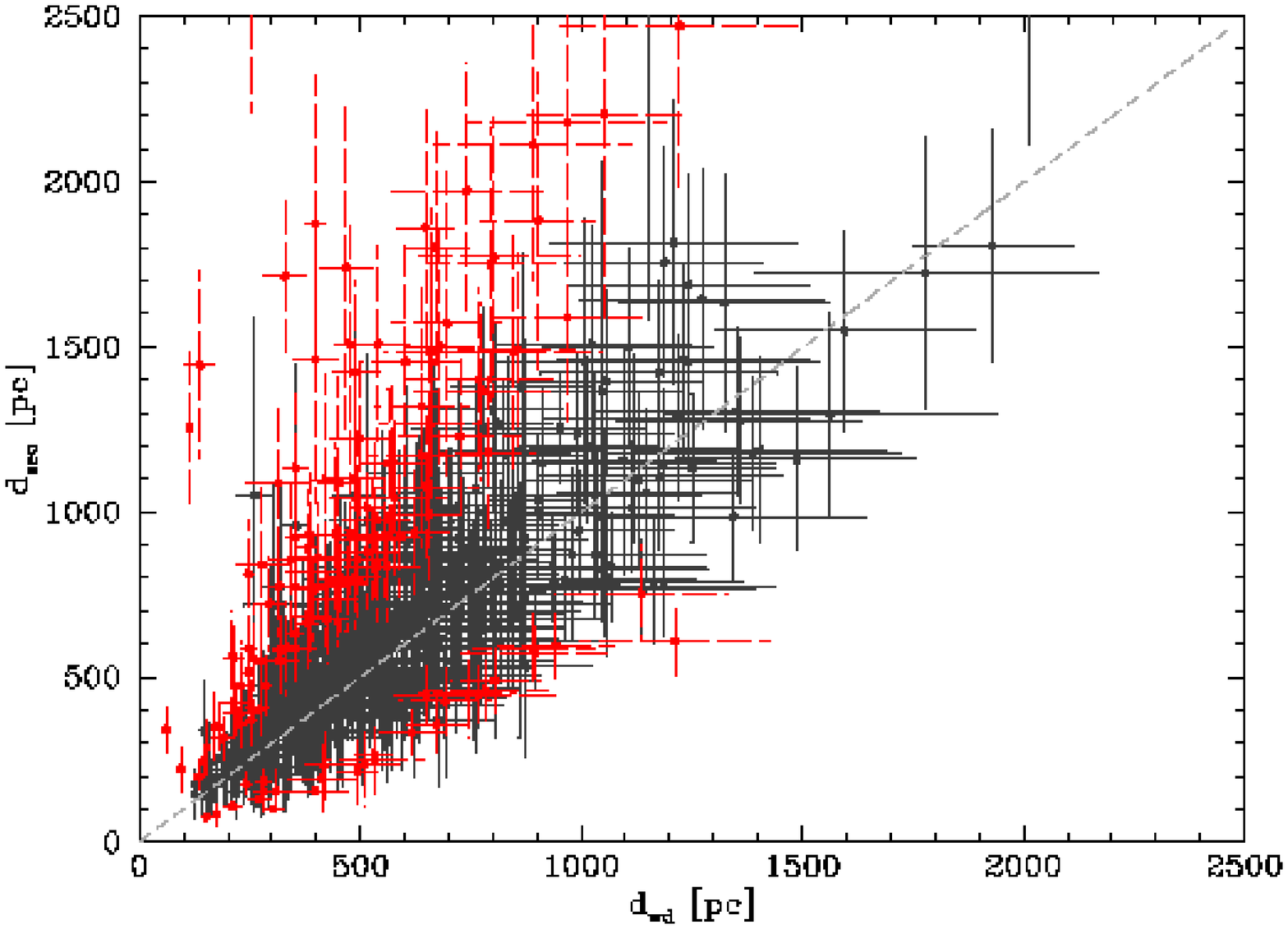}
\includegraphics[angle=-90,width=\columnwidth]{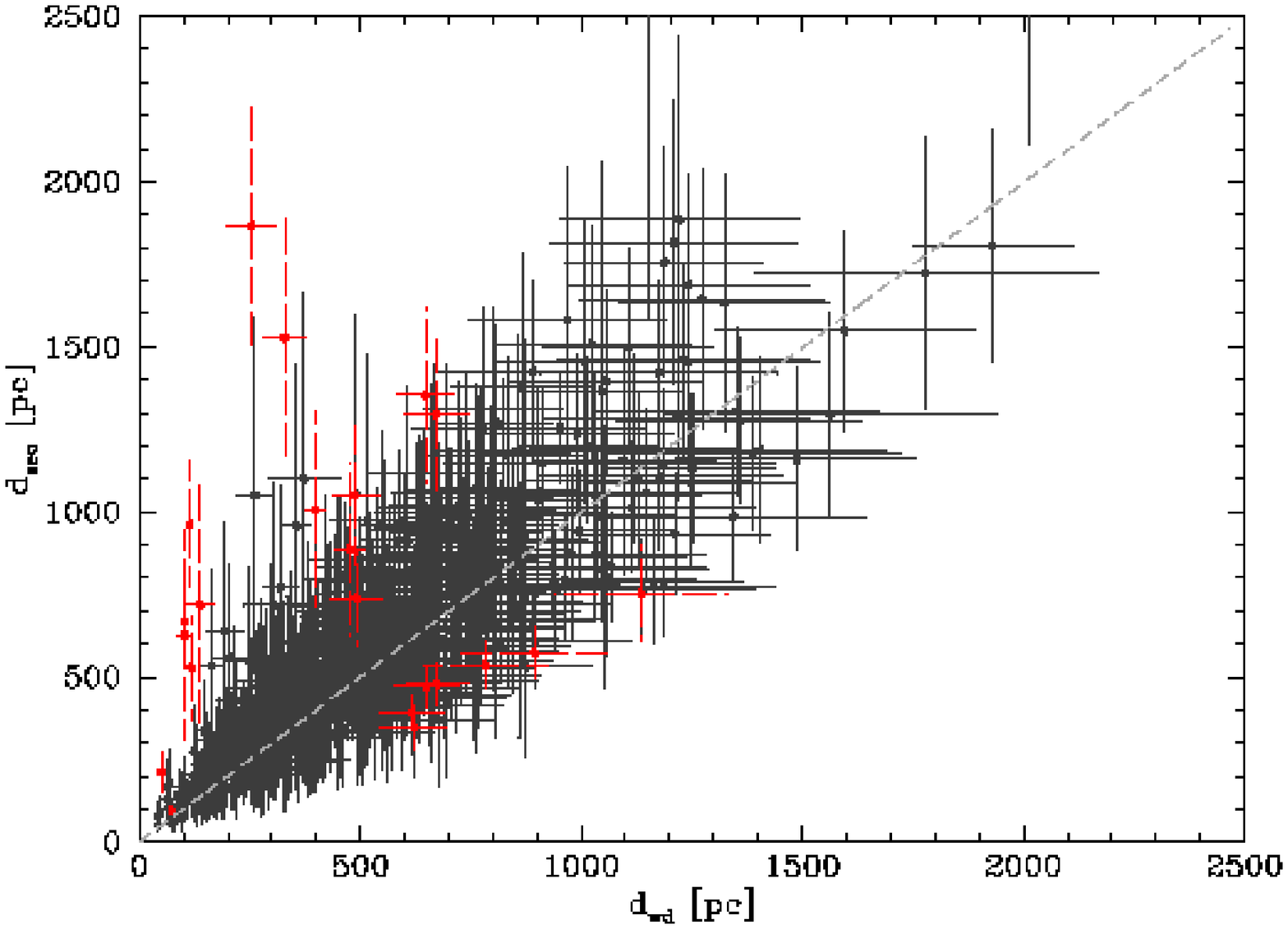}
\caption{Left    panel:    Comparison    of    $d_\mathrm{sec}$    and
$d_\mathrm{wd}$  obtained from  our spectral  decomposition  and white
dwarf fits to  the SDSS spectra. Approximately 1/5  of the systems are
$d_\mathrm{sec}  > d_\mathrm{wd}$  outliers by  more  than 1.5$\sigma$
(red dots).   Right panel: the  spectral types of the  secondary stars
were    adjusted    by    1--2    spectral    classes    to    achieve
$d_\mathrm{wd}=d_\mathrm{sec}$.}
\label{f-distcat}
\end{center}
\end{figure*}

In Sect.\,\ref{s-param} we  derived two independent distance estimates
for the  WDMS binaries in our  catalogue.  We compare  these values in
Fig.\,\ref{f-distcat}.   We here only  consider systems  with relative
errors in the white dwarf  distance less than 25\%. The relative error
in  $d_\mathrm{sec}$  is dominated  by  the  scatter  in the  spectral
type-radius  relation  provided in  \citet{rebassa-mansergasetal07-1}.
Hence,  it   represents  an   intrinsic  uncertainty  rather   than  a
statistical error related to the fit and we therefore do not apply any
cut in $d_\mathrm{sec}$.  Moreover,  we excluded from the analysis any
WDMS  binary  containing a  white  dwarf  of  less than  12000\,K,  as
spectral fits tend to overestimate  the surface gravity, and hence the
mass \citep{koesteretal08-1}, leading to an underestimate of the white
dwarf radius, and hence the  distance.  If more than one SDSS spectrum
is available, we use averaged distances.  This procedure resulted in a
sample of 603 WDMS binaries.

In the top panel of Fig.\,\ref{f-distcat} black dots represent systems
in which the two distances  agree within 1.5$\sigma$ ($\sim$3/4 of the
total  sample)  level, while  the  differences  in  the two  distances
obtained for the  objects in red exceed this limit.   Most of the $\ge
\,1.5\sigma$  outliers  are  above  $d_\mathrm{sec}  =  d_\mathrm{wd}$
($\sim$1/5  of our  WDMS binaries),  while we  find  $d_\mathrm{sec} <
d_\mathrm{wd}$ for only $\sim$ 5\% of the entire sample.  Hence, there
seems  to be  a systematic  effect  that leads  to overestimating  the
secondary star distances.   As in \citet{rebassa-mansergasetal07-1} we
assume that magnetic activity raises the temperature of the inter-spot
regions  in active  stars  that  are heavily  covered  by cool  spots,
leading to  a blue-shift of  the optical colours compared  to inactive
stars.  However, to finally  evaluate this interpretation one needs to
perform  a   detailed  activity  analysis  of  our   sample  based  on
H$_{\alpha}$  emission.   This would  further  require to  distinguish
between the different sub-samples  forming the WDMS binary population,
i.e.   PCEBs and  wide systems,  as the  fraction of  active  stars is
expected to depend on the rotation  rate.  This is beyond the scope of
this  paper  but  will  be  presented  in  a  forthcoming  publication
(Rebassa-Mansergas et al.  2009b, in preparation).

As a first simple test, we adjust the spectral type of the secondaries
to bring into agreement the two  distances. It turns out that a change
of 1--2 spectral subclasses is enough for the majority of cases.  Only
ten systems in the right  panel of Fig.\,\ref{f-distcat} need a change
of  more  than two  spectral  subclasses  to  reach $d_\mathrm{sec}  =
d_\mathrm{wd}$.  We  have inspected these  objects in more  detail and
the large discrepancy might be explained as follows:

\begin{itemize}
\item    Six    objects     contain    hot    white    dwarfs,    i.e.
SDSSJ\,003221.86+073934.4,                   SDSSJ\,032510.84-011114.1,
SDSSJ\,080229.99+072858.1,                   SDSSJ\,095719.25+234240.8,
SDSSJ\,101323.90+043946.1,  SDSSJ\,141536.40+011718.2.   If these  are
short orbital period systems, irradiation  of the secondary by the hot
primary may lead to overestimating the distance to the secondary star.
These  WDMS binaries might  therefore be  considered as  candidates to
probe  for   radial  velocity  variations.   SDSS\,J032510.84-011114.1
benefits   from   two    SDSSJ   radial   velocity   measurements   in
Table\,\ref{t-rvsall} but no variation  is detected.  This may reflect
the speculative nature of the just  given argument or be caused by the
SDSS spectroscopy sampling the same orbital phase twice.

\item         Three         systems        (SDSSJ\,025306.37+001329.2,
SDSSJ\,204729.04-064536.7, SDSSJ\,210624.12+004030.2)  are resolved in
their SDSS images. Depending on  the exact placement of the fibre, the
flux contribution of one or both stars is likely to be underestimated.
This   translates   to  underestimated   flux   scaling  factors   and
overestimated distances.

\item  For  one  object,  SDSSJ\,232624.72-011327.2, the  S/N  of  the
corresponding SDSS spectrum is very  low (S/N = 4.2). This is probably
the reason for the discrepancy found in the distances.
\end{itemize}

\section{Comparison with previous studies}
\label{s-comparison}

Since  \citet{smolcicetal04-1}  discovered   the  WDMS  binary  bridge
several SDSS  WDMS binary catalogues have been  presented, and several
fitting  routines  to  determine  the  stellar  parameters  have  been
applied.  In this section we compare in more detail the results of our
spectral decomposition method with those obtained by earlier studies.

\subsection{Comparison with \citet{rebassa-mansergasetal07-1}.}
\label{s-compreb}

\begin{figure*}
\begin{center}
\includegraphics[angle=-90,width=\textwidth]{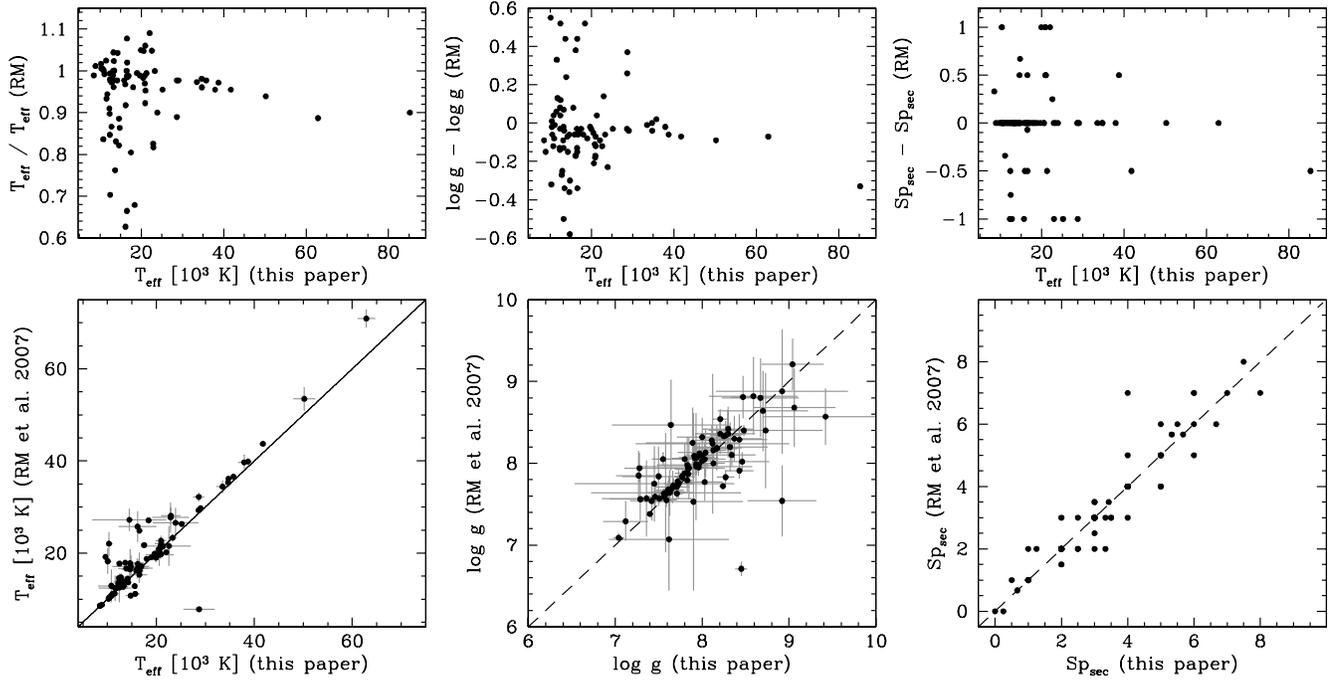}
\caption{Bottom  panels from left  to right:  comparison of  the white
dwarf  effective  temperatures, surface  gravities,  and the  spectral
types  of the secondary  stars determined  in this  work and  those of
\citet{rebassa-mansergasetal07-1}.   Top panels,  from left  to right:
the  white dwarf effective  temperature ratio,  and the  difference in
surface  gravity  and the  secondary's  spectral  types  from the  two
studies as a function of the white dwarf temperature.}
\label{f-dr5-dr6}
\end{center}
\end{figure*}

Given that the  SDSS spectra reduction pipeline was  improved with DR6
\citep{adelman-mccarthyetal08-1},  we decided  to compare  the stellar
parameters    obtained    here    with    those   we    obtained    in
\cite{rebassa-mansergasetal07-1}.   The stellar  parameters  for those
objects     with    multiple     SDSS    spectra     were    averaged.
Fig.\,\ref{f-dr5-dr6}  compares  white  dwarf effective  temperatures,
surface gravities  and secondary  spectral types.  Both  studies agree
within  the errors  in the  majority of  cases, with  average relative
differences of 13.5\% and 2.5\%  in \Teff\, and $\log g$ respectively,
and  an  average  difference  of  0.3 spectral  subtypes.   We  obtain
significantly different values especially for systems containing white
dwarfs with temperatures in the range 10.000-20.000\,K (see top panels
of Fig.\,\ref{f-dr5-dr6}). In  $\sim$50\% of the cases this  is due to
the fact  that we  are making use  of ultraviolet GALEX  magnitudes to
constrain the white dwarf line fit solutions in the present paper (see
Sect.\,\ref{s-param}). In the remaining $\sim$50\% this is most likely
a  consequence  of the  systematic  changes  in  the flux  calibration
pipeline.  The most dramatic  case is SDSS\,J151045.70+404827, a clear
outlier  in the  bottom left  panel of  Fig.\,\ref{f-dr5-dr6},  with a
difference  in  \Teff\,  and  $\log   g$  of  21.000\,K  and  1.7  dex
respectively.  This  was one  of the cases  in which the  solution was
modified by the  use of GALEX magnitudes. The  obtained differences in
spectral  type are  consistent  with the  general  uncertainty of  our
decomposition/fitting  procedure  of   0.5  spectral  subclasses  (see
Sect.\,\ref{s-param}).    Only   for  SDSS\,J173548.36+541424.4   this
difference  exceeds one spectral  subtype. This  system is  very faint
($i\sim20$),  and the  low  S/N  ratio is  causing  the spectral  type
determination to be rather uncertain.

\begin{figure*}
\begin{center}
\includegraphics[angle=-90,width=\textwidth]{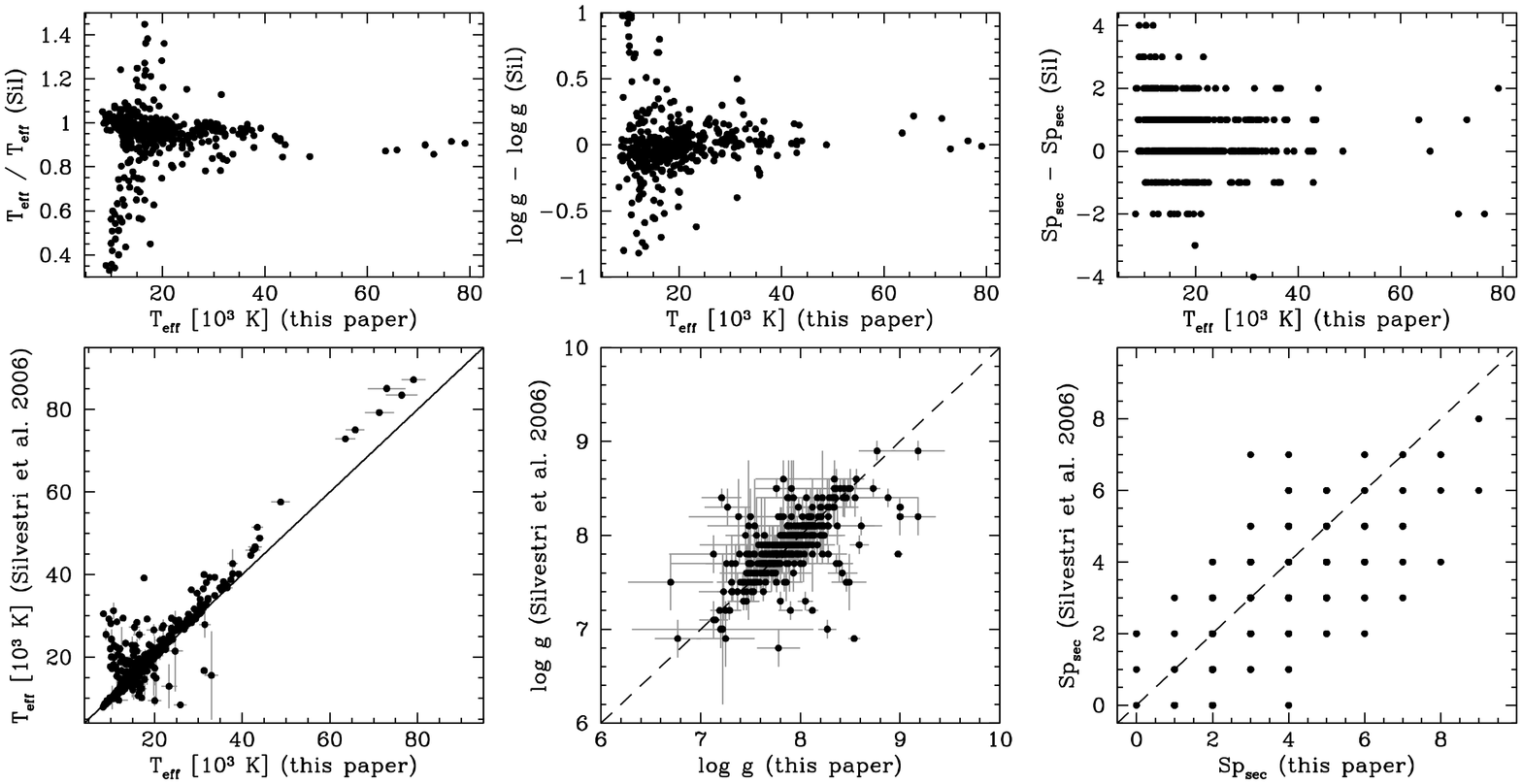}
\caption{\label{f-comp_sil}   Bottom  panels   from  left   to  right:
comparison  of  the white  dwarf  effective  temperatures and  surface
gravities  and the spectral  types of  the secondary  stars determined
from  our fits  and those  of \citet{silvestrietal06-1}.   Top panels,
from left to  right: the white dwarf effective  temperature ratio, and
the difference  in surface gravity and the  secondary's spectral types
from the two studies as a function of the white dwarf temperature.}
\end{center}
\end{figure*}

\begin{figure*}
\begin{center}
\includegraphics[angle=-90,width=\textwidth]{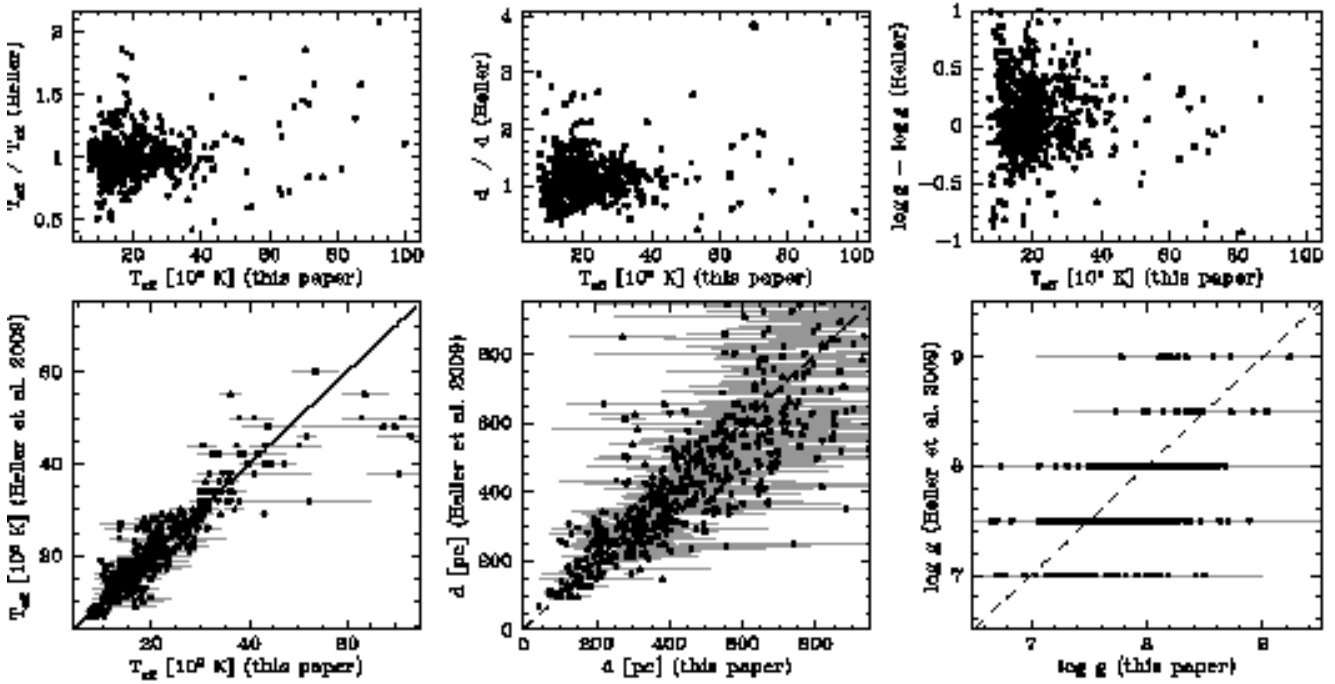}
\caption{\label{f-heller1}   Bottom  panels   from   left  to   right:
comparison of  the white  dwarf effective temperatures,  distances and
surface   gravities   determined   from   our  fits   and   those   of
\citet{helleretal09-1}.   Top panels,  from left  to right:  the white
dwarf effective temperature and distance ratios, and the difference in
surface gravity from the two studies  as a function of the white dwarf
temperature.}
\end{center}
\end{figure*}

\begin{figure*}
\begin{center}
\includegraphics[angle=-90,width=\textwidth]{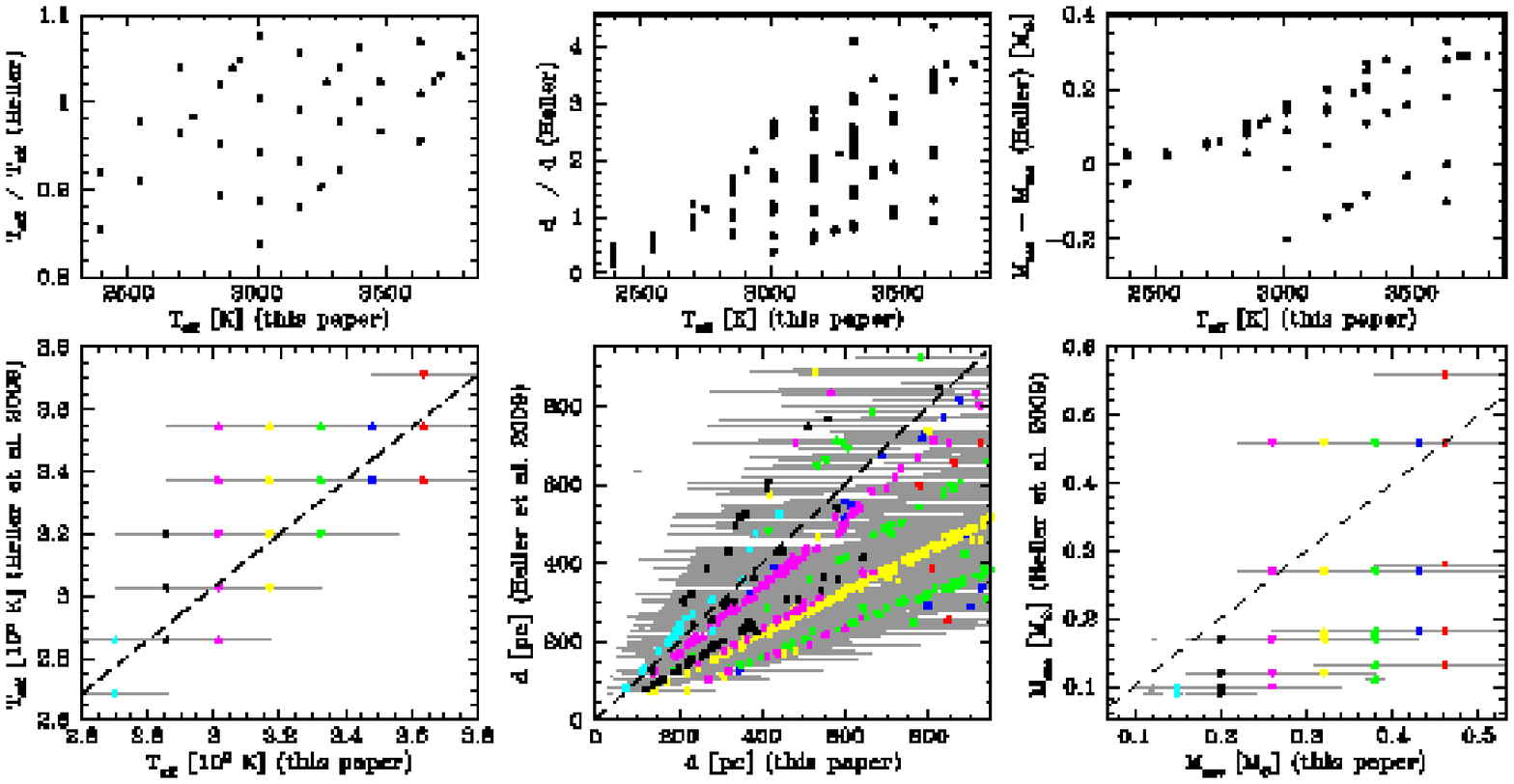}
\caption{\label{f-heller2}   Bottom  panels   from   left  to   right:
comparison of the secondary star effective temperatures, distances and
masses determined  from our fits and  those of \citet{helleretal09-1}.
The colors red, blue, green, yellow, magenta, black, and cyan refer to
spectral types  M1-M7 respectively.  Top  panels, from left  to right:
the secondary star effective  temperature and distance ratios, and the
mass difference from the two studies  as a function of the white dwarf
temperature.}
\end{center}
\end{figure*}

\subsection{Comparison with \citet{vandenbesselaaretal05-1}}
\label{s-besselaar}

We have identified in this work  53 DB/M dwarf binaries, 13 of them in
common  with \citet{vandenbesselaaretal05-1}.  The  stellar parameters
of these  systems are given  in Table\,\ref{t-besselaar}.  Apparently,
the    white   dwarf    temperatures    differ   significantly.     As
\citet{vandenbesselaaretal05-1} used DR3 SDSS spectra and measured all
parameters by  assuming a white dwarf mass  of 0.6 M$_\mathrm{\odot}$,
these discrepancies are not too surprising.  In addition, as stated by
\citet{vandenbesselaaretal05-1},  their derived  effective temperature
is related  to the fitting of  the secondary star,  i.e.  changing the
spectral type  by one  subclass can lead  to differences in  the white
dwarf effective temperatures of $8000$\,K to $10000$\,K.

The distance measurements by \citet{vandenbesselaaretal05-1} are based
on the white  dwarf fitting, while we use  the secondary star spectral
type in case the primary is  not a DA white dwarf. Taking into account
the uncertainties  involved in  both distance measurements,  we obtain
reasonable agreement between both values.

Concerning the spectral type of  the companion stars our values are in
good agreement with those obtained by \citet{vandenbesselaaretal05-1}.
In      all      but      two      (SDSS\,J093645.14+420625.6      and
SDSS\,J113609.59+484318.9)  cases the difference  does not  exceed one
subclass.   SDSS\,J093645.14+420625.6 is  a $i=20$th  magnitude object
with a low S/N spectrum leading to large uncertainties in the obtained
parameters.   SDSS\,J113609.59+484318.9  contains  a hot  white  dwarf
($\gappr30000\,$K) that significantly contaminates the spectrum of the
M-dwarf.

\subsection{Comparison with \citet{silvestrietal06-1}}

We  compare  in   Fig.\,\ref{f-comp_sil}  the  white  dwarf  effective
temperatures  and  surface  gravities,  and secondary  spectral  types
obtained    in   Sect.\,\ref{s-param}    with   those    obtained   by
\citet{silvestrietal06-1}. We  considered a sample of  421 spectra for
which both studies present values for the stellar parameters.

\begin{table}
\caption{\label{t-besselaar}  Effective  temperatures, spectral  types
and distances  obtained from \citet{vandenbesselaaretal05-1}  and this
work.   The first  line on  each  systems corresponds  to the  results
obtained by \citet{vandenbesselaaretal05-1},  the second line provides
our results.}  \setlength{\tabcolsep}{1.8ex}
\begin{center}
\begin{tabular}{lcrccr}
\hline
\hline
SDSS\,J   &             \Teff[k]& err  & Sp&    d[pc] &   err \\
\hline
075235.79+401339.0 &   30252 & 4000 &  3 &  1544 &      \\
                   &   16811 &   56 &  3 &  1295 & 255  \\
080636.85+251912.1 &   24266 & 4000 &  3 &  1045 &      \\
                   &   17439 &   78 &  3 &   852 & 168  \\
093645.14+420625.6 &   15919 & 4000 &  5 &   860 &      \\
                   &       - &    - &  3 &  1949 & 384  \\
100636.39+563346.8 &   14575 & 4000 &  4 &   531 &      \\
                   &   16071 &   46 &  3 &   817 & 161  \\
102131.55+511622.9 &   30252 & 4000 &  4 &   700 &      \\
                   &   17622 &   25 &  5 &   479 & 245  \\
113609.59+484318.9 &   38211 & 4000 &  6 &   354 &      \\
                   &   31324 &  234 &  3 &   896 & 176  \\
134135.23+612128.7 &   30694 & 4000 &  3 &  1054 &      \\
                   &   16051 &   10 &  3 &   886 & 175  \\
143222.06+611231.1 &   36815 & 4000 &  3 &   879 &      \\
                   &   16180 &   47 &  4 &   527 & 155  \\
144258.47+001031.5 &   30694 & 4000 &  3 &   674 &      \\
                   &   31324 &  234 &  4 &   358 & 105  \\
150118.40+042232.3 &   26020 & 4000 &  3 &  1220 &      \\
                   &   18394 &   40 &  3 &  1111 & 219  \\
162329.50+355427.2 &   24266 & 4000 &  3 &   695 &      \\
                   &   15730 &   39 &  3 &   579 & 114  \\
220313.29+113236.0 &   30694 & 4000 &  4 &  1085 &      \\
                   &   19992 &  127 &  4 &  1036 & 305  \\
232438.31-093106.5 &   36815 & 4000 &  3 &   911 &      \\
                   &   19340 &   96 &  4 &   530 & 156  \\
\hline
\end{tabular}
\end{center}
\end{table}

The average relative difference  in effective temperatures and surface
gravities is  reasonably low, i.e.  14.7\% and 2\%  respectively.  For
systems with  white dwarf temperatures  below $\sim$20000\,K, however,
the obtained values  can differ by up to  $22100$\,K and 1.6\,dex.  As
discussed  in  \citet{rebassa-mansergasetal07-1},  we interprete  this
strong disagreement to be caused  by the ambiguity between the hot and
the  cold solutions.   At higher  temperatures ($>$50000\,k)  our fits
tend  to  provide  lower  values   of  \Teff\,  than  those  given  in
\citet{silvestrietal06-1}. This is probably caused by their use of DR4
spectra that were reduced with  a different pipeline.  In the majority
of  cases the  secondary star  spectral types  are in  reasonably good
agreement    (i.e.    the    difference   is    not    exceeding   one
subtype). However, for $\sim$17\%  of the WDMS binaries the difference
is of two or more subtypes, with a maximum difference of four.

We  have  also inspected  the  systems that  \citet{silvestrietal06-1}
failed to  fit and find that in  $\sim$70\% of these we are able to
find     a      solution.      As     previously      discussed     in
\citet{rebassa-mansergasetal07-1}  this indicates  that our  method is
more  robust  if the  S/N  ratio  is low  or  if  one  of the  stellar
components contributes little to the total flux.

\subsection{Comparison with \citet{helleretal09-1}}

\citet{helleretal09-1} studied a sample of 857 WDMS binary candidates.
For 636  of these systems they  provide the results  of an independent
spectroscopic   parameter  fitting   method.   Figure\,\ref{f-heller1}
compares the white dwarf effective temperatures, distances and surface
gravities obtained by  \citet{helleretal09-1} with the values obtained
here.  The  average relative differences between the  two analysis are
12.1\% (white dwarf temperature), 23.9\% (distance) and 3.6\% (surface
gravity). The agreement between both  studies is difficult to asses as
\citet{helleretal09-1} do not provide error estimates.  If one assumes
uncertainties of  2000\,K and 0.3 dex  for their \Teff\,  and $\log g$
values, and a  typical error of 17\% for  their obtained distances, we
find that  only 18\%,  7\% and  16\% of the  given values  (for \Teff,
distance,  and  $\log   g$)  do  not  overlap  at   the  1.5  $\sigma$
level. Hence we conclude that the results obtained in both studies are
in agreement.   Finally, we note  that the horizontal patterns  in the
bottom   right  panel  of   Figure  \ref{f-heller1}   illustrate  that
\citet{helleretal09-1} estimate the surface  gravities by using a grid
with a rather poor resolution of $0.5$\,dex.

\begin{figure}
\begin{center}
\includegraphics[width=\columnwidth]{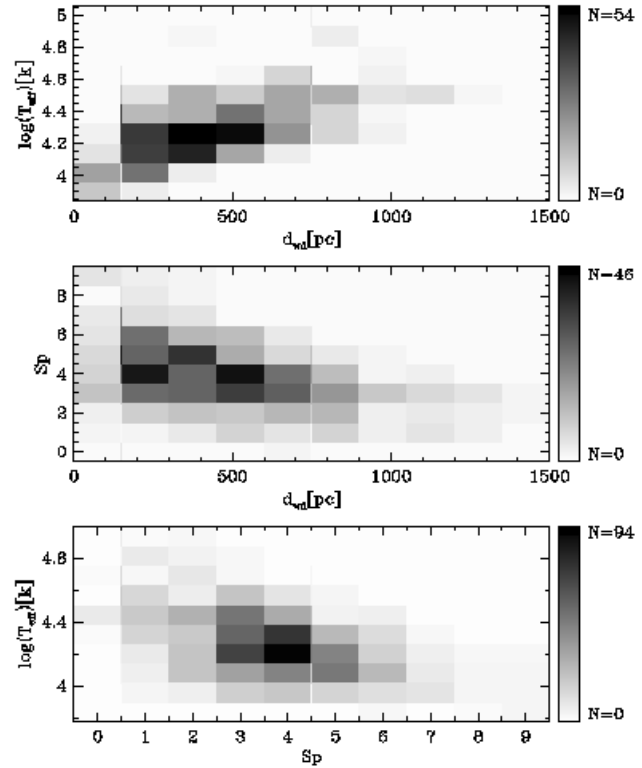}
\caption{Selection effects in SDSS  WDMS binaries can be understood by
analysing  the density  maps obtained  from their  stellar parameters.
From    top   to    bottom    the   ($\log$    \Teff,d$_\mathrm{wd}$),
(d$_\mathrm{wd}$,Sp), and ($\log$ \Teff,Sp) density maps.}
\label{f-effects}
\end{center}
\end{figure}

Fig.\,\ref{f-heller2}  compares the effective  temperatures, distances
and masses of the secondary stars  in both data sets. In this case the
average relative differences are 3.5\% (effective temperature), 42.1\%
(distance) and  42\% (mass).   While we used  M star templates  in our
fitting, \citet{helleretal09-1}  used a grid of  PHOENIX model spectra
\citep{hauschildt+baron99-1}.  The colors  refer to objects containing
secondaries of the same spectral type, i.e.  red, blue, green, yellow,
magenta,  black,  and  cyan  for  spectral  types  M1-M7.   Values  of
effective  temperatures and masses  for the  systems in  our catalogue
were   estimated   using   the   spectral   type-radius-mass-effective
temperature  relations   given  in  \citet{rebassa-mansergasetal07-1}.
Therefore WDMS  binaries containing  secondaries of the  same spectral
type are associated to the  same effective temperature and mass, while
the estimates provided  by \citet{helleretal09-1} cover a considerably
larger range  in both mass  and effective temperature (lower  left and
lower right panels of  Fig.\,\ref{f-heller2}).  According to the lower
right  panel,   \citet{helleretal09-1}  measure  systematically  lower
masses.  A comparison of the obtained distances to the secondary stars
is  shown in the  middle panels  of Fig.\,\ref{f-heller2}.   Since the
observed  flux  is  the same  for  a  given  SDSS spectrum,  the  only
difference between both methods comes  from the radius and the flux at
the stellar  surface. As the grid  of model spectra is  not finer than
our grid of  template spectra both methods differ by  a constant for a
given spectral  type/secondary mass combination.  This  can clearly be
seen  in the bottom  middle panel  of Fig.\,\ref{f-heller2}:  for each
pair of template/model spectrum we obtain one or two straight lines in
the distance-distance  diagram.  It seems that  we overestimate and/or
\citet{helleretal09-1}  underestimate   the  distances.   However,  we
presented     in    \citet{rebassa-mansergasetal07-1}     (see    also
Sect.\,\ref{s-dist}) a reasonable explanation for the $\sim$1/5 of the
WDMS binaries  with $d_\mathrm{sec}>d_\mathrm{wd}$, while  there is no
obvious physical  mechanism that may  account for the  estimated white
dwarf  distances  being  systematically  smaller  than  the  companion
distances as the values of \citep{helleretal09-1} seem to suggest.  We
have to keep in mind  though the uncertainties related to both methods
are quite large and  the activity interpretation of the systematically
larger secondary  distances still needs to be  confirmed. We therefore
conclude  that the  question of  whether  our templates  or the  model
spectra provide more reliable distances remains open.

\begin{figure*}
\begin{center}
\includegraphics[angle=-90,width=\columnwidth]{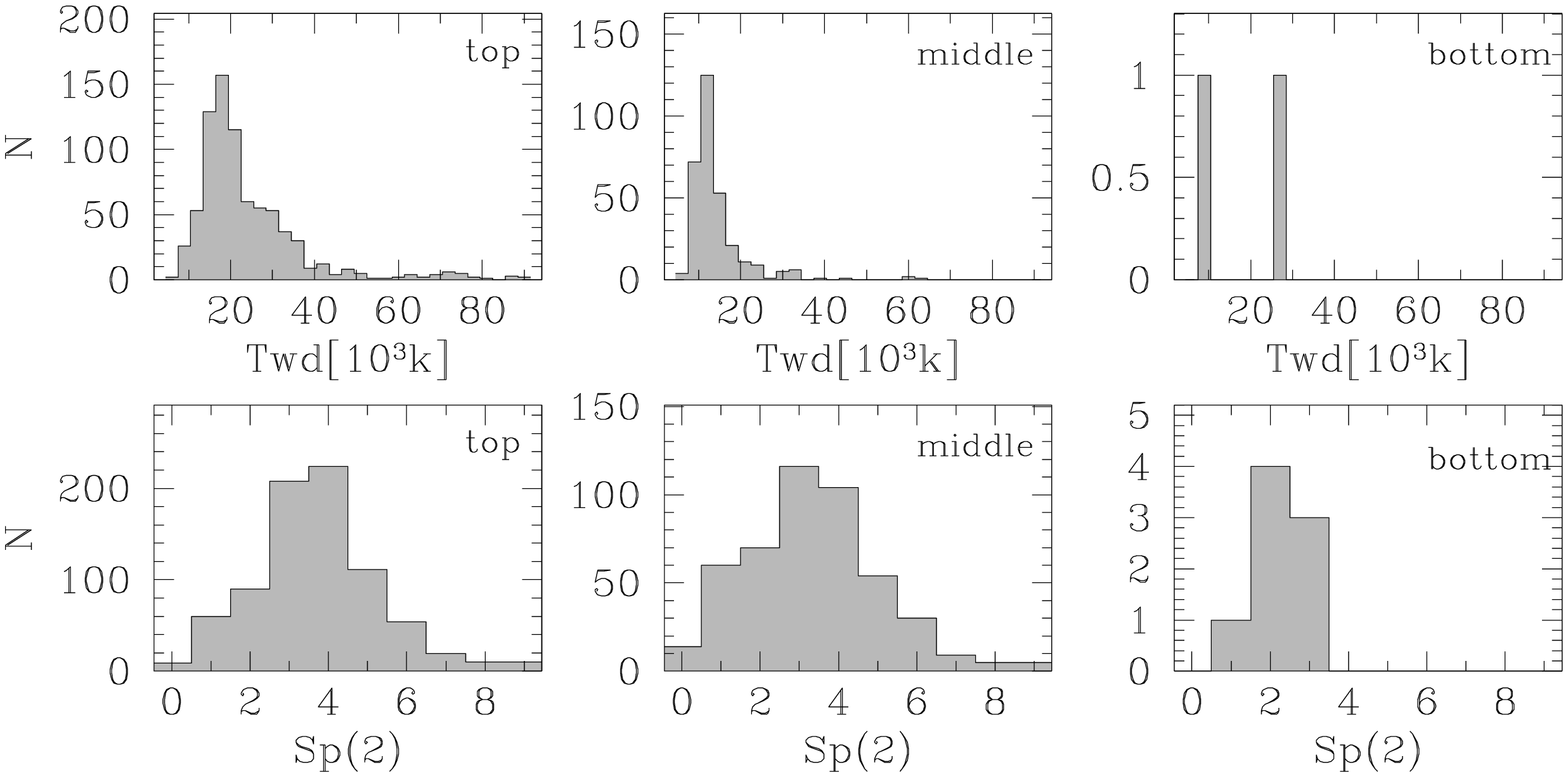}
\vline
\includegraphics[angle=-90,width=\columnwidth]{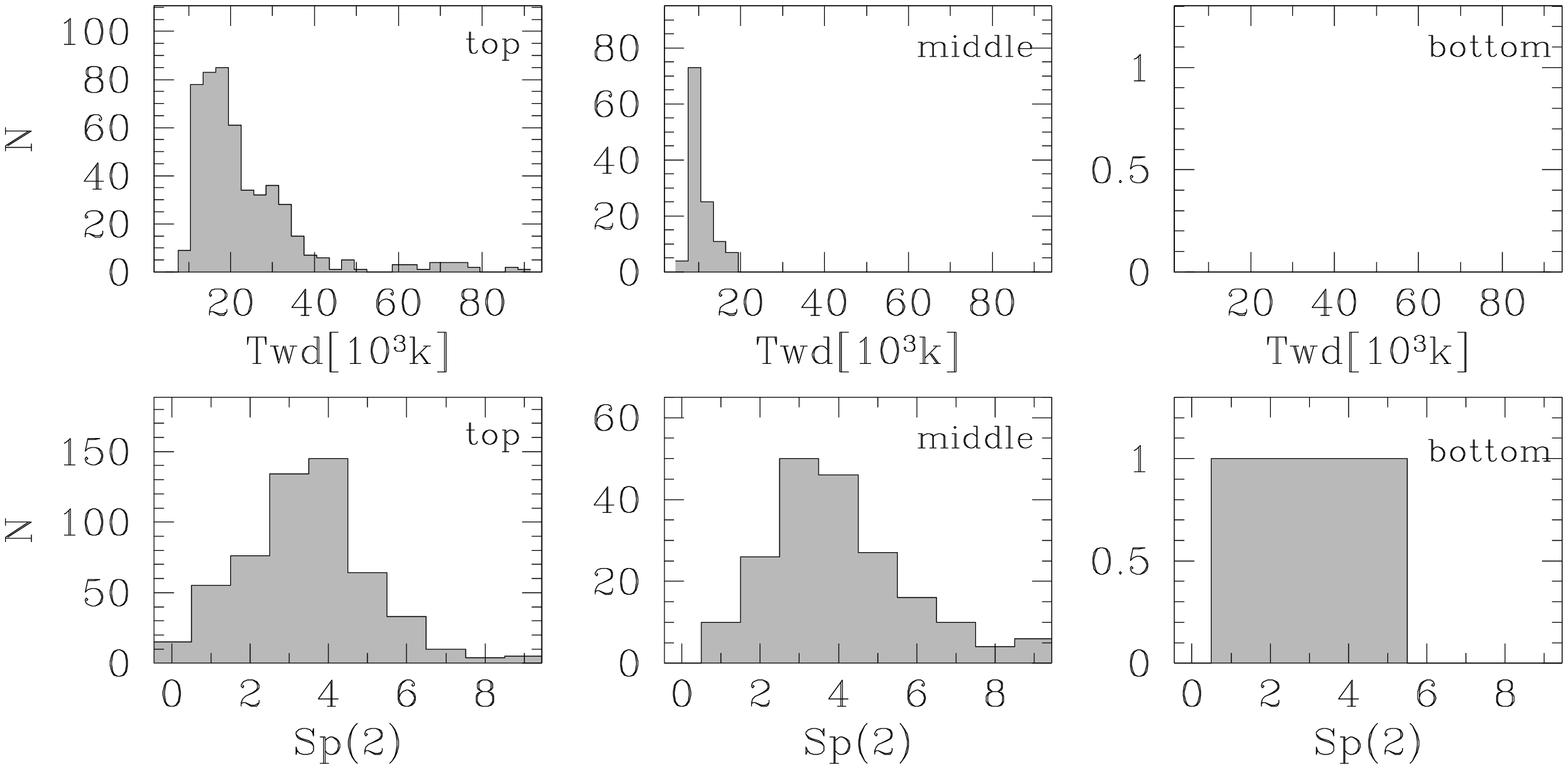}
\includegraphics[angle=-90,width=\columnwidth]{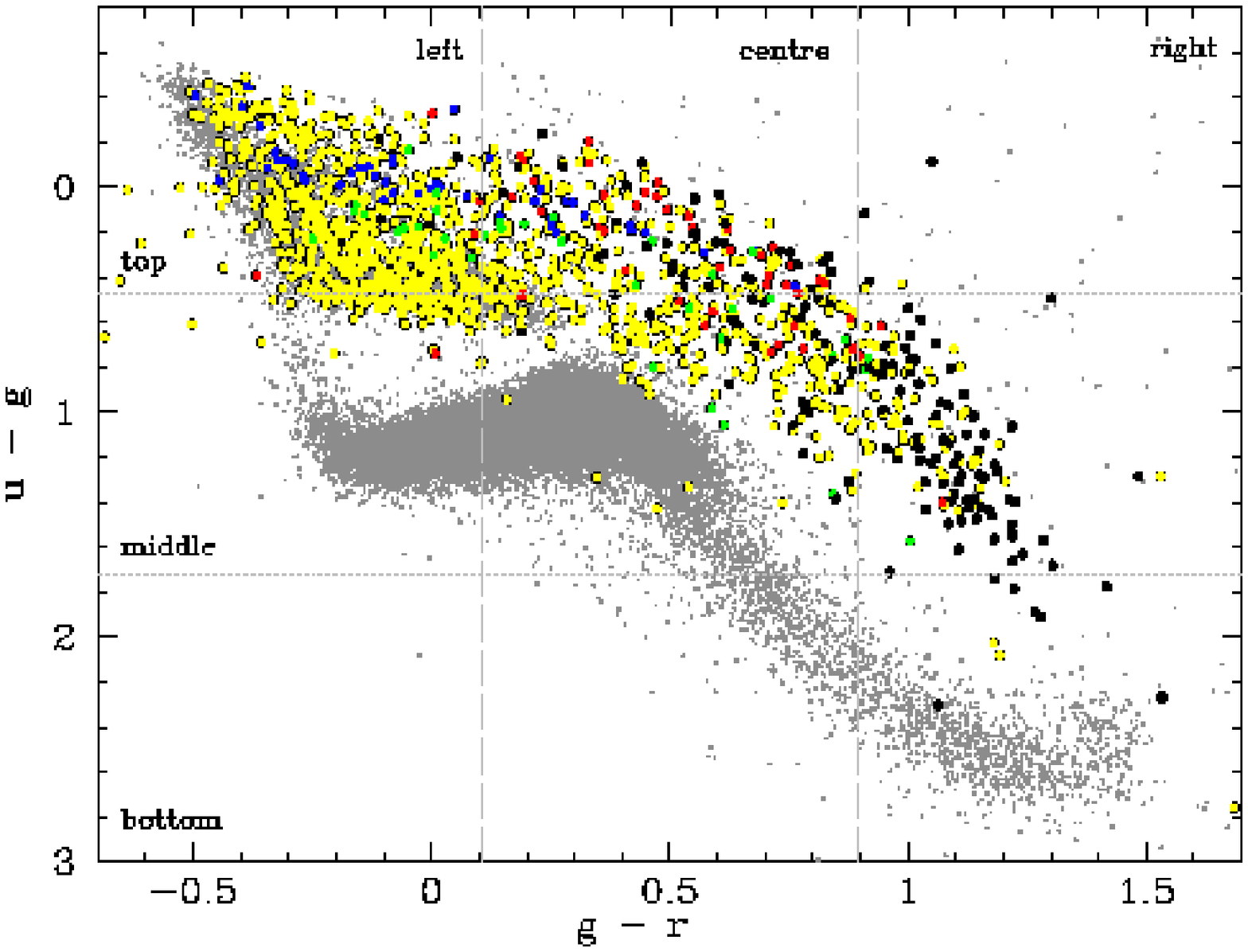}
\vline
\includegraphics[angle=-90,width=\columnwidth]{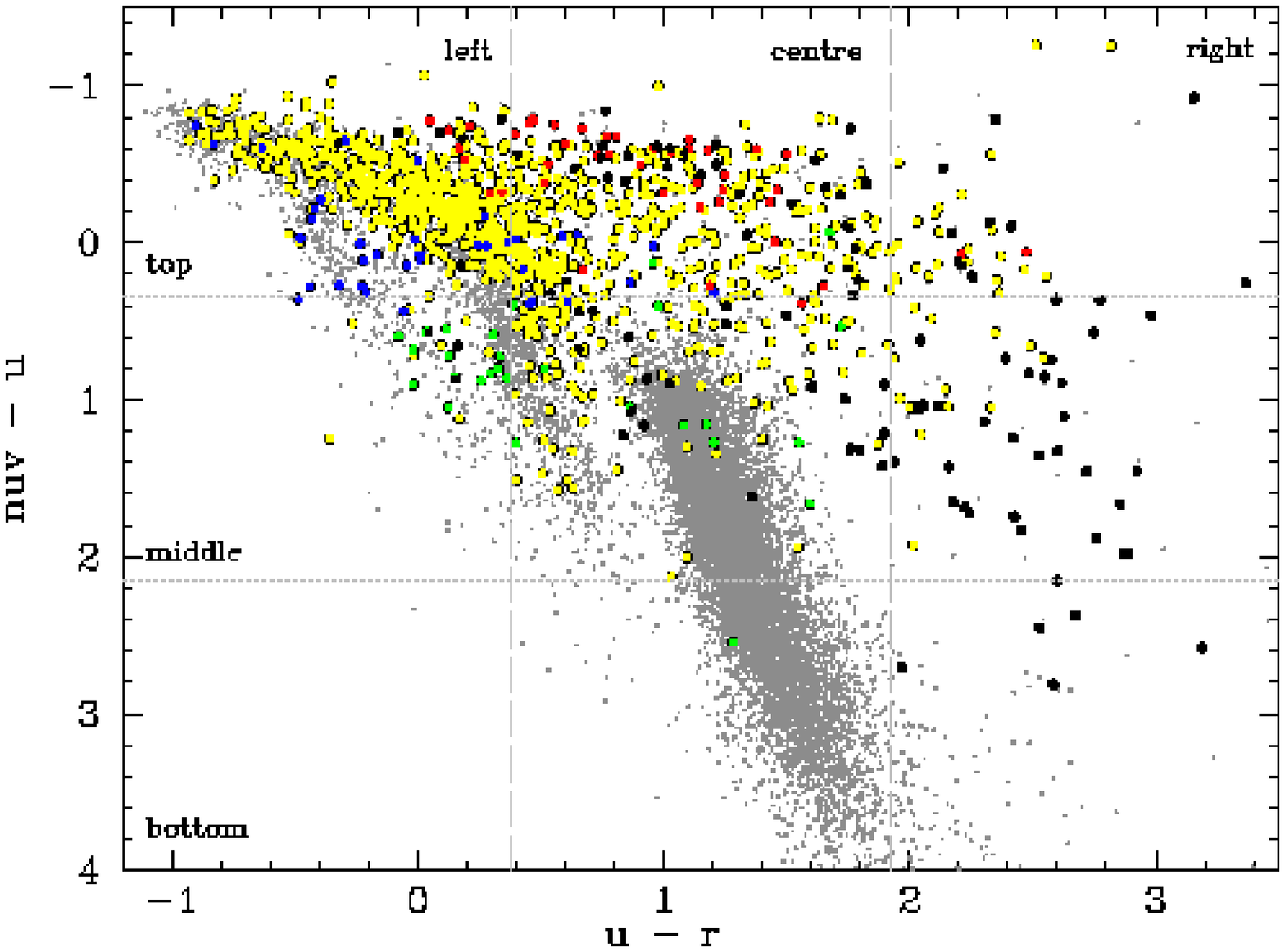}
\includegraphics[angle=-90,width=\columnwidth]{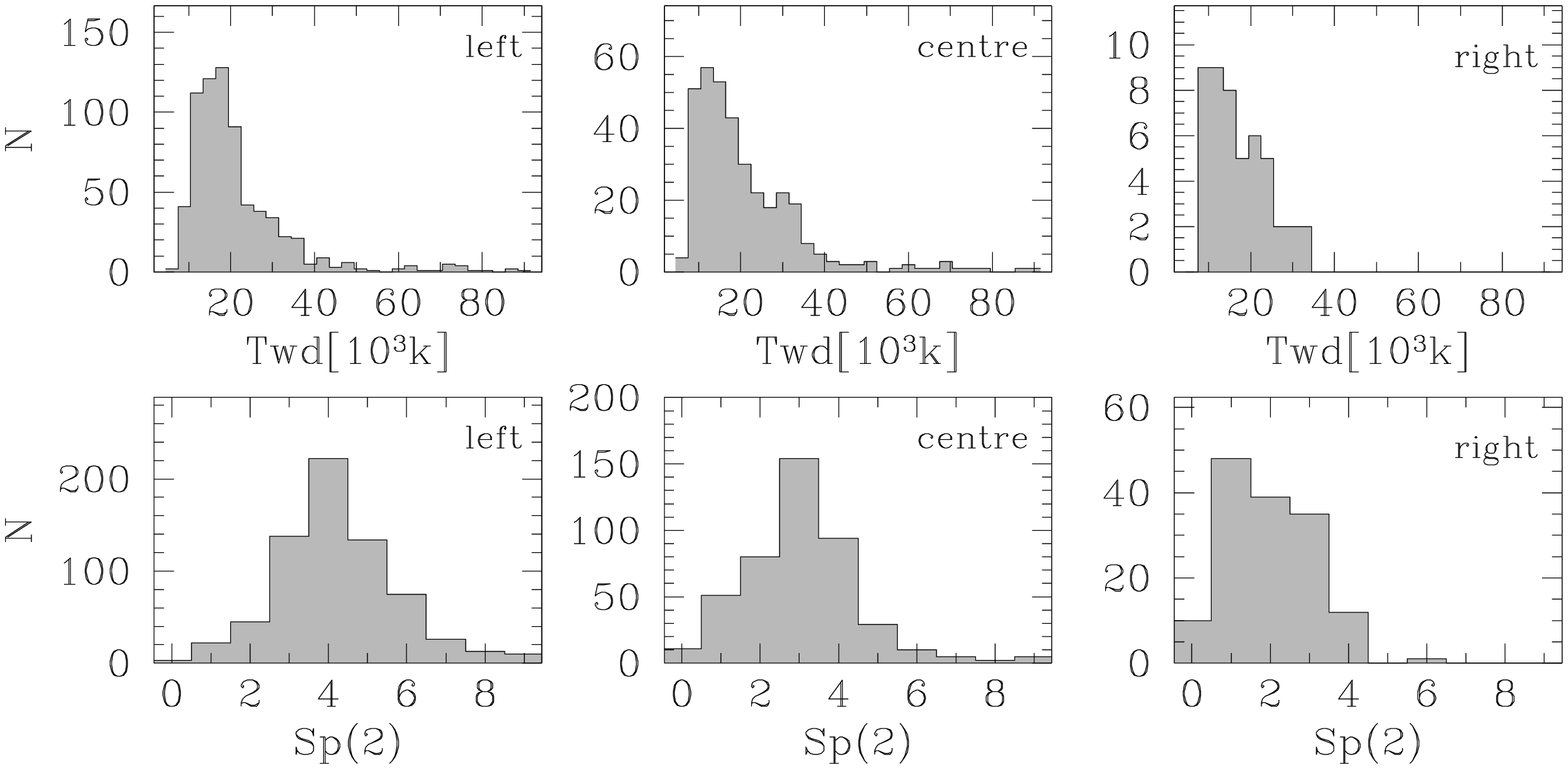}
\vline
\includegraphics[angle=-90,width=\columnwidth]{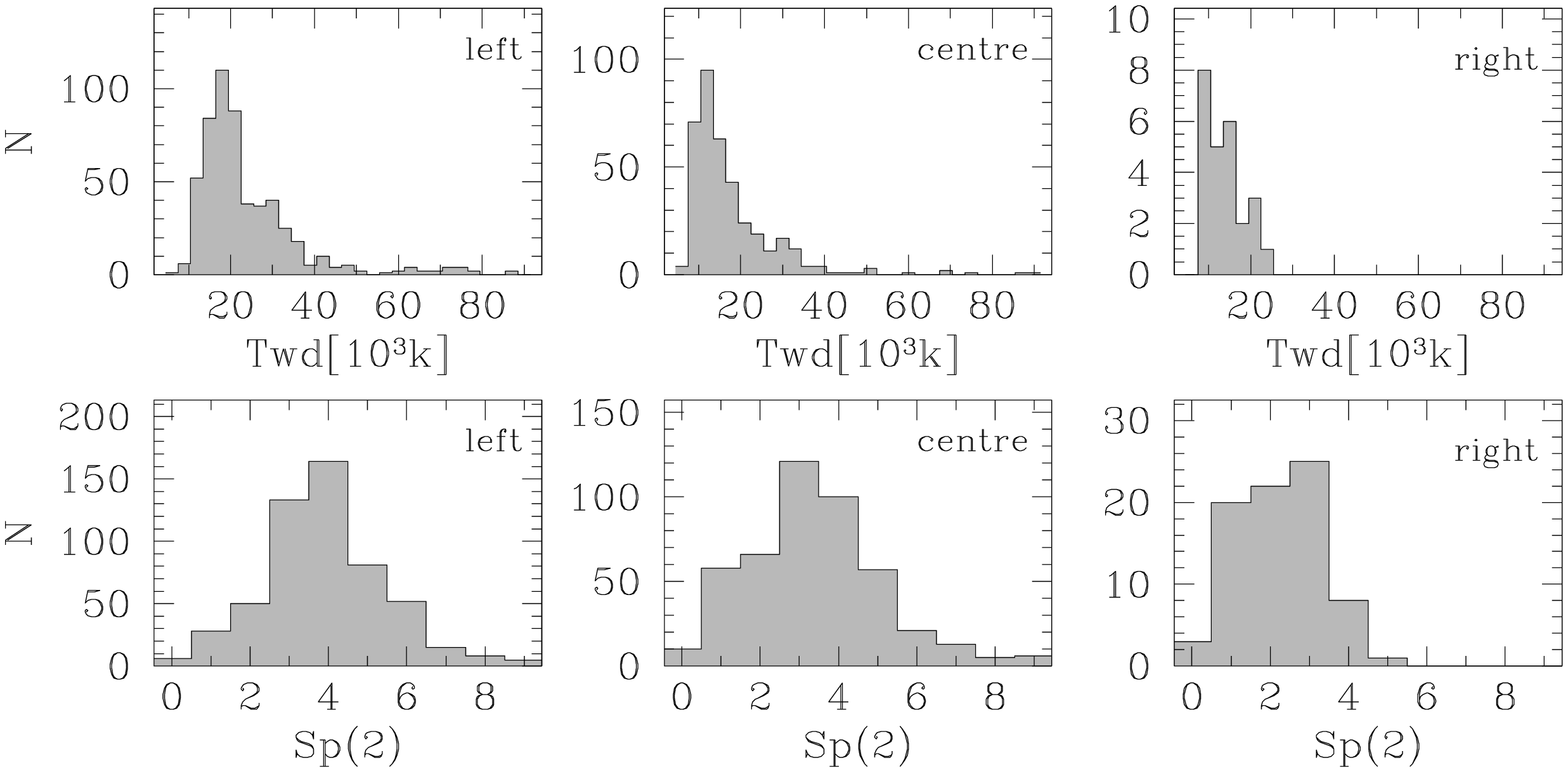}
\caption{Middle  left:  $u-g$ vs  $g-r$  colour-colour diagram.   WDMS
binaries  are  represented according  to  their  binary components  as
follows:  DA/M  binaries in  yellow,  DB/M  in  blue, DC/M  in  green,
DA-DB-DC/K in red, and WD/M in black.  Stellar sources are represented
with gray  dots.  Two  vertical (dashed) lines  divide the  diagram in
three rectangular regions (columns  left, centre, right).  In the same
way,  two  horizontal  (dotted)  lines  divide the  diagram  in  three
different regions  (rows top, middle,  bottom).  Top and  bottom left:
distributions of  white dwarf effective temperature  and spectral type
of  the companions  obtained for  the six  different  regions outlined
above (three  rows and  three columns).  The  right panels  follow the
same  structure,  but  for the  $nuv  -  u$  vs $u  -r$  colour-colour
diagram.}
\label{f-color1}
\end{center}
\end{figure*}

\section{Selection effects}
\label{s-effects}

In  \citet{rebassa-mansergasetal07-1}  we  briefly discussed  possible
selection  effects  that may  affect  the  observed  SDSS WDMS  binary
population.   The   much  larger  sample  presented   here  allows  to
investigate  these  selection  effects   in  more  detail.   To  avoid
contaminations from unreliable stellar parameters we base our analysis
on  systems with mean  relative errors  of less  than $25\%$  in their
white    dwarf    parameters.     As    discussed   in    detail    in
\citet{rebassa-mansergasetal07-1} the distances derived from the white
dwarf parameters  are probably more  reliable than those  derived from
the secondary stars.  We therefore  use the white dwarf distances here
and quote them simply as distances in the following.

Fig.\,\ref{f-effects}  shows three  density maps  that  illustrate the
selection  effects  affecting  the  SDSS  WDMS  sample.   Due  to  the
restrictions in the white dwarf  parameters 597, 692, and 1052 systems
(from   top   to  bottom)   were   considered.    The  $\log   \Teff$,
$d_{\mathrm{wd}}$  density map in  the top  panel shows  that binaries
with white dwarf primaries  cooler than 10\,000\,K are only detectable
at  relatively  short  distances ($d\lappr\,400$\,pc),  while  systems
containing  hotter white  dwarfs have  been detected  at a  much wider
range  of  distances   ($\sim200-1000\,$pc).   In  addition  there  is
obviously  a general  trend of  increasing distance  with  white dwarf
temperature.   This is straight  forward to  understand as  cold white
dwarfs  become too  faint  to  be detected  at  larger distances.   In
contrast hot  white dwarfs are intrinsically brighter,  but also rarer
than cold white  dwarfs, and hence dominate at  larger distances where
the volume surveyed  by SDSS is sufficiently large.   Most objects are
hence  concentrated at $\sim$400-500\,pc,  with white  dwarf effective
temperatures  between  $\sim$15000-25000\,K,  as  also  shown  in  the
effective  temperature   distribution  (Fig.\,\ref{f-histocat}).   The
middle panel  of Fig.\,\ref{f-effects} shows a similar  effect but for
the  secondary   star  spectral  types,  $Sp$.    Early  M-dwarfs  are
relatively hot and consequently exceed the lower SDSS brightness limit
at distances of $\lappr\,300$\,pc.  In contrast, late-type secondaries
are cold enough to  be detected at short distances ($\sim100-200$\,pc)
but too  dim to  be observed at  distances larger  than $\sim500\,$pc.
Finally,  the bottom  panel of  Fig.\,\ref{f-effects} shows  the $\log
\Teff,  Sp$  density  map.   A  clear trend  of  later  spectral  type
companions to colder white dwarf primaries can be seen.  This is again
easy to understand: while late-type companions to hot white dwarfs are
too faint to be detected, cold white dwarf primaries are out-shined by
early spectral type secondaries.

\begin{figure*}
\begin{center}
\includegraphics[angle=-90,width=\columnwidth]{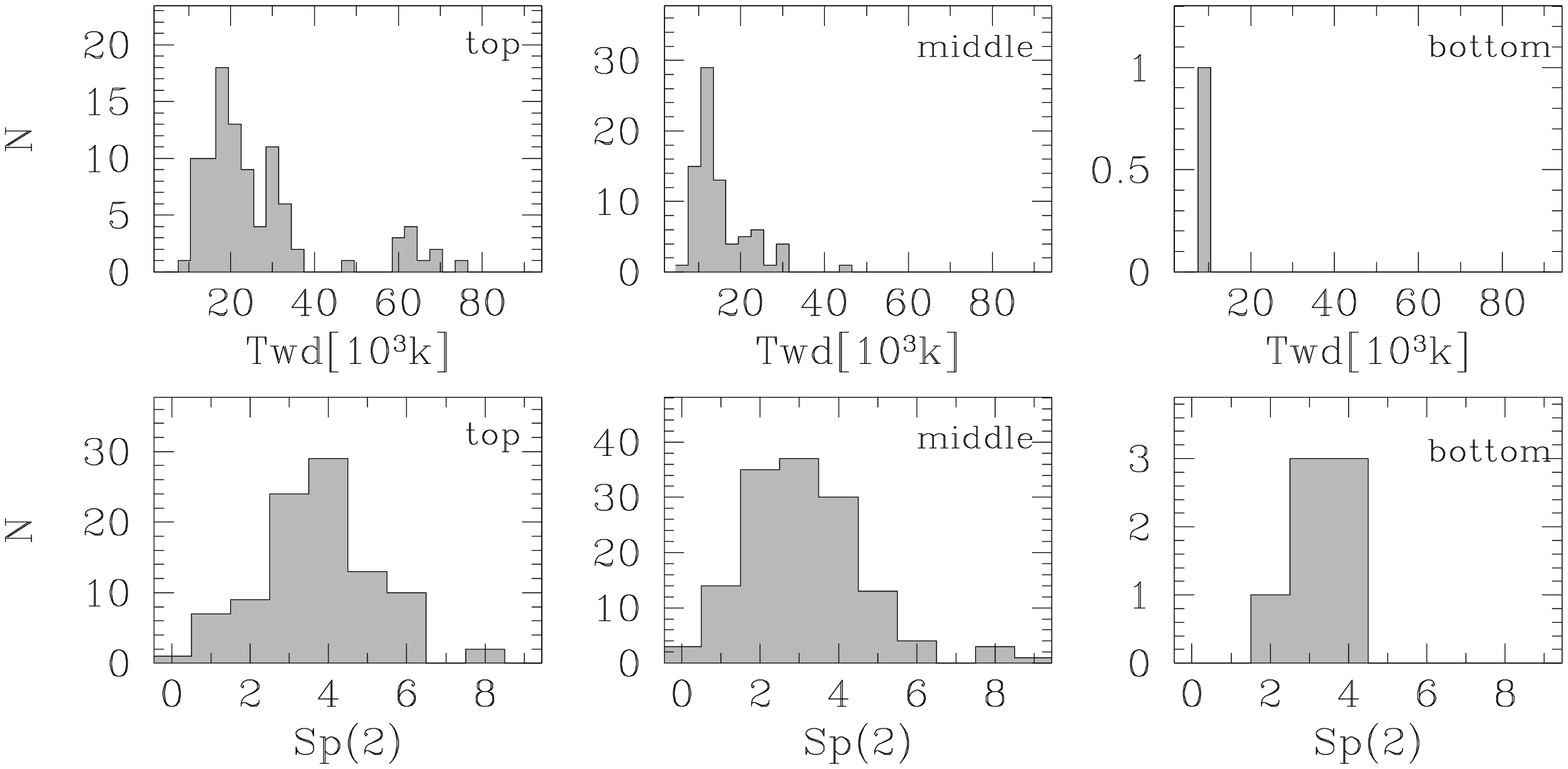}
\vline
\includegraphics[angle=-90,width=\columnwidth]{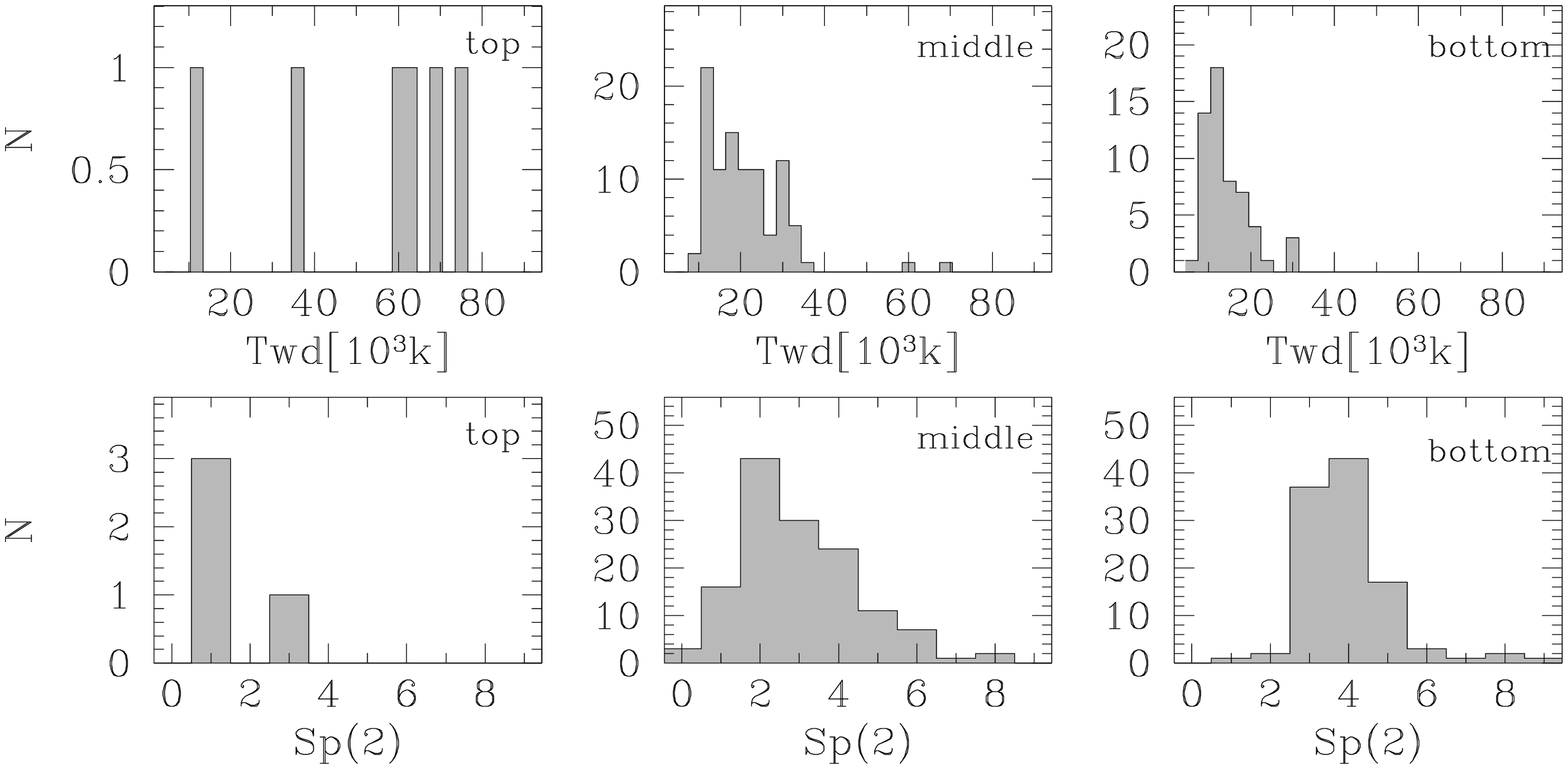}
\includegraphics[angle=-90,width=\columnwidth]{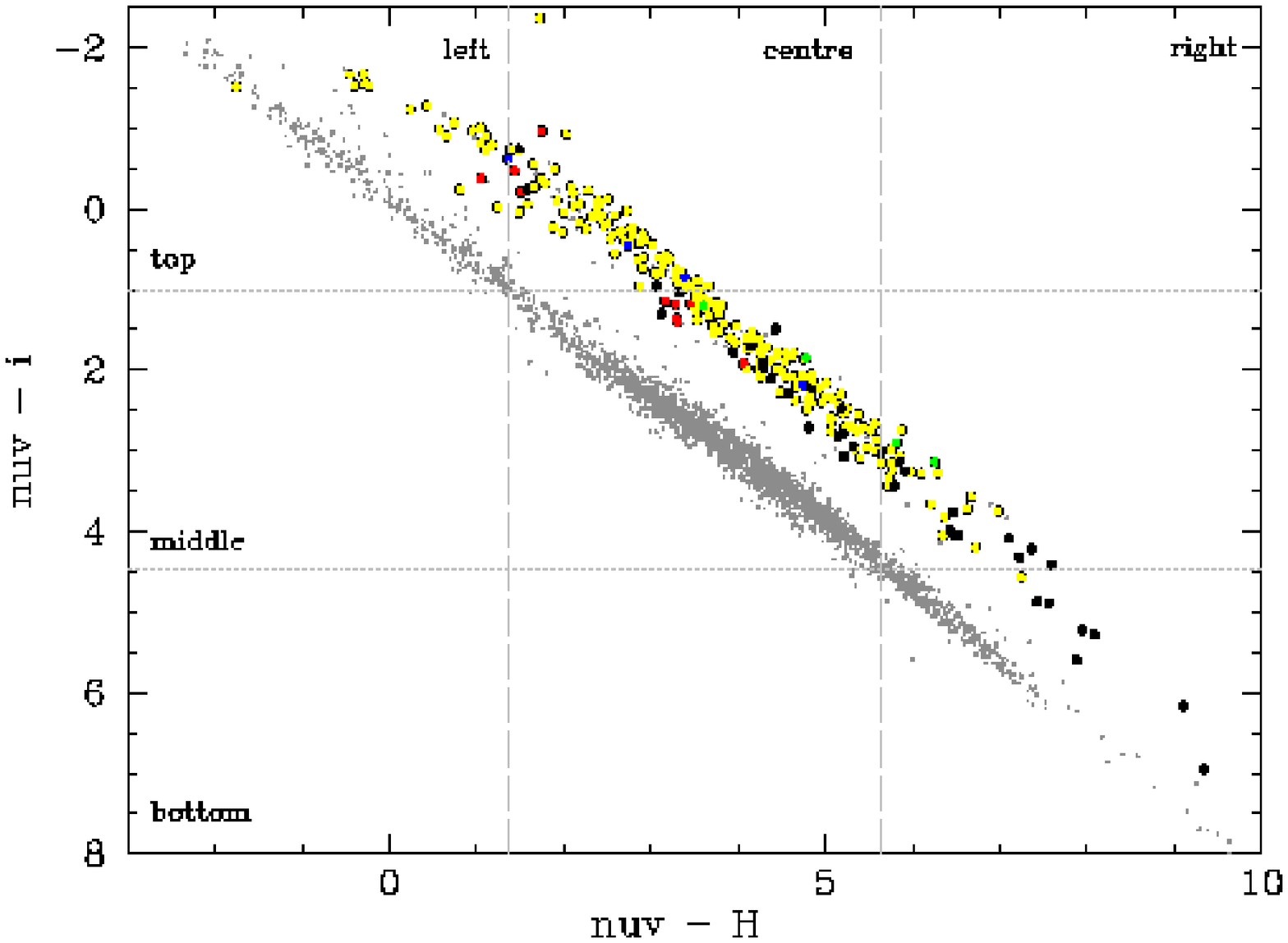}
\vline
\includegraphics[angle=-90,width=\columnwidth]{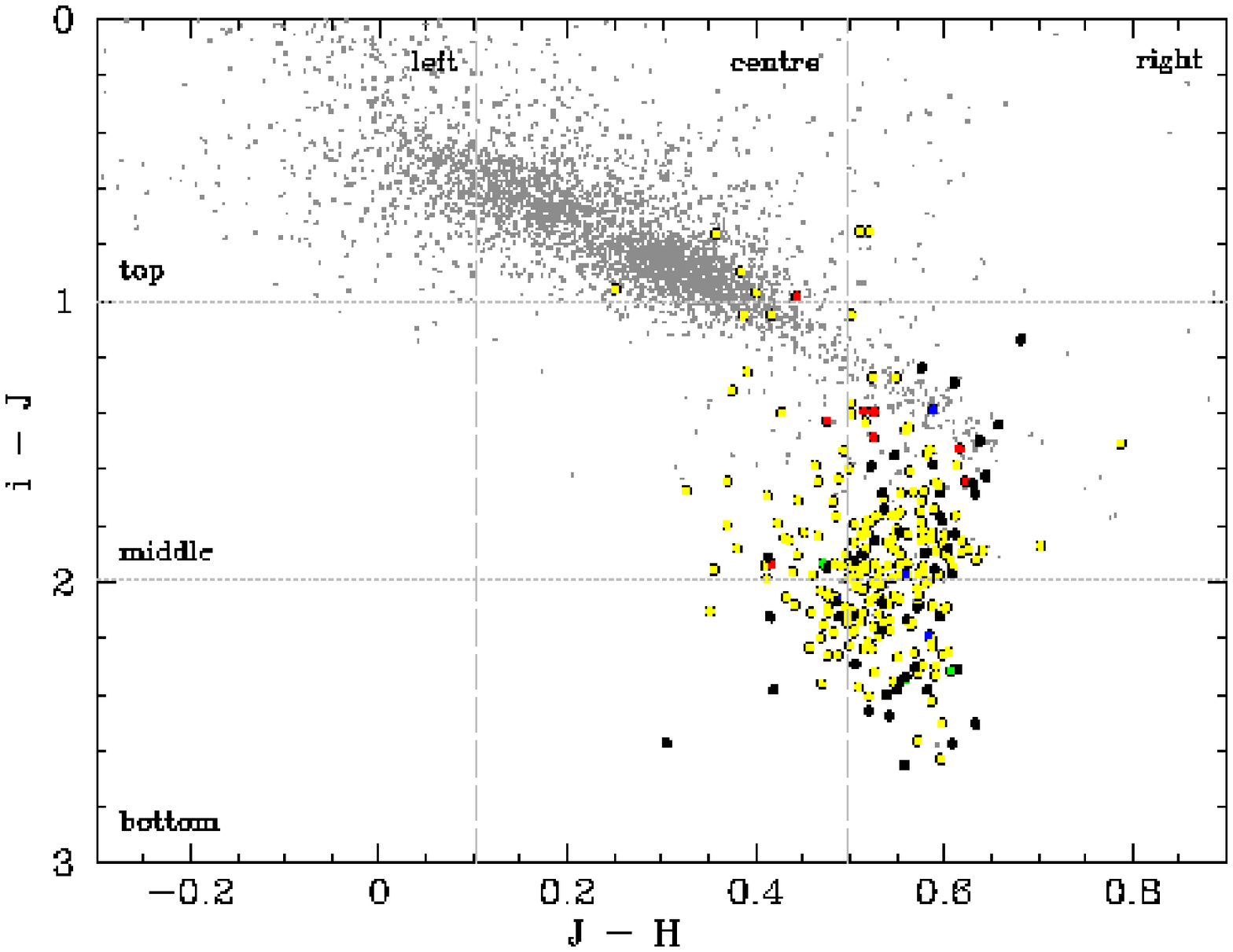}
\includegraphics[angle=-90,width=\columnwidth]{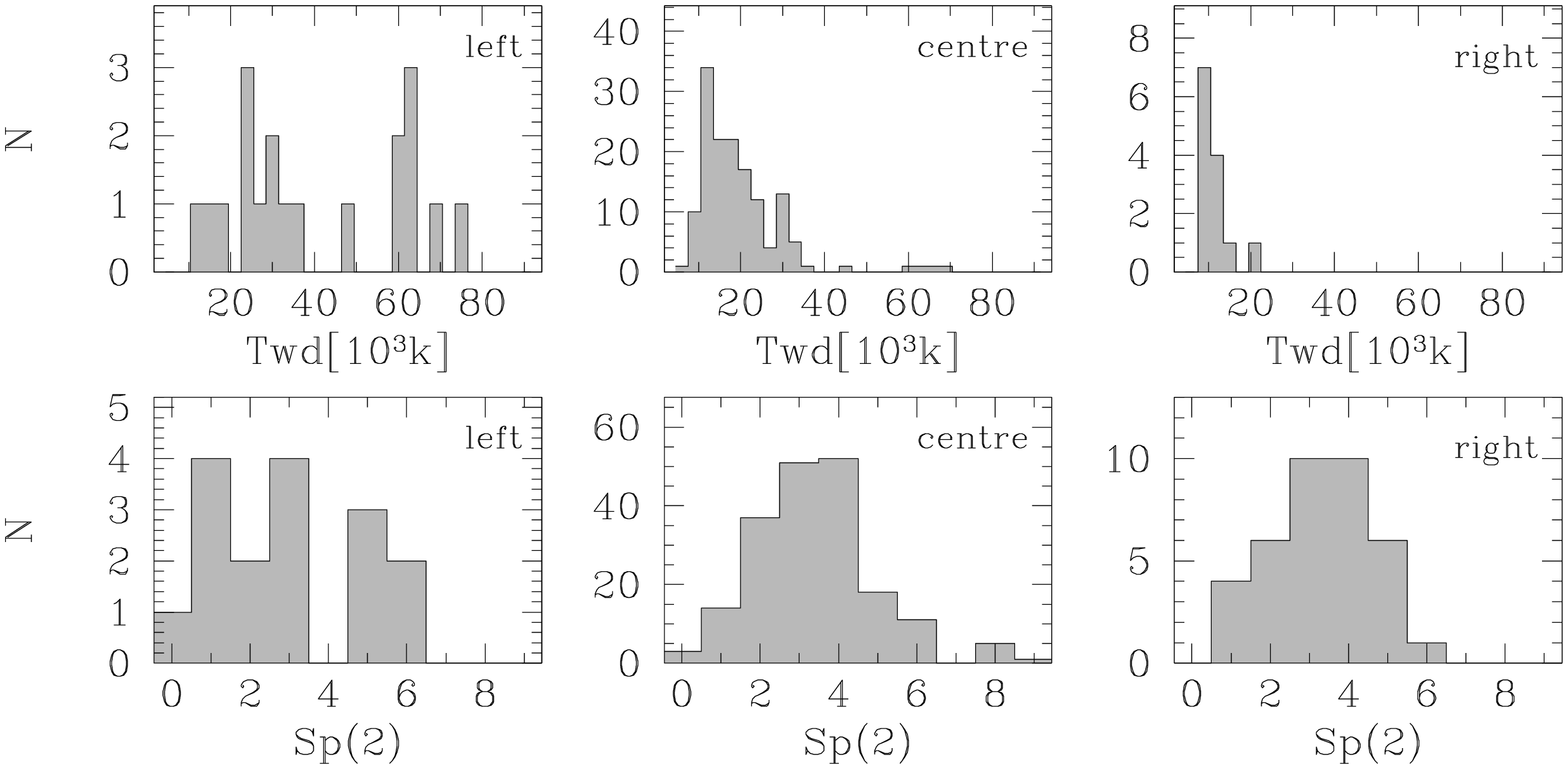}
\vline
\includegraphics[angle=-90,width=\columnwidth]{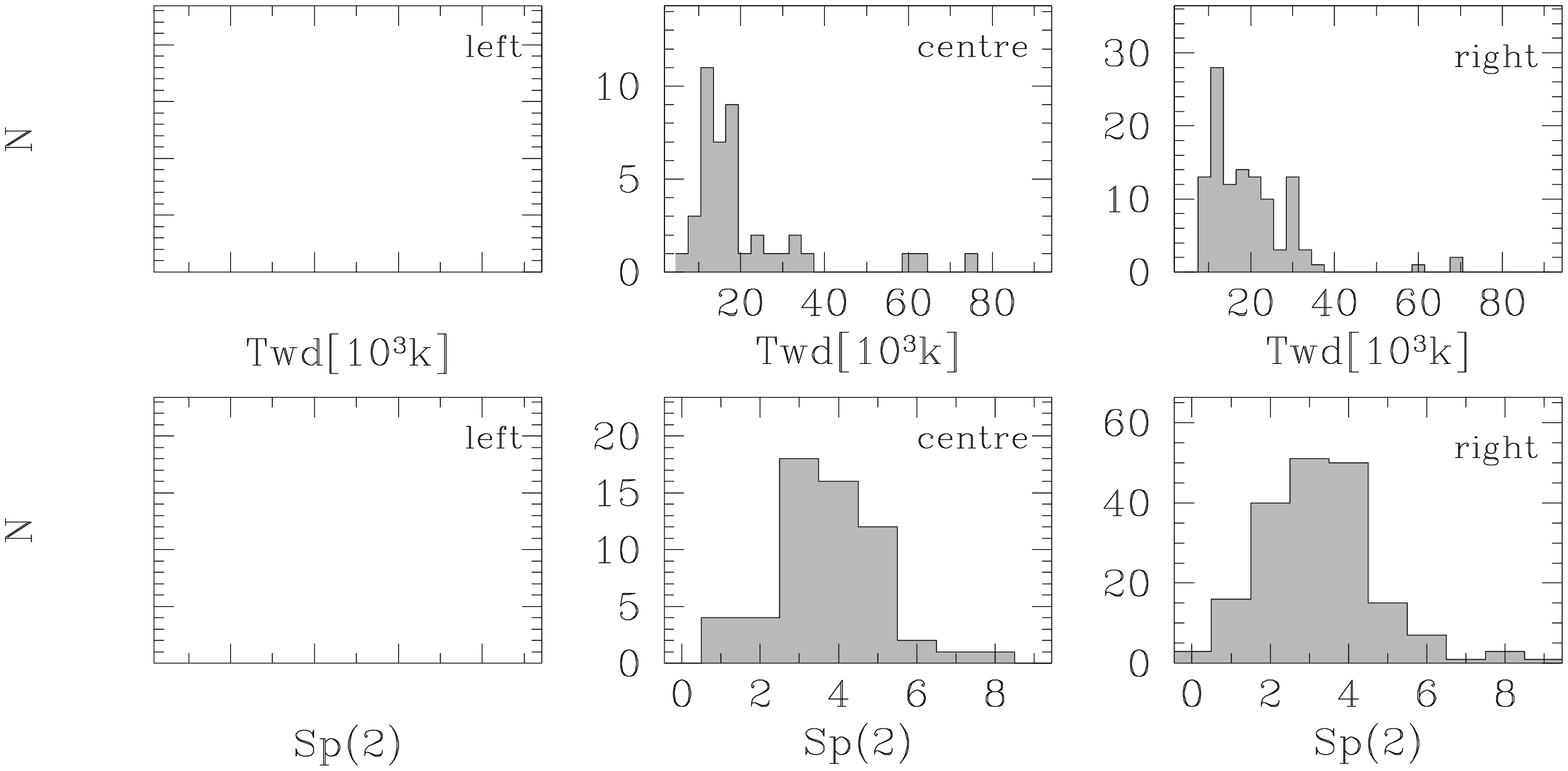}
\caption{Same as in Fig.\,\ref{f-color1} but for the $nuv - i$ vs $nuv
- H$ (left) and $i - J$ vs $J - H$ (right) colour-colour diagrams.}
\label{f-color2}
\end{center}
\end{figure*}
\begin{figure*}
\begin{center}
\includegraphics[angle=-90,width=0.80\textwidth]{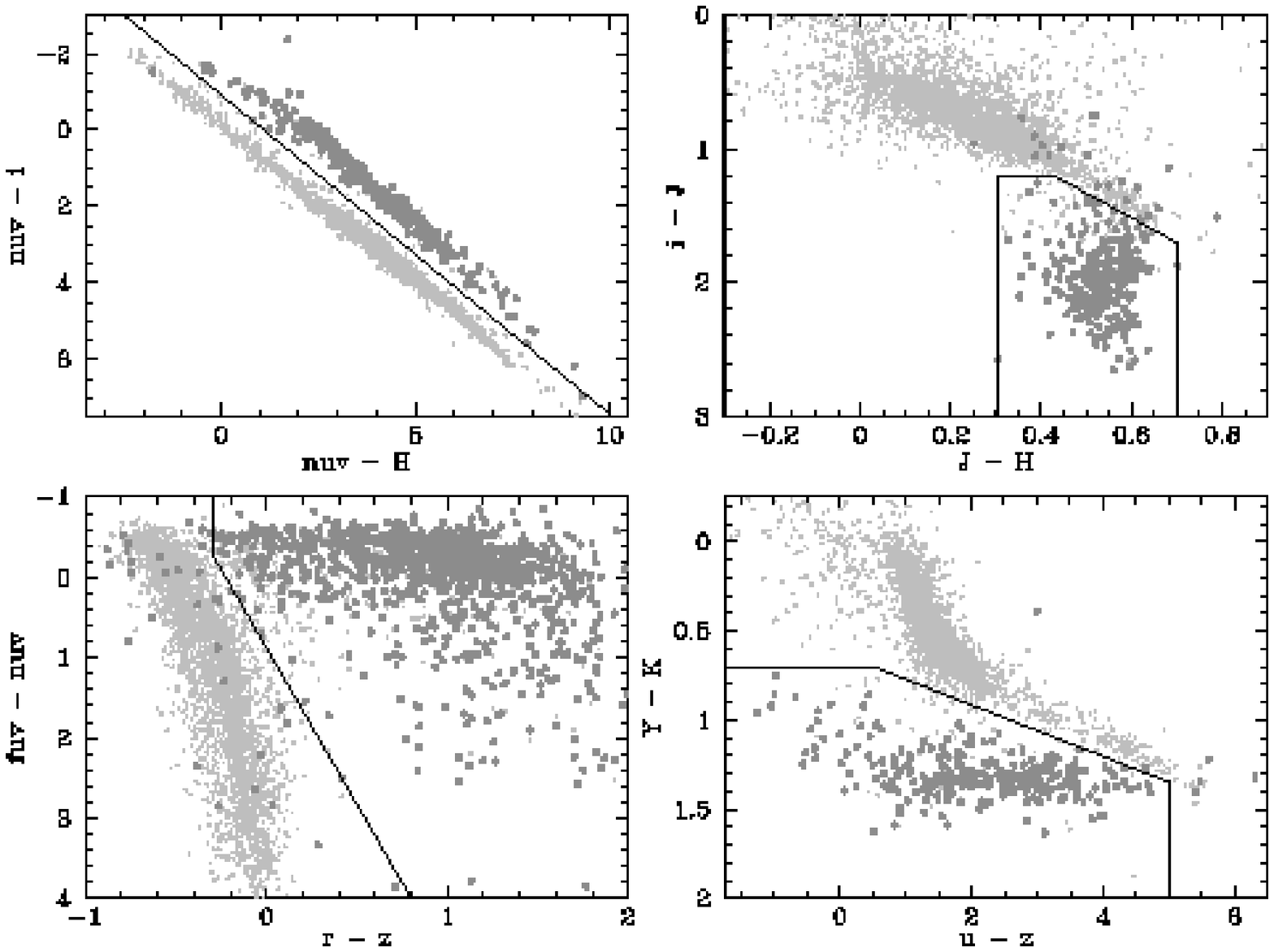}
\caption{Colour-cut selection for WDMS  binaries in $nuv-i$ vs $nuv-H$
(top left),  $i-J$ vs  $J-H$ (top right),  $fuv-nuv$ vs  $r-z$ (bottom
left), and  $y-K$ vs  $u-z$ (bottom right).   Main sequence  stars and
white  dwarfs are  represented in  light gray,  WDMS binaries  in dark
gray, and colour-cuts by black straight lines.}
\label{f-cuts}
\end{center}
\end{figure*}

The selection effects just described  can explain the cut-off at early
spectral types  in the bottom  panel of Fig\,\ref{f-histocat},  as the
white  dwarf primaries are  not detectable.   The scarcity  of systems
with  later-type ($>$M6)  secondaries, however,  is probably  not only
related to  selections effects, as  the spectral type  distribution of
low-mass  field stars also  peaks at  Sp\,$\simeq$ M4--5,  and decline
towards    later    spectral   types    \citep[e.g.][]{farihietal05-1,
reidetal07-1, reidetal08-1}.  The lack of WDMS binaries with late-type
companions  is therefore probably  an intrinsic  property of  the WDMS
binary  population  that  appears  more pronounced  due  to  selection
effects.   We  will  address  this  systematically  in  a  forthcoming
publication.

From  the  analysis  of   Fig.\,\ref{f-effects}  we  conclude  that  a
``typical''   SDSS  WDMS   binary  contains   a  M3--4   companion,  a
$\sim$10000-20000\,K   primary,  and   is  observed   at   a  distance
$\sim$400-500\,pc.  However,  we have to  keep in mind that  a typical
SDSS WDMS binary is not  necessarily a typical WDMS binary. Overcoming
the selection effects just described requires to combine the SDSS with
complementary   magnitude   limited   surveys.    Detecting   binaries
consisting  of a hot  white dwarf  and a  late-type companion  is most
likely to  arise from the  use of infrared  surveys such us  UKIDSS or
2MASS, while  the identification of cool white  dwarfs with early-type
dominated M-dwarfs requires to incorporate ultraviolet surveys such as
GALEX \citep{maxtedetal09-1}.

\section{Colour-colour diagrams}
\label{s-colors}

We have  provided in  the previous sections  of this paper  a detailed
spectroscopic analysis  of WDMS binaries  in SDSS. In this  section we
make use of the  photometric magnitudes given in Table\,\ref{t-magall}
and combine  them with the  stellar parameters measured from  the SDSS
spectra   to  investigate   the   appearance  of   WDMS  binaries   in
colour-colour  space.  Figure\,\ref{f-color1} and  \ref{f-color2} show
four relevant  colour-colour diagrams.   Stellar sources are  shown in
gray, DA/M WDMS  binaries in yellow, the few DB/M are  in blue and the
DC/M binaries in green. Finally,  DA-DB-DC/K systems are shown in red,
while   the   black   dots   represent  the   WD/M-K   binaries   (see
Sect.\,\ref{s-finalcat}).

A general feature evident in all  diagrams is that a certain number of
systems appear to  be outliers from the general  stellar locus of WDMS
binaries,   i.e.   the  WDMS   binary   bridge   described  first   in
\citet{smolcicetal04-1}.   We have inspected  these outliers  and find
the  majority  of  them  being   resolved  in  the  SDSS  images  (see
Sec.\,\ref{s-images}).

To further evaluate the information provided by photometry we show the
white  dwarf effective  temperature and  secondary star  spectral type
distributions in six different regions for each colour-colour diagram,
as indicated  by the horizontal (dotted) and  vertical (dashed) lines.
The   resulting  distributions   are   shown  above   and  below   the
colour-colour diagrams.  As previously, we considered only those white
dwarf effective  temperatures with relative error less  than 25\%.  In
the following  sub-sections we  briefly describe the  main conclusions
that can be drawn from the four colour-colour diagrams.
 
\subsection{u - g vs g - r}

The most commonly used  SDSS colour-colour diagram for stellar sources
in          SDSS          is          $u-g$          vs          $g-r$
\citep{fan99-1,richardsetal02-1,schreiberetal07-1}.    Inspecting  the
left hand side of Fig.\,\ref{f-color1} it can clearly be seen that the
white  dwarf  effective temperature  distribution  is shifted  towards
lower  temperatures if  $u-g$  or $g-r$  increases,  i.e.  for  redder
colors (compare the left, centre, and right as well as the top and the
middle distributions).  Only two WDMS binaries with reliable effective
temperatures  are  found  in  the  bottom region  close  to  the  main
sequence.  As expected, the  distributions of secondary spectral types
contain  more early-type  secondaries if  one moves  to  redder colors
(left to right or top to bottom).

\subsection{nuv - u vs u -r}

The  $nuv -  u$ vs  $u -r$  colour-colour diagram  is provided  in the
middle right panel of Fig.\,\ref{f-color1}.  The white dwarf effective
temperature and spectral type distributions  in the six regions of the
diagram (see top and  bottom right panels in Fig.\,\ref{f-color1}) are
similar  to  those  discussed   in  the  above  subsection,  the  main
difference being the decrease in the number of systems. This is due to
the smaller percentage  of SDSS WDMS binaries that  have been detected
with GALEX. 

\subsection{nuv - i vs nuv - H}

Figure\,\ref{f-color2} (left panels) shows the  $nuv - i$ vs $nuv - H$
colour-colour  diagram and  the corresponding  distributions  of white
dwarf effective  temperatures and secondary  spectral types.  Clearly,
requesting  SDSS-GALEX-UKIDSS magnitudes  reduces the  number  of WDMS
binaries and  the shown  distributions are statistically  less robust.
However,  the general  trend  observed in  the previous  colour-colour
diagrams  remains:  systems  composed  of  white  dwarfs  hotter  than
20000\,K are  generally detected  in the top  and left regions  of the
diagram and there is a clear trend of decreasing white dwarf effective
temperature towards redder colours (top  to bottom and left to right).
The  previously  observed shift  towards  earlier  spectral types  for
redder colours seems still to  be present but is much less pronounced.
The most  striking feature  of the colour-colour  diagram is  the nice
separation of WDMS binaries and single stars.

\subsection{i - J vs J - H}

Finally, we provide in Fig.\,\ref{f-color2} (right panels) the $i - J$
vs  $J   -  H$  red   colour-colour  diagram  and   the  corresponding
distributions.   Again,  due to  the  reduced  number  of systems  the
overall trend of having less  hot white dwarfs and more early spectral
type  secondaries for  redder colours  seems  to be  present but  less
significant.   Nearly  all systems  containing  hot  white dwarfs  are
located in the upper  region. This colour-colour diagram represents an
additional  example  of nicely  separating  WDMS  binaries and  single
stars.  We provide several colour-cuts that can be used to select WDMS
binaries in the next section.

\section{Colour cuts}
\label{s-cuts}

Having studied  in the previous  section the relation  between colours
and stellar  parameters (i.e.   white dwarf effective  temperature and
spectral type)  in four colour-colour  space diagrams, we  define here
four colour-cuts of WDMS binaries.  The fact that we are considering a
total of  11 photometric band passes  (two from GALEX,  five from SDSS
and four  from UKIDSS) increases  considerably the number  of possible
colour-cut  selections.  We here  provide four  examples (two  of them
already  introduced in Sect.\,\ref{s-colors})  in which  WDMS binaries
are clearly  separated from  the locus of  single main  sequence stars
(see Fig.\,\ref{f-cuts}).  To  quantify how complete these colour-cuts
are would imply an analysis  on the different SDSS-I, SEGUE, Stripe 82
areas,  where  different  target  strategies  were  tested.   Such  an
endeavour  is beyond  the  scope of  this  paper and  will be  pursued
elsewhere.
  
The top left panel of  Fig.\,\ref{f-cuts} shows the $nuv-i$ vs $nuv-H$
colour-colour  diagram  introduced above,  which  offers an  excellent
opportunity to  unambiguously isolate  WDMS binaries from  single main
sequence  stars  and  white  dwarfs  with  a  simple  colour-cut.   We
represent in light gray main  sequence stars and white dwarfs, in dark
gray WDMS binaries. With a straight black line, i.e.
\begin{equation}
(nuv-i) < -0.85 + 0.83 \times (nuv-H)
\end{equation}
both populations can be distinguished. 

As  we have seen  in Sect.\,\ref{s-colors},  WDMS binaries  and single
stars also  separate nicely in the  $i - J$  vs $J - H$  colour plane.
However,  the location of  WDMS binaries  containing hot  white dwarfs
overlaps with  those of  single stars.  The  colour-cuts shown  in the
right top panel of Fig.\,\ref{f-cuts}, i.e.
\begin{eqnarray}
0.3 < (J-H) < 0.7\\
(i-J) > 1.2\\
(z-J) > 0.4 + 1.85 \times (J-H)
\end{eqnarray}
will therefore mainly select WDMS binaries composed of cold white dwarfs.

Recently  \citet{bianchietal07-1}   studied  the  properties   of  the
GALEX-SDSS  matched  source   catalogues  and  classified  sources  by
studying their  colours.  Inspired  by their Fig.\,5,  we show  in the
left bottom panel of  Fig.\,\ref{f-cuts} the $fuv-nuv$ vs $r-z$ colour
diagram,  and   provide  colour-cuts  that  should   select  the  main
population of WDMS binaries:
\begin{eqnarray}
(r-z) > -0.3\\
(fuv-nuv) < 0.85 +3.9 \times (r-z)
\end{eqnarray}

Finally, we  provide in the  bottom right panel  of Fig.\,\ref{f-cuts}
the colour-colour  diagram and colour-cuts for WDMS  binaries in $y-K$
vs $u-z$. Again stars are  represented in light gray, WDMS binaries in
dark  gray, and  colour cuts  by  straight black  lines.  This  colour
diagram  has  been already  used  by  \citet{chiuetal07-1} for  quasar
selection. We here only slightly modified their colour-cuts and obtain
\begin{eqnarray}
(u-z) > 5\\
(y-K) > 0.7\\
(y-K) > 0.6 + 0.14 \times (u-z)
\end{eqnarray}
which should successfully select WDMS binaries.

\section{Radial velocities and new PCEB candidates}
\label{s-rvs}

In    this     final    section    of    the     paper    we    follow
\citet{rebassa-mansergasetal07-1}          and         use         the
\Lines{Na}{I}{8183.27,8194.81} absorption doublet and/or the H$\alpha$
emission     to     measure      radial     velocities.      As     in
\citet{rebassa-mansergasetal08-1,schreiberetal08-1}  we  use a  single
width parameter  for both line components in  the \Ion{Na}{I} doublet.
The    radial   velocities   of    1068   systems    with   pronounced
\Lines{Na}{I}{8183.27,8194.81}  absorption  doublet  and/or  H$\alpha$
emission  in their  SDSS spectra  are given  in Table\,\ref{t-rvsall},
where we  also include the HJD of  the observations\footnote{Note that
the HJDs in \citet{rebassa-mansergasetal07-1} were wrong by -0.5\,days
because of  an erroneous  conversion of the  FITS headers of  the SDSS
spectra.}.   As   discussed  in  \citet{rebassa-mansergasetal07-1},  a
significant fraction  of SDSS  spectra are combined  from observations
that  have been taken  in different  nights which  leads to  a reduced
sensitivity to  radial velocity variations.   We indicate this  in the
last column of Table\,\ref{t-rvsall}.

\begin{table}
\caption{\label{t-rvsall}   Radial   velocities   measured  from   the
\Lines{Na}{I}{8183.27,8194.81} doublet and  the H$\alpha$ emission for
1068 systems in our catalogue.  The complete table can be found in the
electronic edition  of the  paper.  In the  last column we  quote with
``y''  and ``n'' those  radial velocity  values obtained  from spectra
that  are, and  are not  combined from  individual exposures  taken on
different nights, respectively.   We use ``-'' to indicate  that no RV
is available.}  \setlength{\tabcolsep}{0.6ex}
\begin{center}
\begin{tabular}{ccccccc}
\hline
\hline
  SDSS\,J & HJD       & RV (Na) & err & RV (H$\alpha$) & err & Com.? \\
          &   245     & [\kms]  &     & [\kms]         &     & \\
\hline
000152.09+000644.7   &  1791.8092  &     0.7  &    21.1  &    24.2 &     16.7 &   n \\
001247.18+001048.7   &  2519.8962  &       -  &       -  &    12.3 &     18.6 &   n \\
001247.18+001048.7   &  2518.9219  &   -14.3  &    30.1  &    30.6 &     14.4 &   n \\
001359.39-110838.6   &  2138.3933  &    28.9  &    16.9  &       - &        - &   y \\
001726.64--002451.2  &  2559.7852  &   -33.7  &    15.5  &   -30.1 &     11.4 &   n \\
001726.64--002451.2  &  2518.9219  &   -19.8  &    17.8  &   -26.5 &     11.8 &   n \\
001733.59+004030.4   &  1794.7737  &    -3.7  &    15.8  &       - &        - &   n \\
001749.25--000955.4  &  2518.9218  &   -18.3  &    17.4  &    -3.2 &     11.6 &   n \\
001749.25--000955.4  &  1794.7737  &   -36.6  &    15.5  &   -22.8 &     10.1 &   n \\
001855.20+002134.5   &  1816.8000  &    41.4  &    27.4  &       - &        - &   n \\
001855.20+002134.5   &  1893.0883  &    15.0  &    22.2  &       - &        - &   y \\
002143.78--001507.9  &  2581.7411  &     1.5  &    14.7  &       - &        - &   n \\
002157.91--110331.6  &  3318.6951  &   148.4  &    15.8  &    -9.1 &     13.3 &   y \\

...                 &    ...       &  ...     &   ...   &     ...    &   ...  &  ...\\
\hline
\end{tabular}
\end{center}
\end{table}

A    comparison    with   the    radial    velocities   obtained    in
\citet{rebassa-mansergasetal07-1} gives an average relative difference
of 19.5\%,  and the measurements generally overlap  within the errors.
The  reasons for  the small  changes  have been  already discussed  in
detail in  \citet{rebassa-mansergasetal08-1} and can  be summarised as
follows:  (1)  we  modified  the  procedure  to  fit  the  \Ion{Na}{I}
absorption doublet  by using  a single width  parameter for  both line
components; and (2) we used DR6 spectra here (instead of DR5).

As described  in the introduction, the orbital  period distribution of
WDMS  binaries is  expected  to be  bimodal,  separating long  orbital
period systems whose stellar  components evolve like single stars from
short  orbital  period  systems  that  suffered  from  mass  transfer
interactions,   mostly  CE  evolution   \citep{willems+kolb04-1}.   As
demonstrated   by  \citet{rebassa-mansergasetal07-1},   multiple  SDSS
spectroscopy of WDMS  binaries can be used to  measure radial velocity
variations thereby eventually identifying candidates for being a PCEB.
Our criterion  for calling a WDMS  binary a strong  PCEB candidate are
radial  velocity  variations  with  $3\sigma$  significance.  This  is
evaluated  by   calculating  the  $\chi^2$  of   the  measured  radial
velocities against their mean value.  The incomplete $\gamma$ function
then gives  the probability for  the measured radial  velocities being
consistent  with  a  constant  value.  If this  probability  is  below
$0.0027$  the  measured   radial  velocity  variations  are  $3\sigma$
significant  and we consider  the corresponding  WDMS binary  a strong
PCEB candidate.

We find here nine and 16 PCEB candidates using the \Ion{Na}{I} doublet
and H$\alpha$  emission respectively.   A comparison with  the results
obtained in \citet{rebassa-mansergasetal07-1} shows that four systems,
i.     e.     SDSS\,J030904.82-010100.9,    SDSS\,J113800.35-001144.5,
SDSS\,J173727.27+540352.2,  and  SDSS\,J234534.50-  001453.7  are  not
anymore  considered PCEBs,  while we  find five  new  PCEB candidates,
namely      SDSS\,J033301.51+005418.5,      SDSS\,J074329.62+283528.0,
SDSS\,J145300.99+005557.1        SDSS\,J231874.73+003403.3,        and
SDSS\,J234638.76+434041.7.  Apparently, the  differences in the method
of determining  radial velocities and/or the re-reduction  of the SDSS
data can cause a given system  to move either way across our criterion
\citep{rebassa-mansergasetal08-1}.   The \Ion{Na}{I}  absorption lines
from the  secondary star are a  more robust probe  for radial velocity
variations,  and we  consequently consider  that  additional follow-up
spectroscopy is  necessary for  those PCEB candidates  identified from
H$\alpha$ radial  velocity variations.  Upper  limits to the  new PCEB
candidate  orbital periods  have been  estimated  in the  same way  as
described  in  \citet{rebassa-mansergasetal07-1}.   These  values  are
given in Table\,\ref{t-newpcebs}.

\begin{table}
\begin{small}
\caption{\label{t-newpcebs} Upper limits to the orbital periods of the
five  PCEB candidates identified  in Sect.\,\ref{s-rvs}.   White dwarf
masses   are    taken   from   Table\,\ref{t-paramcat}    except   for
SDSS\,J2346+4340, where  we assume of mass of  0.5 M$_{\bigodot}$ (see
Fig.\,\ref{f-histocat}).   Secondary star  masses  are estimated  from
Table\,5   in   \citet{rebassa-mansergasetal07-1}.    K$_\mathrm{sec}$
values  are  obtained from  Table\,\ref{t-rvsall},  where  we use  the
\Ion{Na}{I}   doublet   radial   velocities   for   SDSS\,J2346+4340.}
\setlength{\tabcolsep}{1ex}
\begin{center}
\begin{tabular}{lccccc}
\hline
\hline
  SDSS\,J  &  0333(*) &  0743 &    1453 &  2318 &  2346 \\
           & +0054 & +2835 &  +0055 & +0034 & +4340  \\
\hline
  \Porb[d] &  575  &   121 &   8 &    5  &    2      \\
\end{tabular}
\end{center}
\end{small}
\begin{minipage}{\columnwidth}
(*) The effective temperature  is below 12000\,K, and consequently the
white dwarf mass is likely overestimated.
\end{minipage}
\end{table}

\section{Summary}
\label{s-conc}

We  have  presented  a  catalogue  of  1602  WDMS  binaries  from  the
spectroscopic  SDSS   DR6.   We  have   used  a  decomposition/fitting
technique  to measure the  effective temperatures,  surface gravities,
masses  and distances to  the white  dwarfs, as  well as  the spectral
types and distances to the companions in our catalogue.  Distributions
and density maps obtained from these stellar parameters have been used
to study both the general properties and the selection effects of WDMS
binaries in SDSS.  A comparison  between the distances measured to the
white    dwarfs    and   the    main    sequence   companions    shows
$d_\mathrm{sec}>d_\mathrm{wd}$ for $\sim$1/5  of the systems.  We also
have  made use  of  GALEX, SDSS  and  UKIDSS magnitudes  to study  the
distribution  of  WDMS binaries  in  colour-colour  space and  present
simple colour-cuts  that allow to clearly separate  WDMS binaries from
other stellar objects. Finally, we have measured radial velocities for
1068  WDMS binaries  measured from  the \Lines{Na}{I}{8183.27,8194.81}
absorption  doublet and/or  the  H$\alpha$ emission  line.  Among  the
systems  with  multiple  SDSS  spectroscopy,  we find  five  new  WDMS
binaries  showing significant  radial velocity  variations identifying
them as  PCEB candidates.  The  here presented new, updated,  and most
complete catalogue of WDMS binaries  from the SDSS represents a superb
data  base  for  future   follow-up  studies  that  may  significantly
contribute  to a  better understanding  of close  compact  binary star
evolution.

\section*{Acknowledgements.}

ARM acknowledges financial support from ESO, and Gemini/Conicyt in the
form of grant  number 32080023.  MRS thanks for  support from FONDECYT
(1061199).  We  thank the anonymous  referee for his  suggestions that
helped improving the quality of the paper. We also thank Pierre Maxted
for useful discussions.

\end{document}